\documentclass[preprint,12pt, authoryear]{elsarticle}
\usepackage{xr}
\usepackage[margin=1.0in]{geometry}
\usepackage{amsmath}
\usepackage{siunitx}
\usepackage{caption}
\usepackage{subcaption}
\usepackage{tcolorbox}
 \newcommand{\commentb}[1]{{\color{black}{#1}}}
 \newcommand{\commentc}[1]{{\color{black}{#1}}}
 \newcommand{\commente}[1]{{\color{black}{#1}}}
 \newcommand{\commentf}[1]{{\color{black}{#1}}}
 \newcommand{\commentg}[1]{{\color{black}{#1}}}
 \newcommand{\commenth}[1]{{\color{black}{#1}}}
    
\newcommand{\subfigimg}[3][,]{%
  \setbox1=\hbox{\includegraphics[#1]{#3}}
  \leavevmode\rlap{\usebox1}
  \rlap{\hspace*{-10pt}\raisebox{\dimexpr\ht1+0.2\baselineskip}{#2}}
  \phantom{\usebox1}
}
\usepackage{wrapfig}

\usepackage{ragged2e}
\usepackage{booktabs, multirow, tabularx}
\DeclareMathOperator*{\argmax}{arg\,max}

\usepackage{amssymb}
\usepackage{pdfpages}

\journal{}

\begin{document}

\begin{frontmatter}

\title{How Retroactivity Affects the Behavior of Incoherent Feed-Forward Loops}

\author{Junmin Wang\fnref{label1}}
\address{The Bioinformatics Graduate Program, Boston University, Boston, MA, USA}
\author{Calin Belta}
\address{The Bioinformatics Graduate Program, Boston University, Boston, MA, USA}
\author{Samuel A. Isaacson}
\address{Department of Mathematics and Statistics, Boston University, Boston, MA, USA}

\fntext[label1]{corresponding author, lead contact (email: dawang@bu.edu)}

\begin{abstract}
An incoherent feed-forward loop (IFFL) is a network motif known for its ability to accelerate responses and generate pulses. Though functions of IFFLs are well studied, \commente{most} previous computational analysis of IFFLs \commente{used} ordinary differential equation (ODE) models where retroactivity, the effect downstream binding sites exert on the dynamics of an upstream transcription factor (TF), was not considered.   It remains an open question to understand the behavior of IFFLs in contexts with high levels of retroactivity, e.g., \commente{in cells transformed/transfected with high-copy plasmids, or in} eukaryotic cells where a TF binds to numerous high-affinity binding sites \commente{in addition to one or more} functional target sites.  Here we study the behavior of IFFLs by simulating and comparing ODE models with different levels of retroactivity.  \commente{We find} that increasing retroactivity in an IFFL can increase, decrease, or keep the network's response time and pulse amplitude constant.  \commente{This suggests that increasing retroactivity, traditionally considered as an impediment to designing robust synthetic systems, could be exploited to improve the performance of IFFLs.}  We compare the behaviors of IFFLs  \commente{to} \commente{negative autoregulatory loops,} another sign-sensitive response-accelerating network motif, \commente{and find that} increasing retroactivity in a negative autoregulated circuit can only slow the response.  \commente{The inability of a negative autoregulatory loop to flexibly handle retroactivity may have contributed to its lower abundance in eukaryotic relative to bacterial regulatory networks, a sharp contrast to the significant abundance of IFFLs in both cell types.}  

\end{abstract}


\end{frontmatter}

\graphicspath{{./pics/}}
\section{INTRODUCTION}
\label{S:1}

Living cells sense and respond to the environment via a large \commente{variety} of mechanisms.  How do \commente{diverse} biochemical networks, which are at the core of the \commente{process by which cells sense and respond to signals}, yield and maintain specific functional behaviors? A widely held hypothesis in systems biology is that recurring network sub-structures, also known as network motifs, play important roles therein.  Network motifs capable of performing biological functions are preserved over the course of evolution, resulting in a rate of occurrence higher than \commente{if nodes and edges were connected at random (\cite{alon})}.

\begin{figure}[!ht]
\centering
\captionsetup{justification = justified}
\subfigimg[height=4cm]{\textbf{a)}}{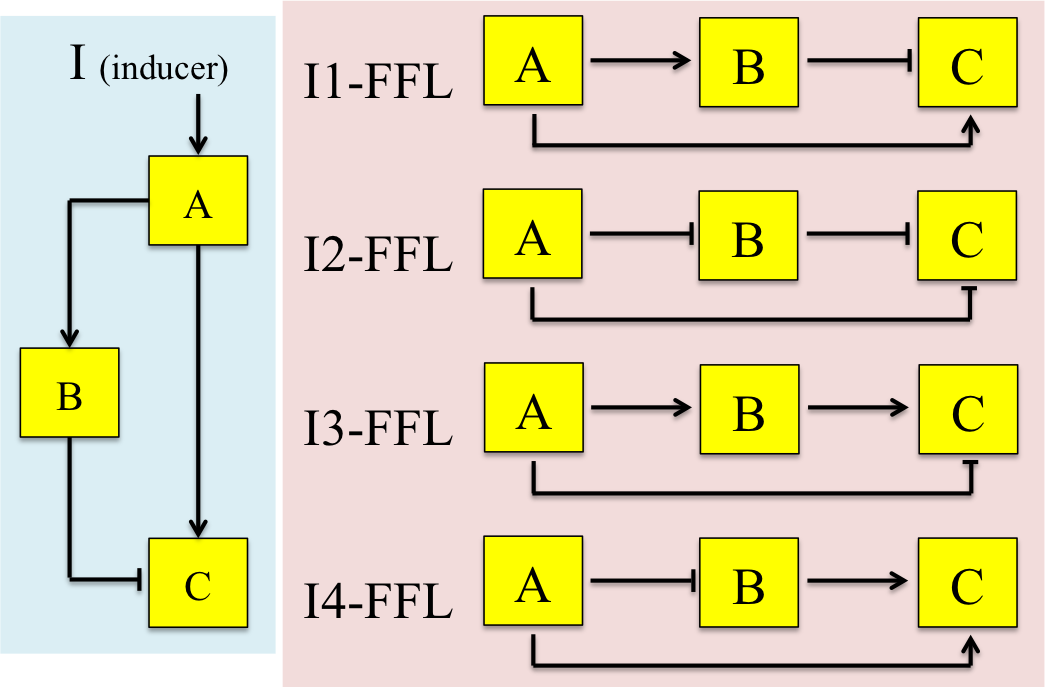} \hspace{0.9cm}
\subfigimg[height=4cm]{\textbf{b)}}{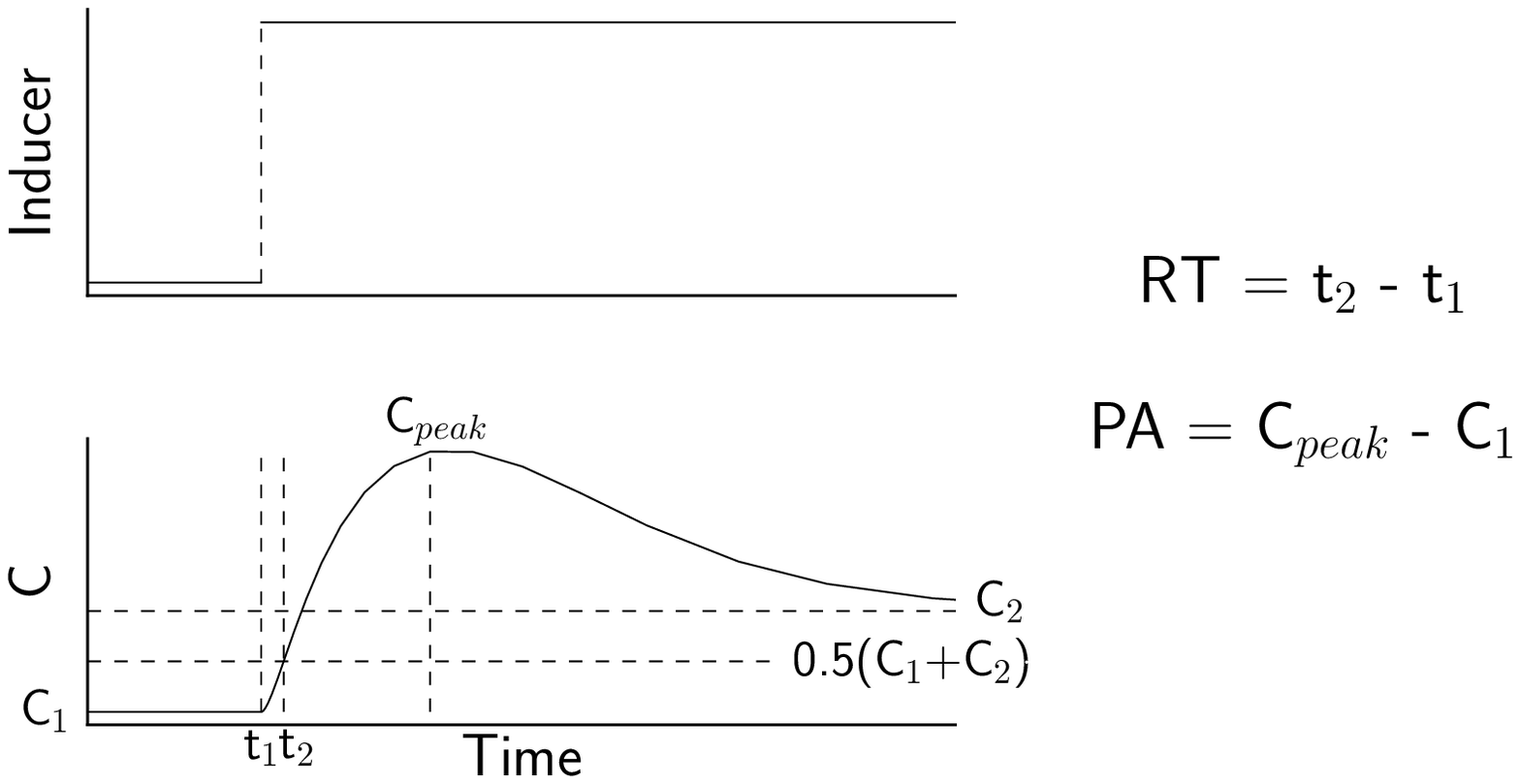}  \\ \vspace{0.3cm}
\subfigimg[height=5cm]{\textbf{c)}}{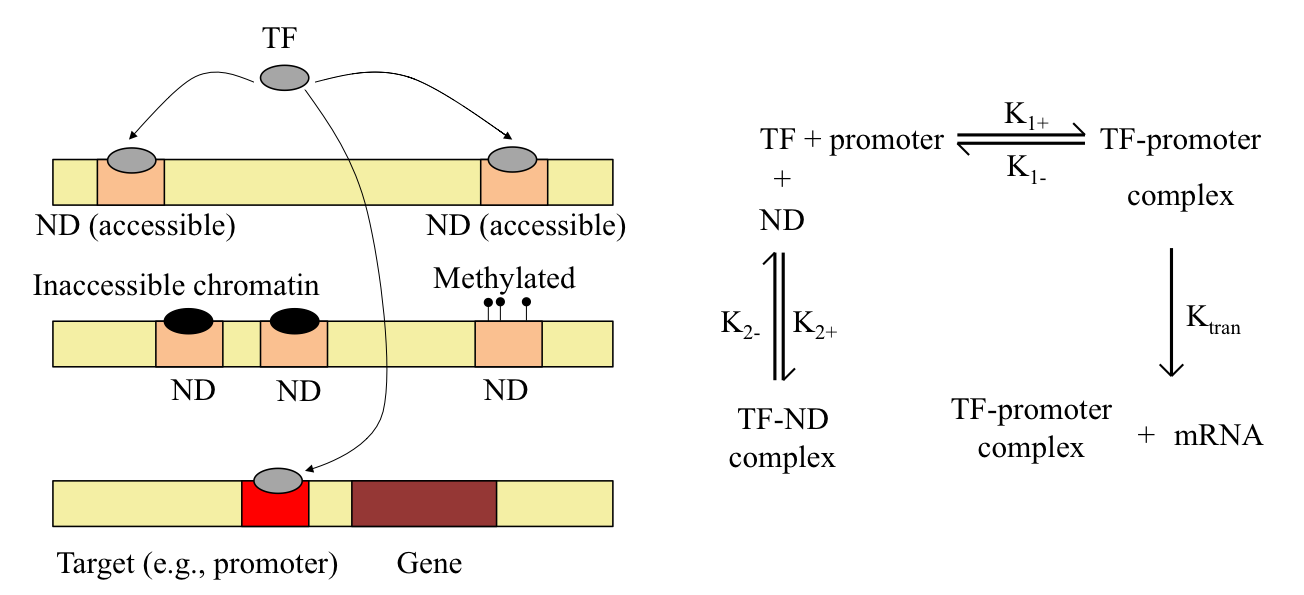}  \\ \vspace{0.3cm}

\caption{(a) Graphical representations of four types of IFFL: I1-FFL, I2-FLL, I3-FFL, and I4-FFL. An IFFL is a three-node network motif, where the input A, stimulated by an external inducer I, regulates the output C in two opposing directions.  Arrows indicate activation, and edges with bars at the ends, inhibition. (b) Definitions of response time and pulse amplitude.  Response time, abbreviated as RT, is defined as the time needed to reach the midpoint between the pre-induction and \commente{the post-induction} steady states (t$_2$-t$_1$), whereas pulse amplitude, abbreviated as PA, is defined as the difference between the pre-induction steady state and the peak concentration (C$_\text{peak}$-C$_1$). (c) \commentf{The effect of accessible ND binding sites on the dynamics of the TF}.   \commentf{Non-functional NDs} sequester some of the upstream TF, so only a fraction of the upstream TF molecules are available to bind to functional target sites (e.g., promoters and enhancers).}
\label{fig:definitions}
\end{figure}

One of the most common three-gene network motifs in transcriptional regulatory networks (TRN) is the incoherent feed-forward loop (IFFL), where a transcription factor (TF) activates and inhibits a downstream gene directly and indirectly (Figure \ref{fig:definitions}(a)).  In a pioneering study guided by ordinary differential equation (ODE) models, \cite{mangan} established IFFLs as a sign-sensitive response accelerator and pulse generator (Figure \ref{fig:definitions}(b)).  Subsequent efforts in synthetic biology supported the findings of \cite{mangan} with compelling experimental evidence.  Using the gal system in \textit{Escherichia coli} (\textit{E. coli}), \cite{alon} showed that compared to simple regulation, IFFLs can accelerate the response times of \commente{a} target gene.  \cite{basu} demonstrated the feasibility of creating synthetic pulse-generating IFFL circuits under the guidance of ODE models.  \commentf{In addition, IFFLs can provide fold-change detection and buffer noise (\cite{goentoro}). \cite{osella}, \cite{siciliano}, and \cite{grigolon} showed that miRNA-mediated IFFLs confer precision and stability to the target protein level despite fluctuations in upstream regulators.}

\commentc{Although a wealth of literature has shed light on this topic,} it remains an open area of research to understand the full functional capabilities of IFFLs.  In their ODE models,  \cite{mangan}, as well as \cite{basu}, made the simplifying assumption that changes in protein concentrations arise from first-order decay and protein production rates \commentb{regulated by upstream TFs. This assumption aligns with the traditional view of TRNs as modular systems, where the temporal dynamics of a protein depend solely upon the TFs that regulate its expression.  In other words, under this assumption, the dynamics of the protein \commente{are} not affected by the components it regulates even if the protein is also a TF.}  However, growing theoretical and experimental evidence suggests that TRNs \commentb{are not modular but quasi-modular}.  A fraction of the TF molecules are employed to form complexes with downstream binding sites, hence becoming unavailable for \commentb{additional} molecular activities, such as degradation, protein-protein interaction, or regulation of other genes.  Examples of such TFs include p53 (\cite{pariat}) and MyoD (\cite{abu}), both of which become resistant to degradation when bound to DNA.  This phenomenon, where downstream binding sites can alter the dynamics of the upstream system, is known as retroactivity (\cite{delvecchio}). 

 In TRNs, retroactivity is large when the amount of TF is comparable to, or smaller than, the copy number of the downstream bindings sites, or when the affinity of such binding is high (\cite{delvecchio}).  \commente{In synthetic biology, retroactivity is widely recognized as an essential parameter to consider in model-based circuit design (\cite{brophy}).} \commente{In the context of endogenous regulatory networks, retroactivity is seldom discussed,} as the level of retroactivity that arises from TF binding in the genome is typically assumed to be negligible (\cite{jayanthi}).  However, results from ChIP-on-chip and ChIP-seq methods suggest that the validity of this assumption is dependent on the biological context of the network \commente{(\cite{kemme})}.  In particular, genome-wide studies driven by the Encyclopedia of DNA Elements (ENCODE) project have shown that in eukaryotic cells, TFs bind to not only functional sites in the cis-regulatory elements (e.g., promoters and enhancers) but also numerous high-affinity sequence-specific binding sites that are seemingly non-functional (\cite{encode, fisher1, li-xiao-yong}) (Figure \ref{fig:definitions}(c)).  It has been suggested that these high-affinity sequence-specific binding sites can serve as natural decoys (NDs), which compete with functional target sites for TF binding (\cite{burger2010, burger2012, maheshri, liu, zhipeng}).  While the majority of ND sites are inaccessible due to chromatin structure, CpG methylation, or competing proteins, an average TF in the human genome still has approximately $10^4 - 10^5$ accessible ND sites, which typically \commente{have} greater or at least comparable \commente{binding} affinity \commente{compared to} sequence-specific TF binding \commente{sites} (\cite{kemme, esadze, kemme2015}) (Figure \ref{fig:definitions}(c)).  \commente{As such, in studying many eukaryotic TRNs retroactivity must be taken into account (\cite{kemme}).}  

\commente{The goal of our study is} to understand how retroactivity affects response acceleration and pulsing of IFFLs.  \commente{In the simplest case where an input is coupled to a downstream promoter binding region, \cite{delvecchio} demonstrated that retroactivity increases response times and dampens pulse amplitude (Figure \ref{fig:definitions}(d)).  In the context of more complicated topologies, changing retroactivity can lead to more sophisticated, often undesired effects on circuit behaviors (\cite{sepulchre, gyorgy, wang})}.  \commente{This raises the question whether retroactivity is simply an impediment to overcome in designing synthetic IFFL circuits.  
Another natural question is the potential role of retroactivity in motif evolution.  As the levels of retroactivity differ sharply in prokaryotes and \commente{higher} eukaryotes due to the number of accessible ND sites, could the behaviors of a network motif under different levels of retroactivity have affected its abundance, as one progresses from bacterial TRNs to eukaryotic ones?}
\commentc{\commente{Note, we focus on IFFLs in particular because} synthesizing functional IFFLs has proven to be experimentally feasible (\cite{basu, bleris}), \commente{making our predictions experimentally testable in controlled synthetic systems.}} 

\cite{gyorgy} developed a systematic modeling framework that accounts for retroactivity in TRNs. Using this framework, we study IFFL networks by simulating, comparing, \commentg{and mathematically analyzing} ODE models with varying levels of retroactivity.  Similar to previous computational studies (\cite{shi, ma, castillo}), we performed time course simulations of IFFLs repeatedly with kinetic parameters representing different regions of parameter space.  \commente{We} quantified the response time, as well as the pulse amplitude, for each parameter set (see Figure \ref{fig:definitions}(b) for the definitions of response time and pulse amplitude).  \commente{Building from these} simulations, we compared the dynamics of the corresponding ODE systems in order to understand how retroactivity affects the behavior of IFFLs. \commentg{To demonstrate that our findings are parameter-independent, we carried out mathematical proofs where model parameters can take arbitrary positive values.} 

We find that increasing retroactivity can \commente{increase, decrease,} or keep the response time \commente{and the pulse amplitude} constant in an IFFL.  \commente{This suggests} that \commente{in contrast to the traditional perception of retroactivity as an impediment to circuit design (\cite{delvecchio})}, \commente{increasing} retroactivity could actually be \commente{harnessed} to improve the performance of IFFLs.  \commente{Our results predict that the introduction of synthetic decoy binding sites into a synthetic IFFL system would affect its response time and pulse amplitude, and the magnitude of this effect would depend on kinetic parameters (e.g., Hill coefficients) and circuit topologies (e.g. I1-FFLs).  Hence, retroactivity should be considered in connection with circuit parts to optimize the behavior of IFFL circuits.}
\commente{Our observations of IFFLs led us to examine a few other motifs capable of sign-sensitive response-acceleration.} Comparing the behavior of IFFLs to that of negative autoregulation, we found that increasing retroactivity in a negative autoregulated circuit can only decelerate the response. Interestingly, we \commente{observed} that IFFLs are conserved in bacteria, mouse, and human networks, whereas negative autoregulatory loops \commente{are only present in significant numbers} in bacteria.
The functional versatility of IFFLs at increasing levels of retroactivity, thus, may have provided IFFLs a selective advantage over negative autoregulation in cases where decreasing or keeping the response time constant \commente{was beneficial}.  

\section{RESULTS}
\subsection{Modeling \commentf{Transcriptional Regulatory Networks}}

\commentg{In this section, we describe our approach to modeling the effect of retroactivity on TRNs.}  A TRN can be mapped to a graph, where each node represents a gene/protein, each edge transcriptional regulation, and the direction of an edge the direction of the regulation; activation or inhibition.  The time evolution of each node can be described by an ODE, where the time derivative represents the rate of change of the protein concentration contributed by protein production and first-order decay.  Mathematically, the rates of changes of proteins in the network can be expressed as:
\begin{align}
\frac{d \vec{x}}{dt} &= h(\vec{x}),
\end{align}
where 
\begin{align}
h(\vec{x}) = \left(
\begin{array}{c}
 \beta_1 \cdot \left[ \left(1 - \gamma_1  \right) H_1(\vec{p_1}) + \gamma_1\right]- \delta_1 x_1 \\
\beta_2 \cdot \left[ \left(1 - \gamma_2  \right) H_2(\vec{p_2}) + \gamma_2 \right]- \delta_2 x_2 \\
... \\
\beta_n \cdot \left[\left(1 - \gamma_n \right) H_n(\vec{p_n}) + \gamma_n \right] - \delta_n x_n 
\end{array} \right),
\label{eq:dynamics}
\end{align}
where $x_i$ denotes the concentration of the $i$-th protein, and $\delta_i$, the decay rate.  $\vec{p_i}$, the concentration of the parent(s) of the $i$-th protein, is a subset of $\vec{x}$.   $\beta_i$ represents the maximal production rate of the $i$-th protein, and $\gamma_i$, the basal fraction of the promoter that is active.  $H_i$ is the Hill function describing the transcriptional regulation of x$_i$ by its parent(s).  If the $i$-th protein species x$_i$ has only one parent species p$_i$, then Hill function $H_i(p_i)$ can be expressed as:
\begin{equation}
  H_i(p_i) = \left \{
  \begin{aligned}
    & \frac{1}{1 + \left(\frac{p_i}{K_i}\right)^{h_i}}, && \text{if species p$_i$ is an inhibitor }  \\
    & \frac{\left(\frac{p_i}{K_i}\right)^{h_i}}{1 + \left(\frac{p_i}{K_i}\right)^{h_i}}, && \text{if species p$_i$ is an activator, }  \\   
  \end{aligned} \right.
  \label{eq:hill}
\end{equation}  
where $H_i(p_i)$ accounts for the fraction of the promoter that is active, $K_i$ is the dissociation constant, and $h_i$ is the Hill coefficient.  Co-regulation by multiple TFs can be modeled by simple logic models.  Unless otherwise specified,  throughout this work we consider an AND logic, where the regulated gene is turned on only when all activators are abundant and all inhibitors are scarce \commentf{(see Supplemental Information Section \ref{suppsec:hill} for Hill functions describing co-regulation)}. 

\commentf{To account for retroactivity, we adopt the framework developed by \cite{gyorgy}. The major assumptions needed to apply this framework are that 1) there is a separation of time \commentg{scales} between protein production/degradation and  reversible binding reactions between TFs and DNA, and 2) the corresponding quasi-steady state is locally exponentially stable (\cite{gyorgy}).  The first assumption is valid \commentg{as} protein turnover and binding reactions typically occur on different time scales \commentg{(\cite{milo})}.  The second assumption is implicit in our use of the Hill-function-based models, \commentg{and its validity is explained in} \cite{gyorgy}.}  Under these assumptions, the rates of changes of protein concentrations with retroactivity considered can be described as:
\begin{align}
\frac{d \vec{x}}{dt} &= \left[I+R(\vec{x})\right]^{-1} h(\vec{x}),
\end{align}
where $R(\vec{x})$, known as the retroactivity matrix (\cite{gyorgy}), can be calculated as:
\begin{align}
R(\vec{x}) = \left \{
\begin{aligned}
&\Sigma_{i | x_i \in \Phi} V_i^T R_i (\vec{p_i}) V_i &\text{  if $\Phi \neq \phi$}, \\
&0_{N \times N} &\text{  if $\Phi = \phi$}.
\end{aligned}\right.
\end{align}
Here $V_i$ is a binary matrix, containing as many rows as the length of $\vec{p_i}$ and as many columns as the number of nodes in the network. The element in the $j$-th row and $k$-th column of $V_i$ is 1 if the $j$-th parent of node $i$ is node $k$, and 0 otherwise.  AND logic is a special case of independent binding, in which case $R_i(\vec{p_i})$ is a diagonal matrix (see Supplemental Information Section \ref{suppsec:retro_matrix} for calculation of $R_i(\vec{p_i})$).  This in turn implies that $V_i^T R_i(\vec{p_i}) V_i$ is also a diagonal matrix (Supplemental Information Section \ref{suppsec:diagonal}). Hence, $R(\vec{x})$ is also diagonal. More details about retroactivity, including its derivation, can be found in \cite{gyorgy}.  Models of IFFL networks with and without retroactivity are given in Supplemental Information Sections \ref{suppsec:iffl_models}  and \ref{suppsec:non_dimensionalized_iffl_models}.  

\subsection{Simulation of IFFLs}
\commentg{In this section, we describe the specific IFFL models in which we study the effect of retroactivity, and outline our simulation protocol.} 
IFFLs are known to be sign-sensitive response accelerators and pulse generators: they accelerate or delay responses to stimulus steps only in one direction (\cite{alon, mangan}).  Considering sign-sensitivity of IFFLs, we separated four IFFL motifs into two groups, one group (i.e., I1-FFL and I4-FFL) capable of response acceleration and pulse generation in response to an ON step (i.e., inducer level $x_I$ changes from 0 to $\infty$) and the other (i.e., I2-FFL and I3-FFL) \commente{capable of response acceleration and pulse generation in response to} an OFF step (i.e., inducer level $x_I$ changes from $\infty$ to 0).  Here we focused on I1-FFLs and I4-FFLs, as similar analysis could be performed for I2-FFLs and I3-FFLs.   We constructed non-dimensionalized ODE models for I1-FFLs and I4-FFLs (details of non-dimensionalization can be found in Supplemental Information Section \ref{suppsec:non_dimensionalization}), and simulated each model using the DifferentialEquations.jl package version 5.3.1 in Julia version 1.1.0 (\cite{rackauckas, bezanson}).  We connected genes A, B, and/or C of the IFFL to additional downstream binding sites denoted by D$_X$ (X=A, B, or C).  The degree of retroactivity arising from additional downstream binding sites was allowed to vary, with the retroactivity coefficient $\tilde{\eta}_{AD_A}$ ($\tilde{\eta}_{BD_B}$, $\tilde{\eta}_{CD_C}$) set to 0, 1.0, 10.0, and 100.0 (see Supplemental Information Section \ref{suppsec:non_dimensionalization} for definition of $\tilde{\eta}_{XD_X}$ (X=A, B, or C)).  By contrast, we assumed that genes A, B, and C themselves are single-copy genes, and hence, retroactivity that arises from binding of A, B, or C to the functional target site(s) (e.g., promoter that controls the expression of B and C) is negligible.  Note, a model without retroactivity is equivalent to a model where $\tilde{\eta}_{XD_X}$ equals zero. 

As an example, the non-dimensionalized model of an I1-FFL (Figure \ref{fig:definitions}(a)) without retroactivity is given here:

\begin{align}
\begin{aligned}
\frac{d\tilde{x}_A}{d\tau} &=f_{\tilde{A}} = \left( 1 - \gamma_A \right) \frac{\left( \frac{x_I}{K_{IA}} \right)^{h_{IA}}}{1 + \left( \frac{x_I}{K_{IA}} \right)^{h_{IA}}} + \gamma_A - \tilde{x}_A \\
\frac{d\tilde{x}_B}{d\tau} &=f_{\tilde{B}} = \left( 1 - \gamma_B \right) \frac{\left( \frac{\tilde{x}_A}{\tilde{K}_{AB}} \right)^{h_{AB}}}{1 + \left( \frac{\tilde{x}_A}{\tilde{K}_{AB}} \right)^{h_{AB}}} + \gamma_B - \tilde{x}_B \\
\frac{d\tilde{x}_C}{d\tau}  &=f_{\tilde{C}} = \left( 1 - \gamma_C \right) \frac{ \left(\frac{\tilde{x}_A}{\tilde{K}_{AC}}\right)^{h_{AC}}}{\left(1 + \left( \frac{\tilde{x}_A}{\tilde{K}_{AC}} \right)^{h_{AC}} \right) \left( 1 + \left( \frac{\tilde{x}_B}{\tilde{K}_{BC}} \right)^{h_{BC}} \right)} + \gamma_C - \tilde{x}_C.
\end{aligned}
\label{no_retro_eq}
\end{align}

With retroactivity applied on all three nodes, the non-dimensionalized model of an I1-FFL becomes:
\begin{align}
\begin{bmatrix} 
\frac{d\tilde{x}_A}{d\tau} \\[5pt]
\frac{d\tilde{x}_B}{d\tau} \\[5pt]
\frac{d\tilde{x}_C}{d\tau}
\end{bmatrix}
= 
\begin{bmatrix}
\frac{1}{a} & 0 & 0 \\[5pt]
0 & \frac{1}{b} & 0 \\[5pt]
0 & 0 & \frac{1}{c}
\end{bmatrix}
\begin{bmatrix}
f_{\tilde{A}} \\[5pt]
f_{\tilde{B}} \\[5pt]
f_{\tilde{C}}
\end{bmatrix} 
=
\begin{bmatrix}
\frac{1}{r_{AD_A} + 1} & 0 & 0 \\[5pt]
0 & \frac{1}{r_{BD_B} + 1} & 0 \\[5pt]
0 & 0 & \frac{1}{r_{CD_C} + 1} 
\end{bmatrix}
\begin{bmatrix}
f_{\tilde{A}} \\[5pt]
f_{\tilde{B}} \\[5pt]
f_{\tilde{C}}
\end{bmatrix},
\label{retro_eq}
\end{align}
where 
\begin{align}
r_{AD_A} &= \tilde{\eta}_{AD_A}h_{AD_A}^2\left(\frac{\tilde{x}_A}{\tilde{K}_{AD_A}} \right)^{h_{AD_A}-1} \left(1 + \left(\frac{\tilde{x}_A}{\tilde{K}_{AD_A}} \right)^{h_{AD_A}} \right)^{-2} \nonumber \\
r_{BD_B} &= \tilde{\eta}_{BD_B}h_{BD_B}^2\left(\frac{\tilde{x}_B}{\tilde{K}_{BD_B}} \right)^{h_{BD_B}-1} \left(1 + \left(\frac{\tilde{x}_B}{\tilde{K}_{BD_B}} \right)^{h_{BD_B}} \right)^{-2}  \label{reduction} \\
r_{CD_C} &= \tilde{\eta}_{CD_C}h_{CD_C}^2\left(\frac{\tilde{x}_C}{\tilde{K}_{CD_C}} \right)^{h_{CD_C}-1} \left(1 + \left(\frac{\tilde{x}_C}{\tilde{K}_{CD_C}} \right)^{h_{CD_C}} \right)^{-2}. \nonumber
\end{align} 

In Equations (\ref{no_retro_eq}) and (\ref{retro_eq}), $\tilde{x}_A$, $\tilde{x}_B$, and $\tilde{x}_C$ are the nondimensionalized concentrations of proteins A, B, and C, whereas $\tau$ is the nondimensionalized time.  $f_{\tilde{A}}$, $f_{\tilde{B}}$, and $f_{\tilde{C}}$ are the sums of regulated protein production and protein decay (Supplemental Information Section \ref{suppsec:iffl_models}).  $a$, $b$, and $c$, defined as the reduction factors of $\frac{d\tilde{x}_A}{d\tau}$, $\frac{d\tilde{x}_B}{d\tau}$, and $\frac{d\tilde{x}_C}{d\tau}$ due to retroactivity, are equal to 1 if retroactivity is not considered.  \commentc{Note that for I4-FFLs, \textit{the only changes in Equations (\ref{no_retro_eq}) and (\ref{retro_eq})  are in the definitions of $f_{\tilde{A}}$, $f_{\tilde{B}}$, and $f_{\tilde{C}}$ \commente{due to the different regulatory interactions}, i.e., $r_{AD_A}$, $r_{BD_B}$, and $r_{CD_C}$ are still given by the same Equation (\ref{reduction})}.  Note also that \textit{retroactivity does not affect steady-state values of $\tilde{x}_A$, $\tilde{x}_B$, and $\tilde{x}_C$}.}

In terms of model simulation, we selected parameters based on values chosen by \cite{mangan}, exploring \commente{several orders of magnitude} of parameter space.  \commente{Specifically we considered Hill coefficients $h_i$ less than, equal to, and larger than 1 (Equation (\ref{eq:hill})).  If $h_i$ is non-integer, then the underlying reaction between the promoter and the TF is likely the resultant of several mechanisms, such as chain reactions (\cite{boekel}).  In this scenario, $h_i$, which is also the reaction order, can be considered an approximation of the detailed mechanisms (\cite{boekel}).  An $h_i$ larger than, equal to, and less than 1 stands for positive, zero, and negative cooperativity, respectively.  }Details of the parameters can be found in Supplemental Information Section \ref{suppsec:non_dimensionalization}.  

\commentc{In the absence of regulatory interactions, we assumed that only \commentb{the expression of} gene A is \commentb{modulated} by an external inducer while genes B and C are constitutively expressed}.  \commentb{We \commente{initialize our models} at a steady state corresponding to a fixed inducer concentration, and \commentc{\commente{subsequently induce} changes in concentrations of proteins A, B, and C} via a sudden increase in the inducer's concentration.}  \commentf{In the case of an ON (OFF) step, the inducer level $x_I$ changes from 0 ($\infty$) to $\infty$ (0).  By integrating the ODEs \commentf{until} \commente{solutions reached a new} steady state, we obtained one trajectory of proteins A, B, and C for each set of kinetic parameters we sampled.}  
  
\subsection{Retroactivity Changes Behaviors of IFFLs}
\label{sec:param}

\commentc{\commentg{We begin by studying how varying levels of retroactivity on just one gene of an IFFL can alter its behaviors.  That is,} we allowed retroactivity on one and only one gene of the IFFL to vary, \commente{keeping retroactivity on the rest of the genes equal to zero}.  The response time of gene C was \commente{then} calculated for each parameter set at each different level of retroactivity.  \commente{Our} results \commentg{show} that changing retroactivity on each node has different effects on response times, as each node of the IFFL serves a different function (Supplemental Information Sections \ref{suppsec:rt_retro} and \ref{suppsec:rt_retro_A}).  While increasing $\tilde{\eta}_{CD_C}$ expectedly slows the response \commente{time} of gene C (Table \ref{supptable:rt_C}), we observed \commente{that the response time of gene C decreases as $\tilde{\eta}_{BD_B}$ increases, most notably for $h_{BD_B} \leq 1$ (Figure \ref{fig:i1_dimension}(a); see Table \ref{supptable:rt_B} for data). To generalize our observation, we proved that increasing $\tilde{\eta}_{BD_B}$ shortens the response time of gene C regardless of the values of \commentg{any} other parameters for I1-FFLs} (see Supplemental Information Section \ref{suppsec:retro_proof} for the mathematical proof).  Serving as the regulatory node in the network, gene B controls the time gap between the opposing forces of regulation exerted on gene C.  In response to an ON step, the expression level of protein B monotonically increases.  Increasing the level of retroactivity $\tilde{\eta}_{BD_B}$ \commente{in turn slows the approach of B to steady state}. \commente{Consequently}, it takes protein B a longer time to effectively repress gene C, allowing protein C to reach the half point over a shorter period of time (Figure \ref{fig:i1_dimension}(a)).  Thus, we find that increasing $\tilde{\eta}_{BD_B}$ shortens the response time of gene C.  \commente{As Table \ref{supptable:rt_B} indicates, the magnitude by which the response time shortens depends on $h_{BD_B}$ as well as the IFFL topology.  A detailed discussion of the underlying association can be found in Section \ref{sec:i1_more_common}.}

\begin{figure}[!ht]
\centering
\subfigimg[height=80mm]{\textbf{a)}}{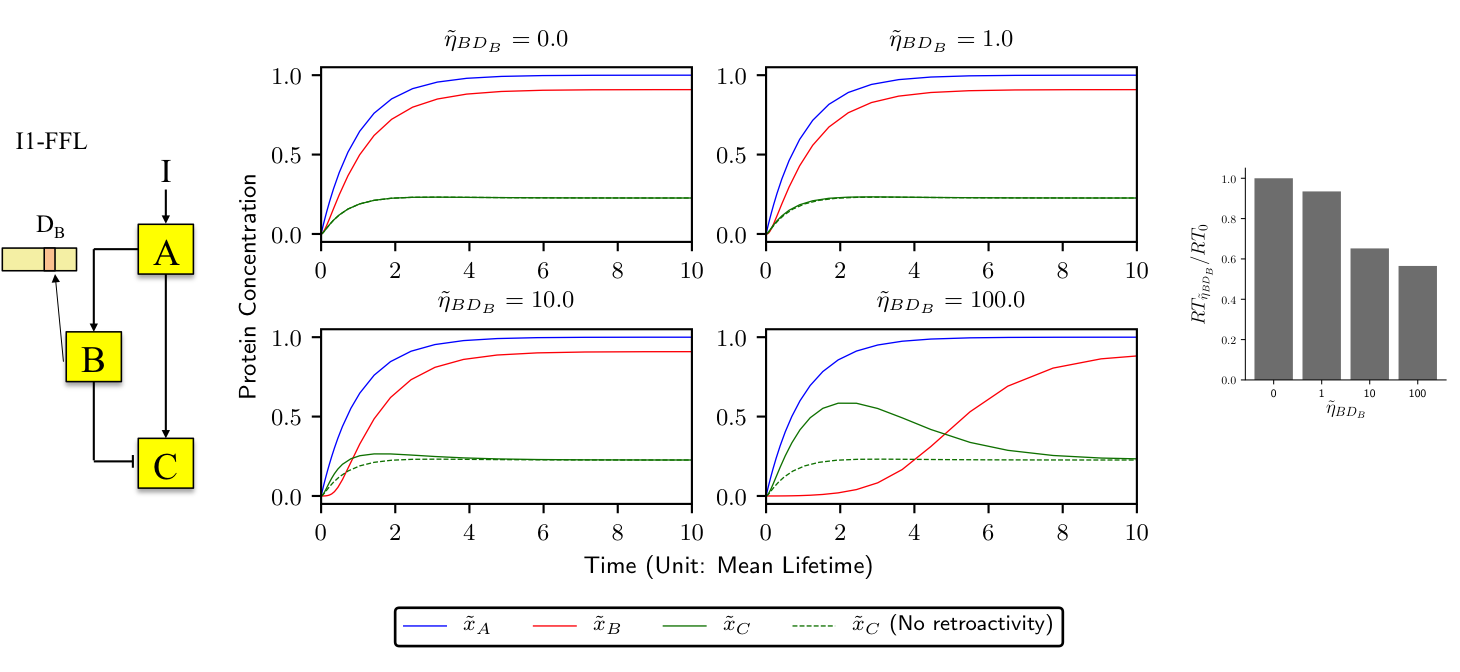} \\
\subfigimg[height=80mm]{\textbf{b)}}{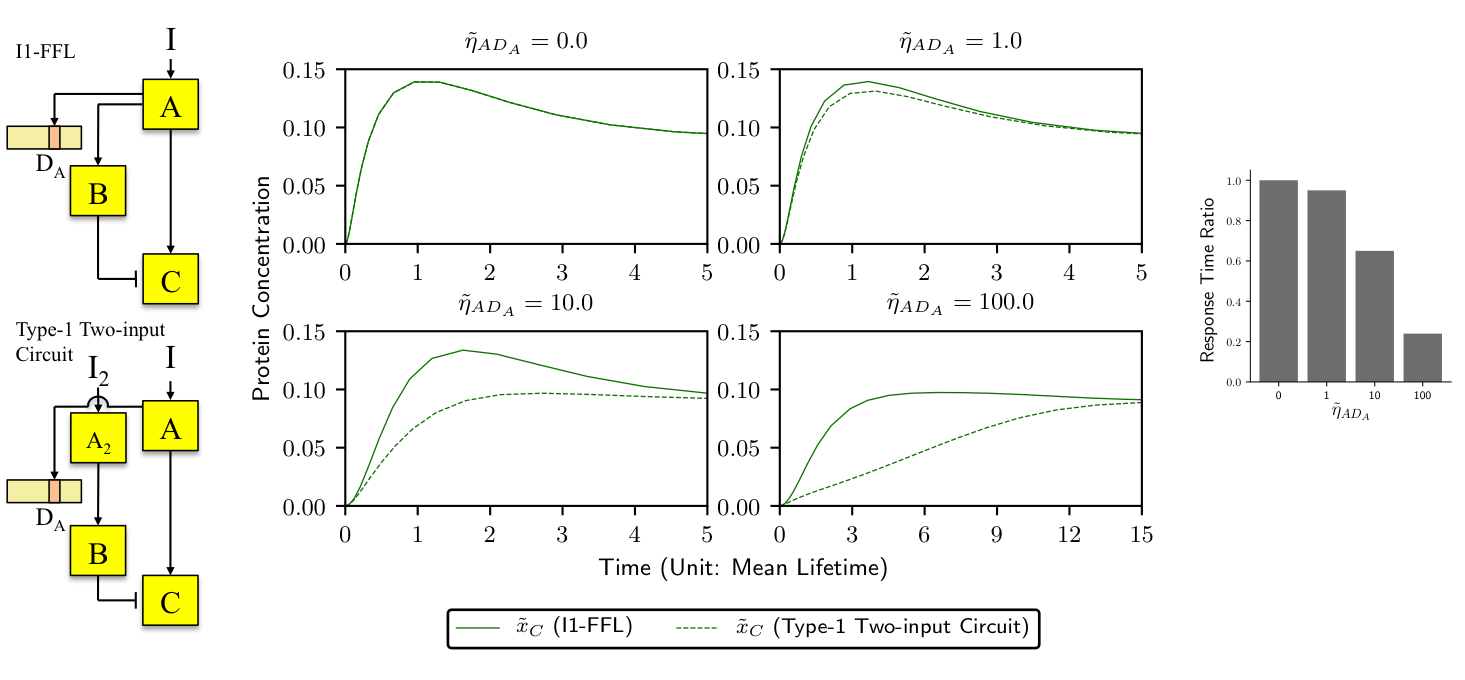}
\caption{(a) Shortened response time due to increasing $\tilde{\eta}_{BD_B}$ in an I1-FFL. $\tilde{\eta}_{BD_B}$ increases in the order of top left, top right, bottom left, and bottom right.  \commente{Values of the other parameters are: $\tilde{K}_{AB} = \tilde{K}_{AC} = \tilde{K}_{BC} = \tilde{K}_{BD_B} = 0.1$, $h_{AB}  = h_{AC} = 1.0$, $h_{BC} = h_{BD_B} = 0.5$.}  For comparison, the green dashed curve represents the trajectory of $\tilde{x}_C$ when $\tilde{\eta}_{BD_B}$ equals 0.  \commente{The bar plot shows the response time for different $\tilde{\eta}_{BD_B}$ compared to the response time without retroactivity.}  (b) Shorter response time in an \commente{I1-FFL} than in a type-1 two-input circuit at different levels of $\tilde{\eta}_{AD_A}$.  \commente{Values of the parameters are: $\tilde{K}_{AB} = \tilde{K}_{A_2B} = \tilde{K}_{AC} = \tilde{K}_{AD_A} = \tilde{K}_{BC} = 0.1$, $h_{AB} = h_{A_2B} = h_{AC} = h_{AD_A} = h_{BC}  = 1.0$.}  \commente{The bar plot shows the ratio of the response time in an I1-FFL to the response time in a type-1 two-input circuit.} \label{fig:i1_dimension}}
\end{figure}

\commentc{When the repressor (activator) B has a strong inhibitory (activating) effect on the production of the target protein, the dynamics of C exhibit a pulse-like shape (\cite{alon}). In addition to response times, we examined how retroactivity affects pulse amplitude, \commentg{when} for a given set of parameters the IFFL generates a pulse.  Increasing $\tilde{\eta}_{CD_C}$ expectedly slows down the response of gene C, resulting in a lower pulse amplitude for all parameters (Table \ref{supptable:pa_C}).}  In contrast, we observed and \commente{subsequently} proved that increasing $\tilde{\eta}_{BD_B}$ always increases the pulse amplitude (Figure \ref{suppfig:i1_pa_dimension}; see Table \ref{supptable:pa_B} for data and Supplemental Information Section \ref{suppsec:retro_proof} for the proof).  The underlying mechanism can \commentg{again} be traced back to the delayed response of $\tilde{x}_B$ due to increased $\tilde{\eta}_{BD_B}$.  While a decreased \commente{initial rate of growth of B} shortens the response time of gene C, it also causes protein B to take a longer time to effectively repress gene C, allowing protein C to develop a larger response over time (Figure \ref{suppfig:i1_pa_dimension}).  \commentf{In Supplemental Information Section \ref{suppsec:inter_retro_proof}, we extend our analysis by exploring the behavior of an IFFL when \commentg{it is embedded in a larger network} and node B serves as \commentg{an} input to other circuits, \commentg{investigating} the effect of intermodular retroactivity on IFFL behaviors.  Similar \commentg{to} before, we \commentg{find} that increasing intermodular retroactivity on node B decreases the response time and increases the pulse amplitude of node C.}

As the input node of the IFFL, gene A regulates gene C in opposing directions.  Our \commentg{simulations show} that changing $\tilde{\eta}_{AD_A}$ affects the response time and pulse amplitude of gene C in a more \commente{complicated} manner than changing $\tilde{\eta}_{BD_B}$ and $\tilde{\eta}_{CD_C}$.  While increasing $\tilde{\eta}_{AD_A}$ slows down the direct activation of gene C, it counteracts \commente{this} delay by decelerating the activation of gene B, thus attenuating the inhibition of C by B and allowing C a longer time to develop a response.  To demonstrate the counteracting effects, we compared the response time of an IFFL to that of a two-input circuit under different levels of $\tilde{\eta}_{AD_A}$.  In a two-input circuit, gene C is simultaneously activated by gene A, which is induced by inducer I, and inhibited by gene A$_2$, which is induced by a separate inducer I$_2$ (Figure \ref{fig:i1_dimension}(b)).  To facilitate a meaningful comparison between an IFFL and a two-input circuit, we assumed that genes A and A$_2$ have the same production rates upon induction\commente{,} but only allowed retroactivity of gene A (not A$_2$) to vary (see Supplemental Information Section \ref{suppsec:non_dimensionalized_other_models} for the model).  \commentg{We find} that because of the counteracting effects, increasing $\tilde{\eta}_{AD_A}$ leads to a smaller increase in response time and a smaller decrease in pulse amplitude in an IFFL than in a two-input circuit where gene A regulates C with no feed-forward mechanism (see Tables \ref{supptable:Type_I_rt_A}, \ref{supptable:Type_IV_rt_A}, \ref{supptable:Type_I_pa_A}, and \ref{supptable:Type_IV_pa_A} for data).}

\begin{figure}[t]
\centering
\subfigimg[width=50mm]{\textbf{a)}}{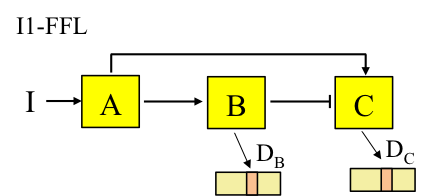}  \hspace{5mm}  \quad
\subfigimg[width=45.13mm]{\textbf{b)}}{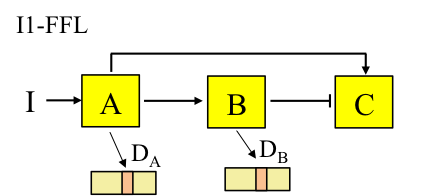}  \hspace{3mm}  \quad
\subfigimg[width=45mm]{\textbf{c)}}{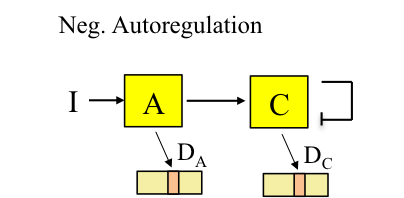} \\
\subfigimg[width=50mm]{}{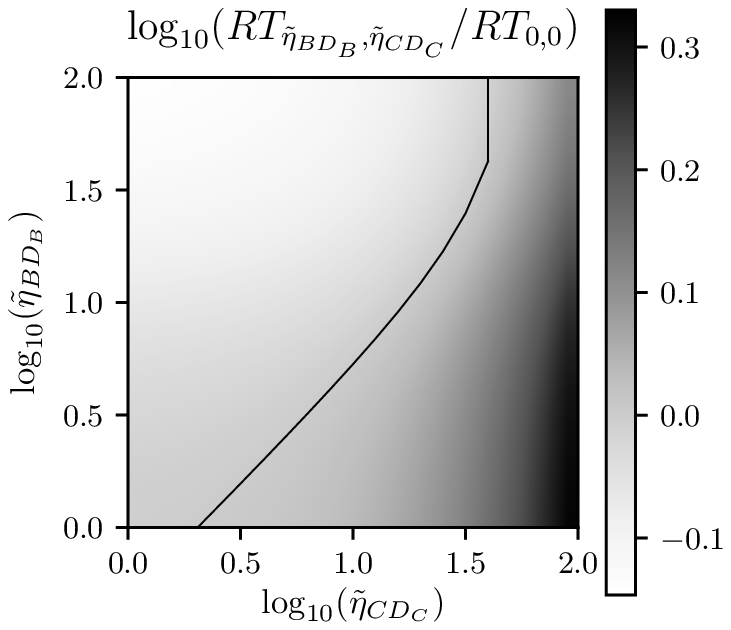} \quad
\subfigimg[width=50mm]{}{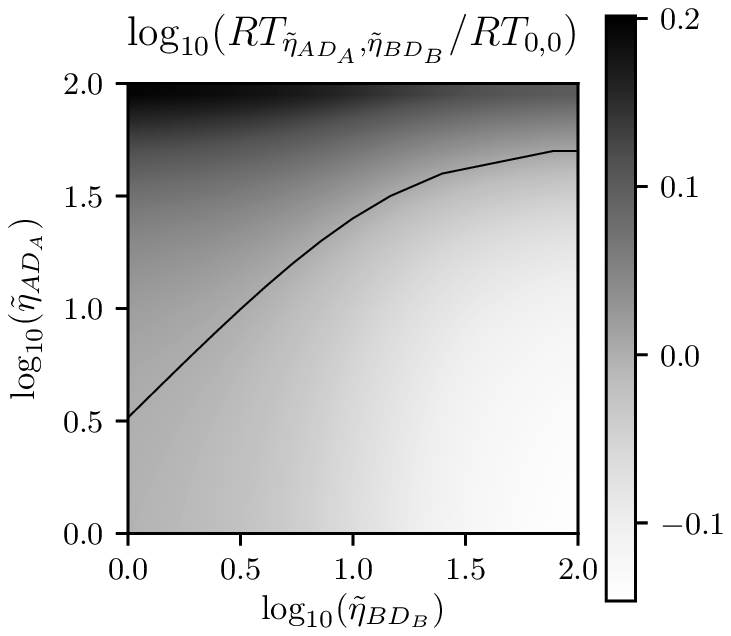} \quad
\subfigimg[width=50mm]{}{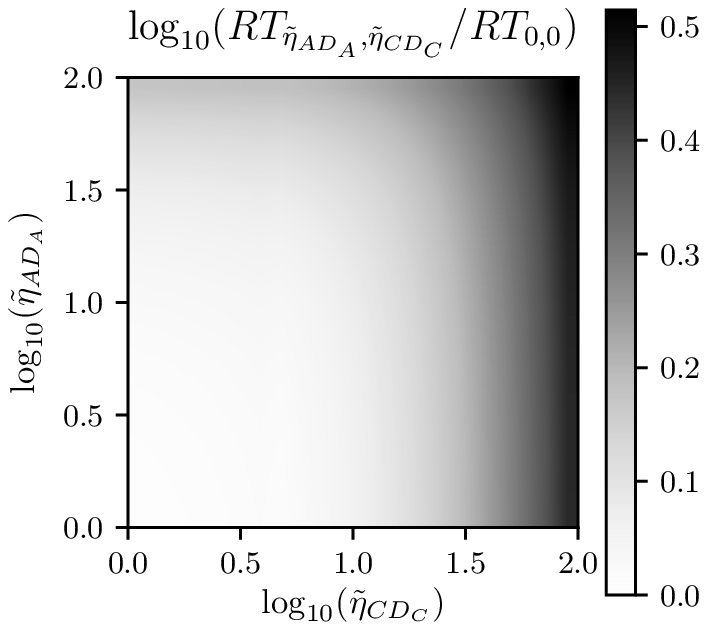}
\caption{(a) Response times of the I1-FFL model at different levels of $\tilde{\eta}_{BD_B}$ and  $\tilde{\eta}_{CD_C}$ compared to that of the model with no retroactivity.  (b) Response times of the I1-FFL model at different levels of $\tilde{\eta}_{AD_A}$ and  $\tilde{\eta}_{BD_B}$ compared to that of the model with no retroactivity.  \commente{(c) Response times of the negative autoregulated circuit model at different levels of $\tilde{\eta}_{AD_A}$ and $\tilde{\eta}_{CD_C}$ compared to that of the model with no retroactivity.  Values of parameters used for making the plots are: in (a), $\tilde{K}_{AB} = \tilde{K}_{AC} = \tilde{K}_{CD_C} = 0.01$, $\tilde{K}_{BC} = \tilde{K}_{BD_B} = 0.1$, $h_{AB} = h_{AC} = h_{BC} = h_{BD_B} = h_{CD_C} = 0.5$; in (b), $\tilde{K}_{AB} = \tilde{K}_{AC} = \tilde{K}_{AD_A} = 0.01$, $\tilde{K}_{BC} = \tilde{K}_{BD_B} = 0.1$, $h_{AB} = h_{AC} = h_{AD_A} = h_{BC} = h_{BD_B} = 0.5$; in (c), $\tilde{K}_{AD_A} = \tilde{K}_{AC} = \tilde{K}_{CC} = \tilde{K}_{CD_C} = 0.01$, $h_{AD_A} = h_{AC} = h_{CC} = h_{CD_C} = 0.5$.}  We chose $\tilde{\eta}_{XD_X}$ $(X=A, B, C)$ to be the midpoints of the 50 subintervals that we split the interval $[\log_{10}(1.0), \log_{10}(100.0)]$ evenly into.  The magnitude of the ratio is shown by the color.  For values of $\tilde{\eta}_{XD_X}$ $(X=A, B, C)$ that were not chosen for simulation, the ratio was interpolated.  The black curve, which we refer to as the ``iso-response-time" curve, represents values of $\tilde{\eta}_{XD_X}$ $(X=A, B, C)$ at which the response time is the same as the response time of the model with no retroactivity. \commente{Note that in (c) there is no ``iso-response-time'' curve because $\log_{10}\left(\frac{RT_{\tilde{\eta}_{AD_A}, \tilde{\eta}_{CD_C} }}{ \tilde{\eta}_{0, 0} } \right)$ is always larger than 1.} \label{fig:multiple_node_retro}}
\end{figure}

\subsection{\commentf{Joint Increases of Retroactivity Can Keep Response Time Constant}}
\label{sec:joint}
Next, we \commentg{investigated} how joint increases of retroactivity on multiple nodes affect response times by letting $\tilde{\eta}_{BD_B}$ and $\tilde{\eta}_{CD_C}$ ($\tilde{\eta}_{AD_A}$) vary simultaneously.  The I1-FFL model was simulated for different values of $\tilde{\eta}_{BD_B}$ and $\tilde{\eta}_{CD_C}$ ($\tilde{\eta}_{AD_A}$) within the interval of 1.0 and 100.0.  The ratio of the response time under each combination of $\tilde{\eta}_{BD_B}$ and $\tilde{\eta}_{CD_C}$ ($\tilde{\eta}_{AD_A}$) to the response time without retroactivity was then calculated (Figures \ref{fig:multiple_node_retro}(a) and (b)).  \commentg{We find} that in an I1-FFL, if $\tilde{\eta}_{BD_B}$ and $\tilde{\eta}_{CD_C}$ ($\tilde{\eta}_{AD_A}$) increase simultaneously, response time can be increased, decreased, or kept constant depending on the values of $\tilde{\eta}_{BD_B}$ and $\tilde{\eta}_{CD_C}$ ($\tilde{\eta}_{AD_A}$).  This is because increasing $\tilde{\eta}_{BD_B}$ and $\tilde{\eta}_{CD_C}$ ($\tilde{\eta}_{AD_A}$) affects response time in opposing directions, and the resulting counteracting effects can be canceled when the values of $\tilde{\eta}_{BD_B}$ and $\tilde{\eta}_{CD_C}$ ($\tilde{\eta}_{AD_A}$) satisfy a certain relationship (the solid black curves in Figures \ref{fig:multiple_node_retro}(a) and (b), which we call the ``iso-response-time" curves).  

Moreover, we compared the behavior of an IFFL under increasing levels of retroactivity to that of negative autoregulation (Figure \ref{fig:multiple_node_retro}(c)), another motif known for sign-sensitive response acceleration (\cite{rosenfeld}).   We found that in contrast to IFFLs, increasing $\tilde{\eta}_{AD_A}$ and/or $\tilde{\eta}_{CD_C}$ in a negative autoregulatory circuit can only slow down the response \commente{regardless of the values of \commentg{any} other parameters}, as the response time of the model with retroactivity is always larger than that of the model without retroactivity (Figure \ref{fig:multiple_node_retro}(c); see Supplemental Information Section \ref{suppsec:non_dimensionalized_other_models} for the model and Supplemental Information Section \ref{suppsec:retro_proof_2} for the proof).   

\commente{Besides IFFLs and negative autoregulation, our simulations suggest that two-node negative feedback loops (NFBLs) can also act as sign-sensitive response accelerators (Figure \ref{suppfig:nfbl}).  Moreover, we find that if $\tilde{\eta}_{BD_B}$ and $\tilde{\eta}_{AD_A}$ increase simultaneously, then response times of gene A can be increased, decreased, or kept constant depending on the values of $\tilde{\eta}_{BD_B}$ and $\tilde{\eta}_{AD_A}$ (Figure \ref{suppfig:nfbl}).}  

\subsection{\commentf{Effects of Retroactivity Depend on the Motif}}
\label{sec:i1_more_common}
\begin{figure}[!hbt]
\centering
\subfigimg[width=110mm]{\textbf{a)}}{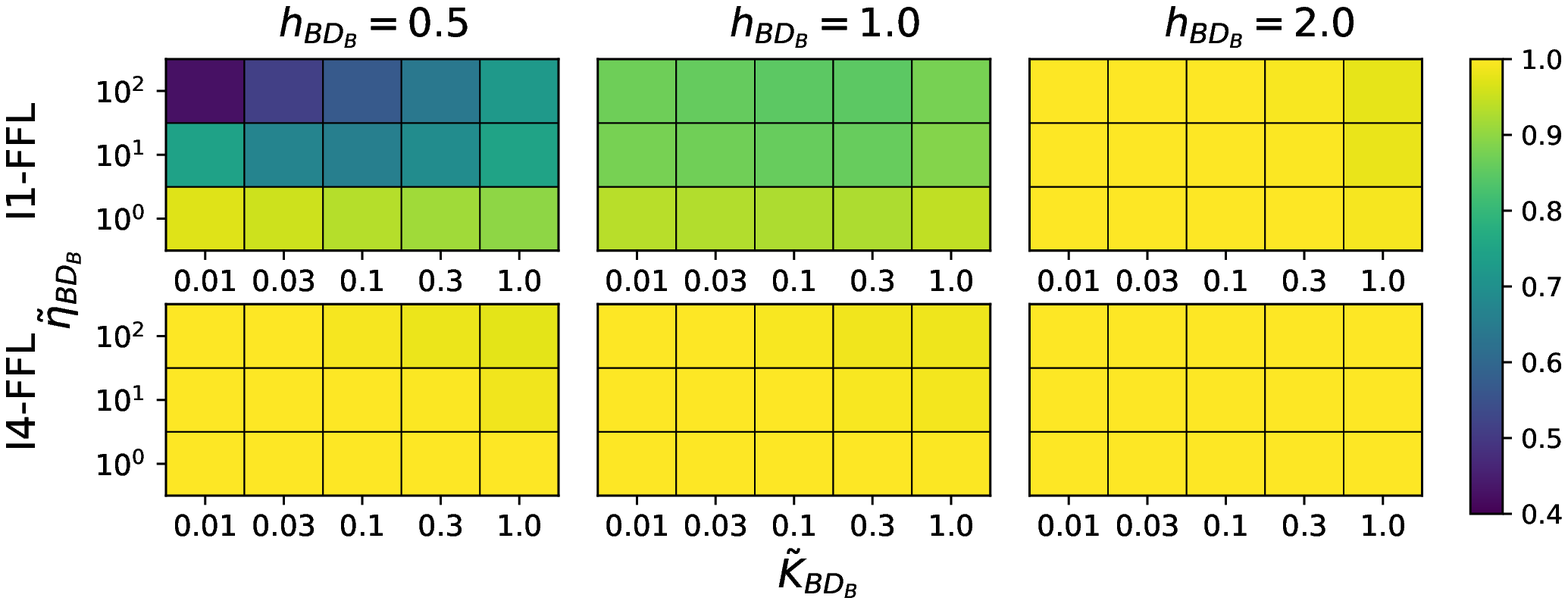} \\
\subfigimg[width=110mm]{\textbf{b)}}{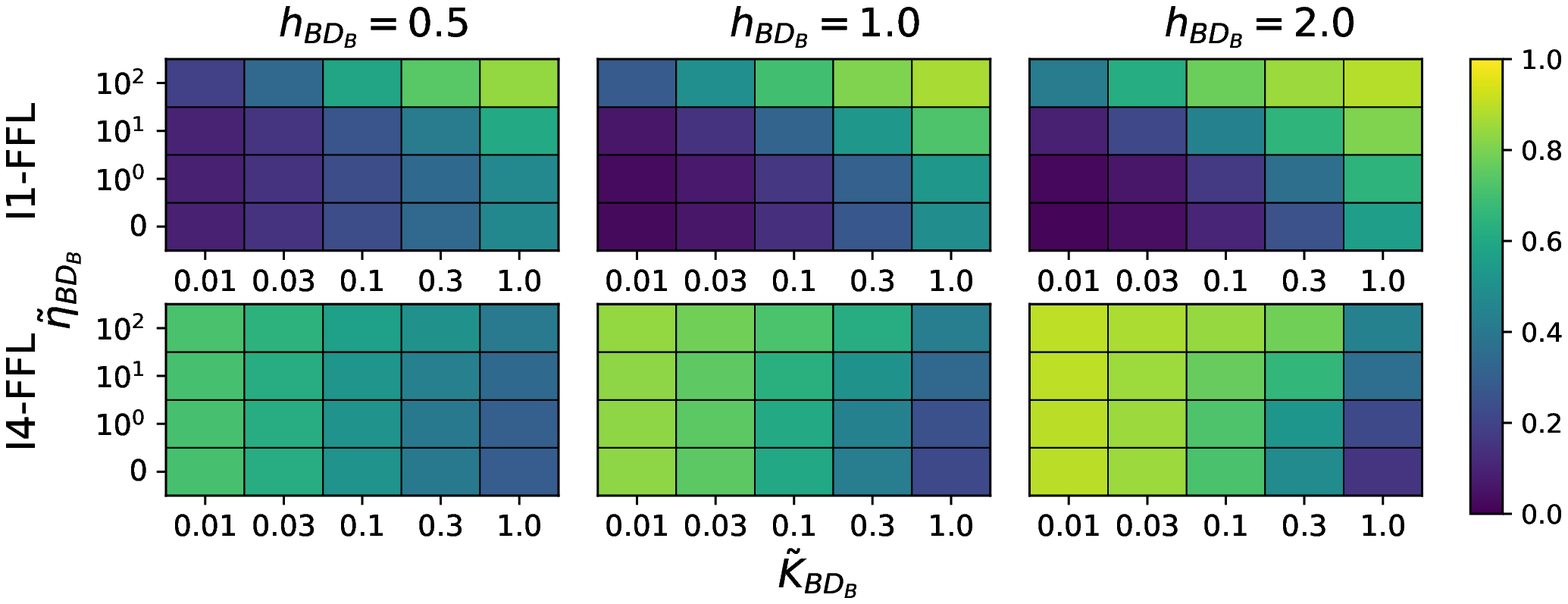} \\
\subfigimg[height=75mm]{\textbf{c)}}{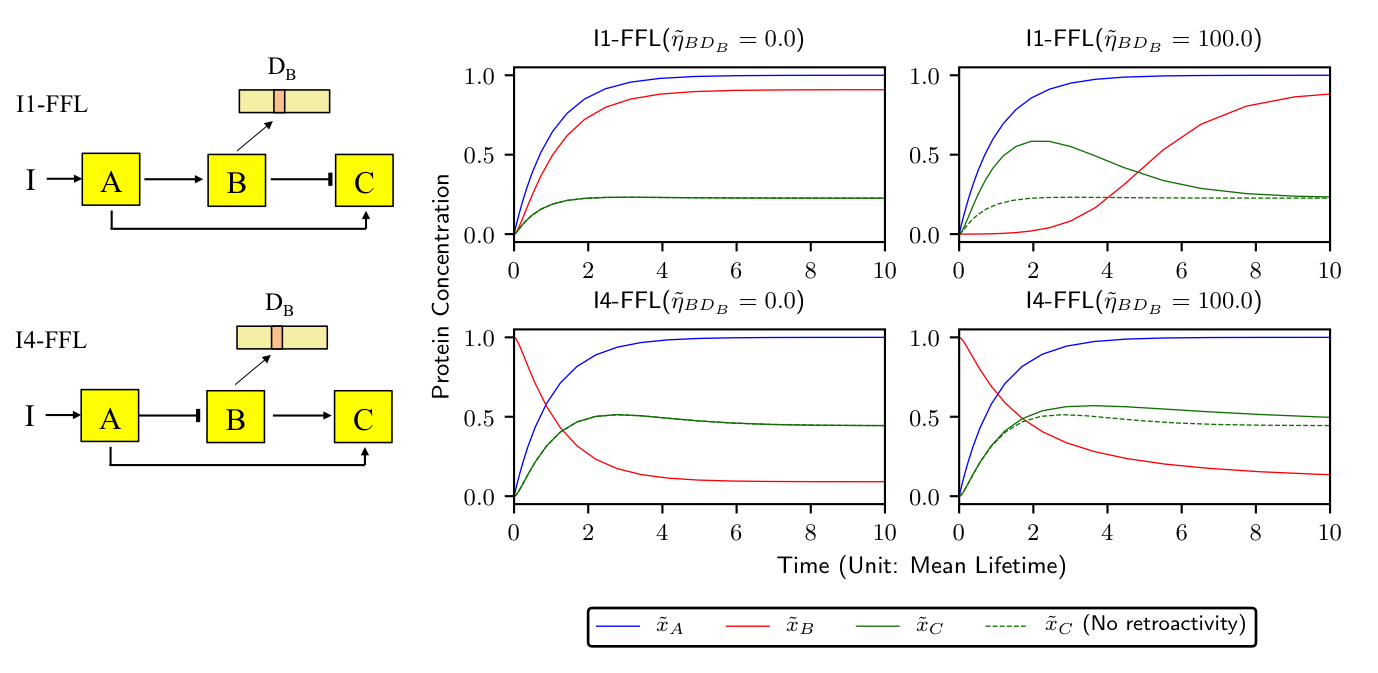} 
\caption{(a) \commente{Relative response time of I1-FFL and I4-FFL models with different values of $\tilde{K}_{BD_B}$.  Here, relative response time} is defined as the ratio of the response time of the model to the response time of the model without retroactivity.  \commente{Values of the parameters are: $\tilde{K}_{AB} = \tilde{K}_{AC} = 0.1$, $h_{AB} = h_{AC} = 1.0$.} (b) \commente{Pulse amplitude of  I1-FFL and I4-FFL models with different values of $\tilde{K}_{BD_B}$.} \commente{Values of the parameters are the same as in (a).}  \commente{Note that pulse amplitudes of I1-FFLs and I4-FFLs should not be compared column-wise, as I1-FFLs generate larger pulses with larger $\tilde{K}_{BD_B}$, while I4-FFLs generate larger pulses with smaller $\tilde{K}_{BD_B}$.}  (c) The effect of retroactivity $\tilde{\eta}_{BD_B}$ on response times is more pronounced in an I1-FFL (top row) than in an I4-FFL (bottom row) for $h_{BD_B}  \leq 1$.  For comparison, the green dashed curves represent the trajectories of $\tilde{x}_C$ when $\tilde{\eta}_{BD_B}$ equals 0.  \commente{Values of the parameters are: $\tilde{K}_{AB} = \tilde{K}_{AC} = \tilde{K}_{BC} = \tilde{K}_{BD_B} = 0.1$, $h_{AB} = h_{AC} = 1.0$, $h_{BC} = h_{BD_B} = 0.5$}.}
 \label{fig:Type-I-IV}
\end{figure}

\commentg{We now examine how varying regulatory logic, e.g., I1- vs I4-FFLs and ``OR" logic, can lead to different responses in the presence of retroactivity.}
As is demonstrated in Section \ref{sec:param}, increasing retroactivity $\tilde{\eta}_{BD_B}$ accelerates the response and increases the pulse amplitude of gene C.  Our simulations also suggest that how much retroactivity affects response time and pulse amplitude depends on the actual type of the IFFL.  In response to an ON step, increasing $\tilde{\eta}_{BD_B}$ accelerates the response times in an I1-FFL \commente{but} not in an I4-FFL for $h_{BD_B} \leq 1$, \commente{and} increases the pulse amplitude more strongly in an I1-FFL than in an I4-FFL (Figures \ref{fig:Type-I-IV}(a)  and (b); see Tables \ref{supptable:rt_B} and \ref{supptable:pa_B} for data). 

The different effects of retroactivity on response time and pulse amplitude in different IFFLs is likely an outcome of how much $\frac{d\tilde{x}_B}{d\tau}$ decreases in different phases of the response.  To explain this argument, we take the derivative of the reduction factor $b$ (Equation (\ref{retro_eq})) with respect to $\tilde{x}_B$:

\begin{align}
\begin{aligned}
\frac{db (\tilde{x}_B)}{d\tilde{x}_B} 
&= \frac{\tilde{\eta}_{BD_B}h_{BD_B}^2 }{\tilde{K}_{BD_B}^{h_{BD_B}-1}} \cdot \tilde{x}_B^{h_{BD_B}-2} \left( 1 + \left(\frac{\tilde{x}_B}{\tilde{K}_{BD_B}} \right)^{h_{BD_B}} \right)^{-2} \\
&\cdot \left[ \left(h_{BD_B}-1\right) - 2h_{BD_B} \cdot \frac{\tilde{x}_B^{h_{BD_B}}}{\tilde{K}_{BD_B}^{h_{BD_B}} + \tilde{x}_B^{h_{BD_B}}} \right].
\end{aligned}
\label{eq:derivative}
\end{align}

If $h_{BD_B}$ is a value between 0 and 1, then $\frac{d b (\tilde{x}_B)}{d \tilde{x}_B}$ is always negative, indicating that $b(\tilde{x}_B)$ is monotonically decreasing on the interval of $(0, 1]$.  In an I1-FFL, $\tilde{x}_B$ transitions from a low pre-stimulus steady state to a high post-stimulus steady state in response to an ON step.  Based on monotonicity of $b(\tilde{x}_B)$, we know that the reduction factor is the largest when $\tilde{x}_B$ is close to 0, which significantly lowers the initial value of $|\frac{d\tilde{x}_B}{d\tau}|$ \commente{relative to the no retroactivity case} (Figure \ref{fig:Type-I-IV}(c)).  \commente{Consequently, $\tilde{x}_B$ increases more slowly, and hence} $\frac{d\tilde{x}_C}{d\tau}$ is significantly increased during the initial response phase.  \commente{This results in a} shortened response time and increased pulse amplitude.  On the other hand, in an I4-FFL, $\tilde{x}_B$ transitions from a high pre-stimulus steady state to a low post-stimulus steady state in response to an ON step (Figure \ref{fig:Type-I-IV}(c)).  Due to monotonicity of $b(\tilde{x}_B)$, the reduction factor \commentg{is} smallest when $\tilde{x}_B$ is close to 1. This means that \commente{initially} $\frac{d\tilde{x}_C}{d\tau}$ is \commente{minimally} affected in an I4-FFL, so the effects of $\tilde{\eta}_{BD_B}$ on response time and pulse amplitude are not as strong in an I4-FFL as in an I1-FFL.  

If $h_{BD_B}$ is larger than 1, then $b(\tilde{x}_B)$ reaches its maximum for some value of $\tilde{x}_B$ between 0 and 1.  Moreover, $b(\tilde{x}_B)$ monotonically increases (decreases)  to the left (right) of $\argmax_{\tilde{x}_B} b(\tilde{x}_B)$.  Setting $\frac{db(\tilde{x}_B)}{d\tilde{x}_B}$ equal to zero, we can obtain the following expression for $\argmax_{\tilde{x}_B} b(\tilde{x}_B)$:
\begin{equation}
\argmax_{\tilde{x}_B} b(\tilde{x}_B) = \left(\frac{h_{BD_B}-1}{h_{BD_B}+1}\right)^{\frac{1}{h_{BD_B}}} \cdot \tilde{K}_{BD_B}.
\label{argmax:x}
\end{equation} 

\commente{The qualitative behavior of the IFFL for $h_{BD_B} > 1$ is similar to the $h_{BD_B}=2$ case.  When $h_{BD_B}$ equals 2, $\tilde{\eta}_{BD_B}$ minimally affects response times in either I1-FFLs or I4-FFLs (Figure \ref{fig:Type-I-IV}(a)).}
This is likely because for $h_{BD_B}$ equal to 2 , $b(\tilde{x}_B)$ reaches its maximum when $\tilde{x}_B$ reaches approximately $50\%$ of $\tilde{K}_{BD_B}$, which happens much later \commente{than when} $\tilde{x}_C$ reaches its half response point even in I1-FFLs (Table \ref{supptable:supp_figure_4_a}).  As a result, retroactivity $\tilde{\eta}_{BD_B}$ barely affects the response time when $h_{BD_B}$ equals 2.  

On the other hand, \commente{an I1-FFL generally experiences a more significant change in pulse amplitude than an I4-FFL as $\tilde{\eta}_{BD_B}$ increases (Figure \ref{fig:Type-I-IV}(b)).}  In order to generate a pulse, $\tilde{x}_B$ often needs to get larger (smaller) than $\tilde{K}_{BC}$ so that it can effectively inhibit C in an I1-FFL (I4-FFL) (Table \ref{supptable:supp_figure_4_b}), which happens after $b(\tilde{x}_B)$ reaches its maximum.  In response to an ON step, $\tilde{x}_B$ transitions from a low state to a high state in an I1-FFL, so according to Equation (\ref{eq:derivative}), the reduction factor is the largest in the initial response phase before $\tilde{x}_B$ becomes large relative to $\tilde{K}_{BC}$, greatly lowering the initial value of $|\frac{d\tilde{x}_B}{d\tau}|$  (Figure \ref{suppfig:Type-I-IV_pa}).  In contrast, in an I4-FFL, because $\tilde{x}_B$ transitions from a high state to a low state, the reduction factor is the largest when $\tilde{x}_B$ becomes small relative to $\tilde{K}_{BC}$ somewhere in the return phase (Equation (\ref{eq:derivative})) (Figure \ref{suppfig:Type-I-IV_pa}).  As a result, $\tilde{\eta}_{BD_B}$ affects pulse amplitude more strongly in an I1-FFL than in an I4-FFL for $h_{BD_B}$ larger than 1, \commente{similar to the $h_{BD_B} \leq 1$ case}.

Under the assumption of OR logic, I1-FFLs and I4-FFLs become sign-sensitive response accelerators in response to an OFF step (inducer level $x_I$ changes from $\infty$ to 0).  OR logic is another special case of independent binding, where either the presence of an activator or the absence of an inhibitor is sufficient to turn on the expression of the regulated gene (see Supplemental Information Section \ref{suppsec:or_logic} for model details).  Though response time is more sensitive to changes in $\tilde{\eta}_{BD_B}$ in an I1-FFL than in an I4-FFL under the assumption of AND logic, the reverse becomes true under the assumption of OR logic: in response to an OFF step, increasing $\tilde{\eta}_{BD_B}$ decreases the response time more strongly in an I4-FFL than in an I1-FFL (Figure \ref{suppfig:i1_vs_i4_OR}).  This is because in response to an OFF step, $\tilde{x}_B$ transitions from a high pre-stimulus steady state to a low post-stimulus steady state in an I1-FFL whereas $\tilde{x}_B$ transitions from a low pre-stimulus steady state to a high post-stimulus steady state in an I4-FFL (Figure \ref{suppfig:i1_vs_i4_OR}; see Supplemental Information Section \ref{suppsec:or_logic} for the data).

\begin{figure}[!ht]
\centering
\subfigimg[width=160mm]{}{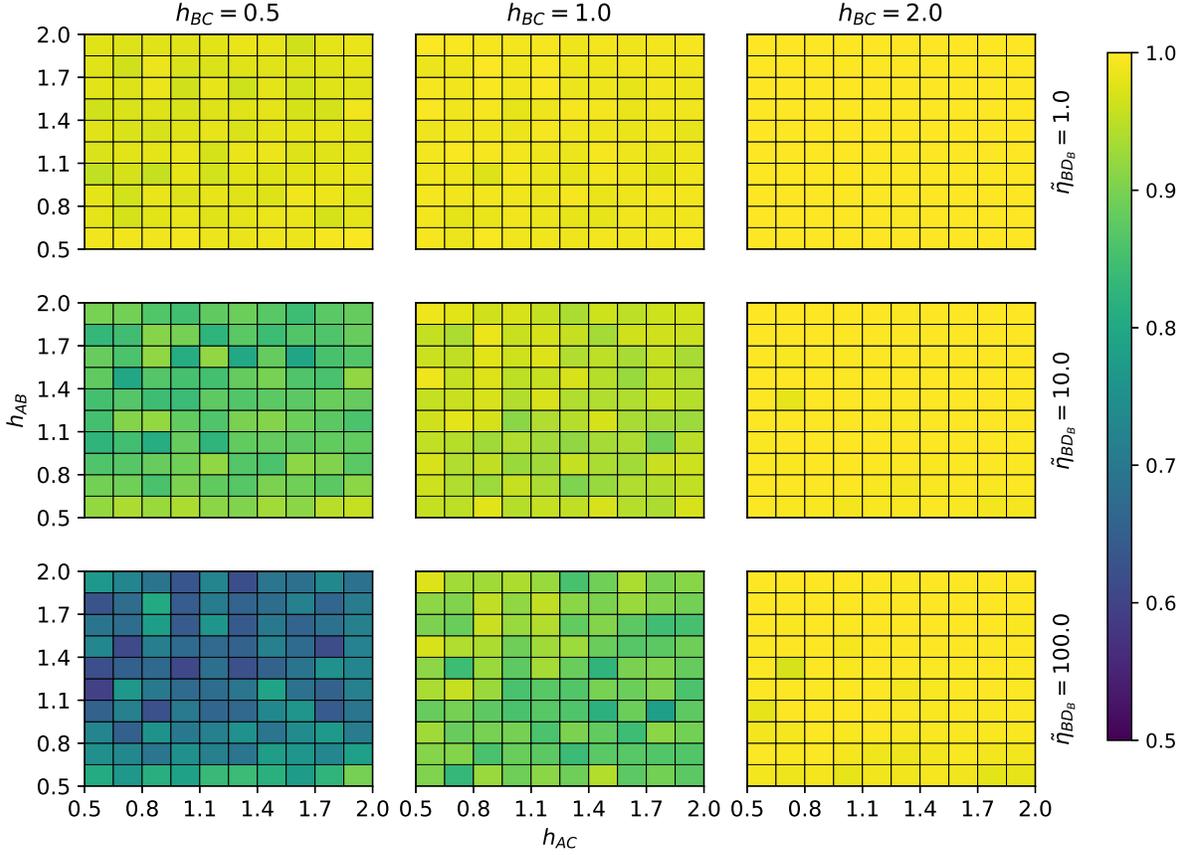}
\caption{\commentf{\commentg{Response acceleration can occur over a wide set of parameters when node B regulates node C through negative or neutral cooperativity.}  Median relative response time of I1-FFLs \commentg{as model parameters are systematically varied, with random sampling of $h_{BD_B}$ and $\tilde{K}_{BD_B}$ within the intervals} $0.8 h_{BC} \le h_{BD_B} \le 1.2 h_{BC}$ and $0.8 \tilde{K}_{BC} \le \tilde{K}_{BD_B} \le 1.2 \tilde{K}_{BC}$.  The trajectories are separated evenly by $h_{AB}$ and $h_{AC}$ into 100 \commentg{voxels}, the color of which represents the median relative response time of the 100 simulated trajectories \commenth{within} that bin, \commenth{where for each trajectory the remaining parameters were chosen via random Latin Hypercube Sampling}.  Here, relative response time is defined as the ratio of the response time of the model to the response time of the model without retroactivity.  \commentg{See Supplemental Information Section \ref{suppsec:ext_sim} for details of parameter sampling.} }}
\label{fig:rt_ext_sim}
\end{figure}

\subsection{\commentf{Effects of Retroactivity Are Independent of Parameter Isometry}}
\label{sec:isometry}
\commentf{To demonstrate the robustness of our findings, we performed extensive simulations on an I1-FFL model, where Hill coefficients, binding affinity, and decay rates were all allowed to vary.  Similar \commentg{to our earlier simulations}, we set $\tilde{\eta}_{BD_B}$ equal to 0, 1.0, 10.0, and 100.0, and $h_{BC}$ equal to 0.5, 1.0, and 2.0, to represent different levels of retroactivity and cooperativity.  The rest of the parameters were sampled from their corresponding ranges via Latin Hypercube Sampling (details of parameter sampling can be found in Supplemental Information Section \ref{suppsec:ext_sim}).  The assumption of isometry, where TFs bind to the functional target site and non-functional decoy sites with equal affinity and cooperativity (i.e., $h_{BC} = h_{BD_B}$ and $\tilde{K}_{BC} = \tilde{K}_{BD_B}$) was also relaxed.  Instead, we assume that $h_{BC}$ ($h_{BD_B}$) and $\tilde{K}_{BC}$ ($\tilde{K}_{BD_B}$) may be unequal but correlated, as the target sites and decoy sites we considered here have the same binding motifs.  To preserve correlation, we sampled $h_{BD_B}$ and $\tilde{K}_{BD_B}$ from the intervals $(1-c) h_{BC} \le h_{BD_B} \le (1+c) h_{BC}$ and $(1-c) \tilde{K}_{BC} \le \tilde{K}_{BD_B} \le (1+c) \tilde{K}_{BC}$, where flexibility coefficient $c$ equals 0, 0.2, or 0.5.  The process of ODE simulation was repeated for 10000 sets of parameters.  

After the simulation was completed, we separated the trajectories by $h_{AB}$ and $h_{AC}$ evenly into 10 x 10 \commentg{voxels}.  Within each \commenth{voxel}, we calculated the median relative response time (Figure \ref{fig:rt_ext_sim}) as well as the \commentg{percent} of trajectories achieving relative response time less than \commentg{$90\%$ of the model in the absence of retroactivity} (i.e., 10$\%$ response acceleration) (Figure \ref{suppfig:rt_ext_sim_frac}).  The results suggest \commentg{that} parameter isometry, which we had assumed earlier (e.g., $h_{AB} = h_{AC}$, $\tilde{K}_{AB} = \tilde{K}_{AC}$, $h_{BC} = h_{BD_B}$, $\tilde{K}_{BC} = \tilde{K}_{BD_B}$, $\delta_A = \delta_B = \delta_C$), \commentg{is not essential to our conclusions}.  Moreover, the model exhibits significant response acceleration in a large region of parameter space (Figure \ref{fig:rt_ext_sim} and Figure \ref{suppfig:rt_ext_sim_frac}), including \commentg{regions} where $h_{AB} > 1$ and/or $h_{AC} > 1$.  Simulation results assuming flexibility coefficient $c$ equal to 0 and 0.5 exhibit similar patterns (Figures \ref{suppfig:rt_ext_sim_0}, \ref{suppfig:rt_ext_sim_frac_0}, \ref{suppfig:rt_ext_sim_0_5}, and \ref{suppfig:rt_ext_sim_frac_0_5}) \commentg{and further corroborate the mathematical proof in Supplemental Information Section \ref{suppsec:retro_proof}}.
}

\section{DISCUSSION}

In this work, we studied how retroactivity affects the behavior of IFFLs \commentg{via simulation and mathematical analysis}.  Our findings can be summarized as follows.  First, in IFFLs, increasing retroactivity of the input node A, $\tilde{\eta}_{AD_A}$, induces counteracting effects on response time and pulse amplitude, slowing both the direct activation and the indirect inhibition of node C.  
Second, increasing retroactivity of the regulatory node B, $\tilde{\eta}_{BD_B}$, \commentg{can shorten} response \commentg{times} \commentf{and increase pulse amplitudes}, particularly in an I1-FFL with AND logic and an I4-FFL with OR logic.  As a result, compared to negative autoregulation, IFFLs \commente{exhibit a larger variety of functional capabilities at high levels of retroactivity.  \commentg{While mathematical proofs \commenth{in Supplemental Information Sections \ref{suppsec:retro_proof} and \ref{suppsec:inter_retro_proof}} demonstrate that our second finding is parameter-independent, \commenth{the} simulations \commenth{systematically exploring parameter space} in Section \ref{sec:isometry} \commenth{show} that the magnitude by which retroactivity affects response times in IFFLs is \commenth{significant} in large regions of parameter space.}} 

\subsection{\commentf{Tuning Retroactivity in Synthetic IFFLs}}
\commente{In synthetic biology,} our work lends novel insights into designing gene circuits. \commente{Most prior studies have focused on the disruptive effects of retroactivity on the intended behavior of circuits, e.g., shrinking the bistable region of a toggle switch (\cite{gyorgy, gardner}).  In contrast, here we showed} that increasing retroactivity may be used as a strategy to improve the behavior of IFFLs, i.e., \commente{creating synthetic IFFLs with shorter response times and larger pulse amplitudes while maintaining the same steady-state behaviors}.  One approach to changing retroactivity in synthetic systems is to mimic NDs by adding synthetic decoy sites, i.e., recombined bacterial plasmids that contain high-affinity sequence-specific binding sites.  Biologically, \commente{the} number of synthetic decoys can be adjusted by changing the \commente{transformation/transfection} protocol, including the plasmid dose, the \commente{transformation/transfection} reagent, and/or the method of transformation/transfection.  Via a mechanism similar to NDs, synthetic decoys can affect the behavior of synthetic circuits by sequestering TFs.  

\commentf{\commentg{Our} work also suggests that topology alone does not constitute the entire solution to circuit design. As shown by Figures \ref{fig:Type-I-IV} and \ref{fig:rt_ext_sim}, retroactivity affects the behaviors of I1-FFLs much more strongly for $h_{BD_B} \leq 1$ \commentg{(negative cooperativity)} than for $h_{BD_B} > 1$ \commentg{(positive cooperativity)}. \commentg{Importantly,} negative cooperativity and non-cooperativity are typical of the synthetic transcriptional repressors/activators used in constructing mammalian gene circuits.   \commentf{\cite{yuchen} constructed four ABA/GA-inducible VPR-Sp/Sa dCas9 gene activators, \commentg{with} Hill coefficients \commentg{ranging} from 0.70 to 0.97 (coefficients fit from source data to Fig. 2 provided by \cite{yuchen}). \cite{yinqing} constructed and characterized a library of 26 transcription activator-like effector repressors (TALERs) that bind designed hybrid promoters. The Hill coefficients of the characterized TALER-binding promoters range from 0.67 to 1.15, while the Hill coefficients of TALERs range from 0.51 to 1.56 (\cite{yinqing}).}} 

\begin{figure}[t]
\centering
\subfigimg[width=155mm]{}{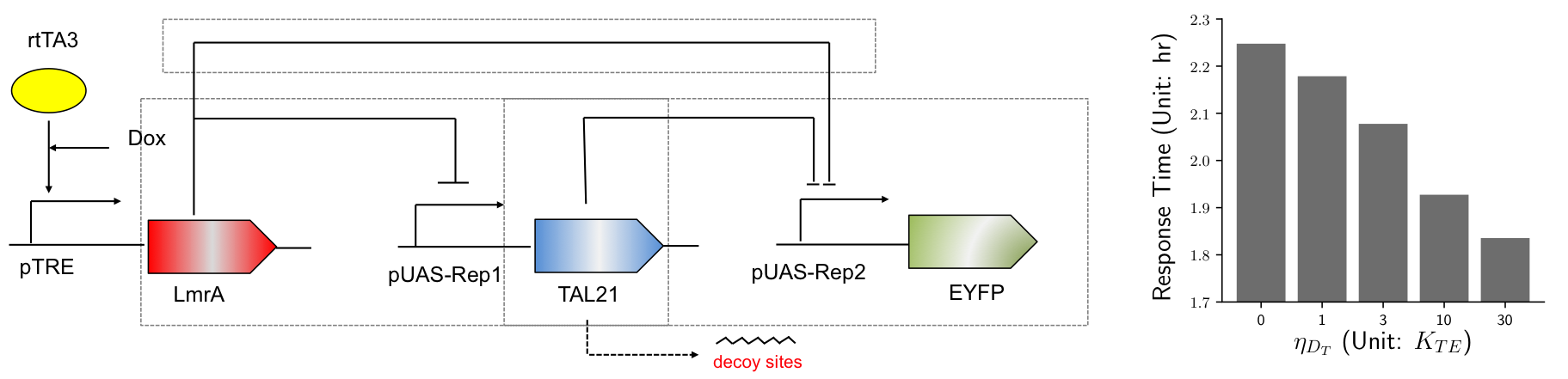}
\caption{\commente{\commentf{\commentg{Tuning retroactivity is predicted by models to increase} the response time of a synthetic IFFL \commentg{construct} (\cite{davidsohn, wang2018, wang_acs})}.  The level of retroactivity may be adjusted via the delivery of plasmids containing synthetic decoy binding sites.  \commentg{See Supplemental Information Section \ref{suppsec:sim_syn_iffl} for details.}}  \label{fig:synbio_procedure}}
\end{figure}

\commentf{As a practical example, we consider a mammalian-cell-based IFFL circuit composed of biological parts built and tested by \cite{davidsohn} (Figure \ref{fig:synbio_procedure}).  Induced by Dox, the pTRE promoter turns on the expression of LmrA, which inhibits EYFP directly and activates EYFP indirectly through TAL21.  As an I2-FFL, this construct mediates response acceleration in response to an OFF step.  Using parameter estimates from \cite{wang2018, wang_acs}, we simulate its behaviors in the absence and presence of retroactivity (see Supplemental Information Section \ref{suppsec:sim_syn_iffl} for details).  As Figure \ref{fig:synbio_procedure} suggests, the response time of EYFP decreases by more than $18\%$ from 2.2 hr to 1.8 hr when the concentration of the decoy sites increases from 0 to 8.7e+6 MEFL (30 times the binding affinity of TAL21 to pUAS-Rep2). Compared to manipulation of kinetic parameters, the strategy we propose may allow more precise control of the circuit as introduction of plasmids containing TAL21 binding sites does not interfere with the circuit's steady state. }

\subsection{\commentf{A Potential Role of Retroactivity in Motif Evolution}}
From an evolutionary perspective, we hypothesize that the behaviors of IFFLs and negative autoregulated circuits under increasing levels of retroactivity may have shaped the relative abundance of sign-sensitive response-accelerating motifs in different organisms.  Using published databases of \textit{E. coli}, mouse, and human TRNs (\textit{E. coli}: RegulonDB v10 developed by \cite{regulondb10}; mouse and human: TTRUST v2 developed by \cite{ttrust}), we compared the number of times an IFFL is observed in the TRN of each organism to the number of times an IFFL is expected in the corresponding randomized Erdos-Renyi (ER) networks.  We observed a total of 1258, 470, and 1171 IFFLs in \textit{E. coli}, mouse, and human TRNs, whereas only 11, 5, and 7 would be expected, respectively if TF-gene interactions were completely randomized (see Supplemental Information Section \ref{suppsec:sig_motif} for details).  
The number of times an IFFL is observed versus expected suggests that IFFLs are conserved in both prokaryotic and eukaryotic organisms.  
In contrast, the occurrence of negative autoregulation differs drastically between prokaryotes and \commente{higher} eukaryotes.
 In agreement with \cite{stewart}, we found that while almost half of all repressors in \textit{E. coli} are negatively self-regulated, approximately only one percent of repressors in the mouse and the human genomes are negatively autoregulated (Supplemental Information Section \ref{suppsec:sig_motif}). 
 
From a general standpoint, \commente{higher eukaryotes have much larger genomes and non-coding genomes than prokaryotes.}
A direct consequence is that while an average \textit{E. coli} TF has 3 - 25 binding sites in the genome (\cite{gao}),  an average human TF, as is mentioned in Section \ref{S:1}, has approximately $10^4 - 10^5$ accessible ND sites.  The degree of retroactivity that arises from accessible ND sites is, thus, expected to be \commente{substantially} higher in \commente{higher} eukaryotes such as mouse and human TRNs than in bacteria TRNs.  

\commentf{Due to technical challenges, \commentg{studies} quantifying the cooperativity of TF-DNA binding \commentg{are} much less common in natural systems than in synthetic systems.  Nevertheless, a ChIP-Seq-based study led by \cite{ghosh} \commentg{examining} the DNA-binding of eight common TFs \commentg{(i.e., Oct4, Nanog, CTCF, IRF2, FoxA1, NFAT, IRF1, and RelA) suggested} that negative cooperativity and non-cooperativity may be a prevalent phenomenon of TF-DNA binding in mammalian cells (Supplementary Figure 5b of \cite{ghosh}).  As \commentg{observed} by \cite{ghosh}, the ChIP-seq signal strength of most of these TFs stays relatively constant or decreases as the number of binding sites increases, which indicates non- and negative cooperativity.  Querying \commentg{these} TFs against the TTRUST v2 database (\cite{ttrust}), we found that three TFs, namely, RelA, IRF1, and CTCF, act as intermediate regulators (node B) in 101 IFFLs in human TRNs.  RelA and IRF1, the latter of which exhibits clear negative cooperativity (Supplementary Figure 5b of \cite{ghosh}),  serve as the input node (node A) and regulatory node (node B) of IFFLs regulating BCL2, CCNB1, CDK4, CDKN1A, and FOXP3, as a part of the interferon pathway (\cite{ttrust, kochupurakkal}).  \commentg{As such, they are physical examples of IFFLs where protein B binds to DNA with negative cooperativity.}}

\commente{As is shown in Section \ref{sec:joint} and Supplemental Information Section \ref{suppsec:retro_proof_2}, higher retroactivity in a negative autoregulatory loop results in a longer response time.  This indicates that a negative autoregulatory loop achieves its minimum response time under the condition of zero retroactivity.  
In contrast, as is shown in Section in \ref{sec:param} and Supplemental Information Section \ref{suppsec:retro_proof}, an IFFL with retroactivity on the regulatory node B can achieve response times shorter than that of an IFFL  with zero retroactivity, \commentf{especially if protein B binds to DNA with negative cooperativity or non-cooperativity.}} \commente{One can speculate} that network motifs that exhibit a larger diversity of functional capabilities under a high level of retroactivity are more likely to be conserved in higher eukaryotes.  This is because the desired outcome of increasing retroactivity, i.e., whether the response time should increase, decrease, or stay constant, depends on the actual biological context, and a network motif that exhibits a larger diversity of functions is more likely to meet the expectation of the context.  
 This suggests that IFFLs could enable organisms to better adapt to a \commente{large} number of accessible ND sites during evolution than negative autoregulation in cases where \commente{a response time shorter than that of the circuit under zero retroactivity} is desired.  Therefore, IFFLs \commente{might} confer \commente{upon} the organism a selective advantage \commente{compared to} negative autoregulation \commente{at high} levels of retroactivity.  
 
 \commente{It is interesting to note that, in contrast to the decreased abundance of autoregulatory loops, we have observed an increased abundance of two-node NFBLs in higher eukaryotic TRNs compared to bacterial TRNs (Supplemental Information Section \ref{suppsec:sig_motif}).  This may be because, similar to IFFLs, two-node NFBLs also exhibit a larger diversity of functional capabilities than negative autoregulatory loops under high levels of retroactivity.}
 
 \subsection{Future Directions} 
\commentf{Our work can be generalized and extended in several directions.   First, it would be interesting to explore how the connection of IFFLs to additional network motifs, such as NFBLs, affects the ability of IFFLs to accelerate responses.  In a modeling study, \cite{joanito} proposed that in \textit{Arabidopsis thaliana}, the CCA1/LHY-PRR9/7(PRR5/TOC1)-CCA1/LHY IFFL circuit serves to break the bistability generated by the double NBFLs between CCA1/LHY and PRR5/TOC1.  Compared to a plain NFBL, the IFFL-NFBL combination allows cells to switch between the two states more rapidly (\cite{joanito}). In addition, \cite{reeves} and \cite{ma} demonstrated that conjoining an NFBL to an IFFL can also increase the robustness of IFFL-mediated adaptation.  That is, perfect or near-perfect adaptation can be achieved over a wider region of parameter space in an IFFL-NFBL combination than in either motif alone.  Studying the effect of retroactivity that arises from the interconnection of an IFFL to an NFBL will serve to inform the design of a robust IFFL-NFBL synthetic system.}

\commentf{More generally, the modeling framework we apply here is based on ODEs.  Other approaches to studying retroactivity include stochastic gene expression models, which take transcriptional bursting into consideration.  Via stochastic simulation, \cite{kim} found that retroactivity can dampen fluctuations and lengthen correlations in the output signal noise when the output of a network is connected to a downstream module.  It will be interesting to study whether retroactivity can further enhance the ability of IFFLs to attenuate the stochastic variation in gene expression.}
 
\section*{Limitations of the Study}
\label{sec:limitations}
This study presents a minimal model of IFFL circuits with and without retroactivity.  In cases where two transcription factors bind to the same promoter, the model excludes binding types other than AND and OR logics, such as the competitive logic.  

\section*{Methods}
\label{sec:methods}
All methods can be found in the ``Transparent Methods'' section of the Supplemental Information.

\commenth{\section*{Data and Code Availability}
\label{sec:code}
A Julia script for implementing and solving the ODEs that model IFFLs, type-1 two-input circuits, and negative autoregulated circuits can be found online at https://github.com/wang-junmin/IFFL.
}

\section*{SUPPLEMENTAL INFORMATION}
Supplemental Information including proof of diagonality of retroactivity matrices, ODE models for IFFLs and other response-acceleration motifs, response time and pulse amplitude of IFFLs and two-input circuits at different levels of retroactivity,  response time of IFFLs with OR logic, proofs of the effects of retroactivity on response time and pulse amplitude, \commentg{simulation results based on systematically exploring parameter space, an example of a simulated synthetic IFFL}, two-node NFBLs, and motif abundance can be found online together with this article.

\section*{ACKNOWLEDGEMENTS}
The authors thank Prof. Domitilla Del Vecchio, Prof. Daniel Segr\`e, and Brian Teague for helpful discussions and constructive feedbacks.  SAI was supported by National Science Foundation awards DMS-1255408 and DMS-1902854.

\section*{AUTHOR CONTRIBUTIONS}
J.W. conceptualized the ideas, took the lead in writing the manuscript, and supervised this work.  J.W., C.B., and S.A.I. developed the computational models.  J.W. performed the simulations, interpreted the results, and wrote the mathematical proofs with input from S.A.I, who provided critical feedback and co-interpreted the results.  All author contributed to the writing of the manuscript.

\section*{DECLARATION OF INTERESTS}
The authors declare no conflict of interest.

\bibliography{sample}

\begin{thebibliography}{50}
\expandafter\ifx\csname natexlab\endcsname\relax\def\natexlab#1{#1}\fi
\expandafter\ifx\csname url\endcsname\relax
  \def\url#1{\texttt{#1}}\fi
\expandafter\ifx\csname urlprefix\endcsname\relax\def\urlprefix{URL }\fi

\bibitem[{Abu~Hatoum et~al.(1998)Abu~Hatoum, Gross-Mesilaty, Breitschopf,
  Hoffman, Gonen, Ciechanover, and Bengal}]{abu}
Abu~Hatoum, O., Gross-Mesilaty, S., Breitschopf, K., Hoffman, A., Gonen, H.,
  Ciechanover, A., Bengal, E., 1998. Degradation of myogenic transcription
  factor myod by the ubiquitin pathway in vivo and in vitro: regulation by
  specific dna binding. Mol. Cell. Biol. 18~(10), 5670--5677.

\bibitem[{Alon(2007)}]{alon}
Alon, U., 2007. An Introduction to Systems Biology - Design Principles of
  Biological Circuits. Chapman and Hall.

\bibitem[{Basu et~al.(2004)Basu, Mehreja, Thiberge, Chen, and Weiss}]{basu}
Basu, S., Mehreja, R., Thiberge, S., Chen, M.-T., Weiss, R., 2004.
  Spatiotemporal control of gene expression with pulse-generating networks.
  Proc. Natl. Acad. Sci. U.S.A 101~(17), 6355--6360.

\bibitem[{Bezanson et~al.(2017)Bezanson, Edelman, Karpinski, and
  Shah}]{bezanson}
Bezanson, J., Edelman, A., Karpinski, S., Shah, V., 2017. Julia: A fresh
  approach to numerical computing. SIAM Rev. 59~(1), 65--98.

\bibitem[{Bleris et~al.(2011)Bleris, Xie, Glass, Adadey, Sontag, and
  Benenson}]{bleris}
Bleris, L., Xie, Z., Glass, D., Adadey, A., Sontag, E., Benenson, Y., 2011.
  Synthetic incoherent feedforward circuits show adaptation to the amount of
  their genetic template. Mol. Syst. Biol. 7, 519.

\bibitem[{Boekel(2009)}]{boekel}
Boekel, M. A. J.~S., 2009. Kinetic modeling of reactions in foods. CRC Press,
  Boca Raton, FL.

\bibitem[{Brophy and Voigt(2014)}]{brophy}
Brophy, J. A.~N., Voigt, C.~A., 2014. Principles of genetic circuit design.
  Nat. Methods 11~(5), 508--520.

\bibitem[{Burger et~al.(2010)Burger, Walczak, and Wolynes}]{burger2010}
Burger, A., Walczak, A.~M., Wolynes, P.~G., 2010. Abduction and asylum in the
  lives of transcription factors. Proc. Natl. Acad. Sci. U.S.A 107~(9),
  4016--4021.

\bibitem[{Burger et~al.(2012)Burger, Walczak, and Wolynes}]{burger2012}
Burger, A., Walczak, A.~M., Wolynes, P.~G., 2012. Influence of decoys on the
  noise and dynamics of gene expression. Phys. Rev. E Stat. Nonlin. Soft Matter
  Phys. 86, 041920.

\bibitem[{Castillo-Hair et~al.(2015)Castillo-Hair, Villota, and
  Coronado}]{castillo}
Castillo-Hair, S.~M., Villota, E.~R., Coronado, A.~M., 2015. Design principles
  for robust oscillatory behavior. Syst. Synth. Biol. 9~(3), 125--133.

\bibitem[{Consortium(2012)}]{encode}
Consortium, E.~P., 2012. An integrated encyclopedia of dna elements in the
  human genome. Nature 489~(7414), 57--74.

\bibitem[{Davidsohn et~al.(2015)Davidsohn, Beal, Kiani, Adler, Yaman, Li, Xie,
  and Weiss}]{davidsohn}
Davidsohn, N., Beal, J., Kiani, S., Adler, A., Yaman, F., Li, Y., Xie, Z.,
  Weiss, R., 2015. Accurate predictions of genetic circuit behavior from part
  characterization and modular composition. ACS Synth. Biol. 4~(6), 673--681.

\bibitem[{{Del Vecchio} et~al.(2008){Del Vecchio}, Ninfa, and
  Sontag}]{delvecchio}
{Del Vecchio}, D., Ninfa, A.~J., Sontag, E.~D., 2008. Modular cell biology:
  retroactivity and insulation. Mol. Syst. Biol. 4, 161.

\bibitem[{Esadze et~al.(2014)Esadze, Kemme, Kolomeisky, and Iwahara}]{esadze}
Esadze, A., Kemme, C.~A., Kolomeisky, A.~B., Iwahara, J., 2014. Positive and
  negative impacts of nonspecific sites during target location by a
  sequence-specific dna-binding protein: origin of the optimal search at
  physiological ionic strength. Nucleic Acids Res. 42~(11), 7039--7046.

\bibitem[{Fisher et~al.(2012)Fisher, Li, Hammonds, Brown, Pfeiffer, Weiszmann,
  MacArthur, Thomas, Stamatoyannopoulos, Eisen, Bickel, Biggin, and
  Celniker}]{fisher1}
Fisher, W.~W., Li, J.~J., Hammonds, A.~S., Brown, J.~B., Pfeiffer, B.~D.,
  Weiszmann, R., MacArthur, S., Thomas, S., Stamatoyannopoulos, J.~A., Eisen,
  M.~B., Bickel, P.~J., Biggin, M.~D., Celniker, S.~E., 2012. Dna regions bound
  at low occupancy by transcription factors do not drive patterned reporter
  gene expression in drosophila. Proc. Natl. Acad. Sci. U.S.A 109~(52),
  21330--21335.

\bibitem[{Gao et~al.(2016)Gao, Xiong, Wong, Charles, Lim, and Qi}]{yuchen}
Gao, Y., Xiong, X., Wong, S., Charles, E.~J., Lim, W.~A., Qi, L.~S., 2016.
  Complex transcriptional modulation with orthogonal and inducible dcas9
  regulators. Nat. Methods 13~(12), 1043--1049.

\bibitem[{Gao et~al.(2018)Gao, Yurkovich, Seo, Kabimoldayev, Dr{\"a}ger, Chen,
  Sastry, Fang, Mih, Yang, Eichner, Cho, Kim, and Palsson}]{gao}
Gao, Y., Yurkovich, J.~T., Seo, S.~W., Kabimoldayev, I., Dr{\"a}ger, A., Chen,
  K., Sastry, A.~V., Fang, X., Mih, N., Yang, L., Eichner, J., Cho, B.-K., Kim,
  D., Palsson, B.~O., 2018. Systematic discovery of uncharacterized
  transcription factors in escherichia coli k-12 mg1655. Nucleic Acids Res.
  46~(20), 10682--10696.

\bibitem[{Gardner et~al.(2000)Gardner, Cantor, and Collins}]{gardner}
Gardner, T., Cantor, C., Collins, J., 1 2000. Construction of a genetic toggle
  switch in escherichia coli. Nature 403, 339--342.

\bibitem[{Ghosh et~al.(2019)Ghosh, Shi, Yang, Reddick, Nikitina, Zhurkin,
  Fordyce, Stasevich, Chang, Greenleaf, and Liphardt}]{ghosh}
Ghosh, R.~P., Shi, Q., Yang, L., Reddick, M.~P., Nikitina, T., Zhurkin, V.~B.,
  Fordyce, P., Stasevich, T.~J., Chang, H.~Y., Greenleaf, W.~J., Liphardt,
  J.~T., 2019. Satb1 integrates dna binding site geometry and torsional stress
  to differentially target nucleosome-dense regions. Nat. Commun. 10~(1), 3221.

\bibitem[{Goentoro et~al.(2009)Goentoro, Shoval, Kirschner, and
  Alon}]{goentoro}
Goentoro, L., Shoval, O., Kirschner, M.~W., Alon, U., 2009. The incoherent
  feedforward loop can provide fold-change detection in gene regulation. Mol.
  Cell 36~(5), 894--899.

\bibitem[{Grigolon et~al.(2016)Grigolon, {Di Patti}, {De Martino}, and
  Marinari}]{grigolon}
Grigolon, S., {Di Patti}, F., {De Martino}, A., Marinari, E., 2016. Noise
  processing by microrna-mediated circuits: The incoherent feed-forward loop,
  revisited. Heliyon 2~(4), e00095.

\bibitem[{Gyorgy and {Del Vecchio}(2014)}]{gyorgy}
Gyorgy, A., {Del Vecchio}, D., 2014. Modular composition of gene transcription
  networks. PLOS Comput. Biol.

\bibitem[{Han et~al.(2018)Han, Cho, Lee, Yun, Kim, Bae, Yang, Kim, Lee, Kim,
  Lee, Kang, Jeong, Kim, Jeon, Jung, Nam, Chung, Kim, and Lee}]{ttrust}
Han, H., Cho, J.-W., Lee, S., Yun, A., Kim, H., Bae, D., Yang, S., Kim, C.~Y.,
  Lee, M., Kim, E., Lee, S., Kang, B., Jeong, D., Kim, Y., Jeon, H.-N., Jung,
  H., Nam, S., Chung, M., Kim, J.-H., Lee, I., 2018. Trrust v2: an expanded
  reference database of human and mouse transcriptional regulatory
  interactions. Nucleic Acids Res. 46~(D1), D380--D386.

\bibitem[{Jayanthi et~al.(2013)Jayanthi, Nilgiriwala, and {Del
  Vecchio}}]{jayanthi}
Jayanthi, S., Nilgiriwala, K.~S., {Del Vecchio}, D., 2013. Retroactivity
  controls the temporal dynamics of gene transcription. ACS Synth. Biol. 2~(8),
  431--441.

\bibitem[{Joanito et~al.(2018)Joanito, Chu, Wu, and Hsu}]{joanito}
Joanito, I., Chu, J.-W., Wu, S.-H., Hsu, C.-P., 2018. An incoherent
  feed-forward loop switches the arabidopsis clock rapidly between two
  hysteretic states. Sci. Rep. 8~(1), 13944.

\bibitem[{Kemme et~al.(2015)Kemme, Esadze, and Iwahara}]{kemme2015}
Kemme, C.~A., Esadze, A., Iwahara, J., 2015. Influence of quasi-specific sites
  on kinetics of target dna search by a sequence-specific dna-binding protein.
  Biochemistry 54~(44), 6684--6691.

\bibitem[{Kemme et~al.(2016)Kemme, Nguyen, Chattopadhyay, and Iwahara}]{kemme}
Kemme, C.~A., Nguyen, D., Chattopadhyay, A., Iwahara, J., 2016. Regulation of
  transcription factors via natural decoys in genomic dna. Transcription 7~(4),
  115--120.

\bibitem[{Kim and Sauro(2011)}]{kim}
Kim, K.~H., Sauro, H.~M., 2011. Measuring retroactivity from noise in gene
  regulatory networks. Biophys. J. 100~(5), 1167--1177.

\bibitem[{Kochupurakkal et~al.(2015)Kochupurakkal, Wang, Hua, Culhane, Rodig,
  Rajkovic-Molek, Lazaro, Richardson, Biswas, and Iglehart}]{kochupurakkal}
Kochupurakkal, B.~S., Wang, Z.~C., Hua, T., Culhane, A.~C., Rodig, S.~J.,
  Rajkovic-Molek, K., Lazaro, J.-B., Richardson, A.~L., Biswas, D.~K.,
  Iglehart, J.~D., 2015. Rela-induced interferon response negatively regulates
  proliferation. PLOS ONE 10~(10), 1--33.

\bibitem[{Lee and Maheshri(2012)}]{maheshri}
Lee, T.-H., Maheshri, N., 2012. A regulatory role for repeated decoy
  transcription factor binding sites in target gene expression. Mol. Syst.
  Biol. 8, 576--576.

\bibitem[{Li et~al.(2008)Li, MacArthur, Bourgon, Nix, Pollard, Iyer, Hechmer,
  Simirenko, Stapleton, Luengo~Hendriks, Chu, Ogawa, Inwood, Sementchenko,
  Beaton, Weiszmann, Celniker, Knowles, Gingeras, Speed, Eisen, and
  Biggin}]{li-xiao-yong}
Li, X.-y., MacArthur, S., Bourgon, R., Nix, D., Pollard, D.~A., Iyer, V.~N.,
  Hechmer, A., Simirenko, L., Stapleton, M., Luengo~Hendriks, C.~L., Chu,
  H.~C., Ogawa, N., Inwood, W., Sementchenko, V., Beaton, A., Weiszmann, R.,
  Celniker, S.~E., Knowles, D.~W., Gingeras, T., Speed, T.~P., Eisen, M.~B.,
  Biggin, M.~D., 2008. Transcription factors bind thousands of active and
  inactive regions in the drosophila blastoderm. PLoS Biol. 6~(2), e27.

\bibitem[{Li et~al.(2015)Li, Jiang, Chen, Liao, Li, Weiss, and Xie}]{yinqing}
Li, Y., Jiang, Y., Chen, H., Liao, W., Li, Z., Weiss, R., Xie, Z., 2015.
  Modular construction of mammalian gene circuits using tale transcriptional
  repressors. Nat. Chem. Biol. 11~(3), 207--213.

\bibitem[{Liu et~al.(2007)Liu, Wu, Szary, Kofoed, and Schaufele}]{liu}
Liu, X., Wu, B., Szary, J., Kofoed, E.~M., Schaufele, F., 2007. Functional
  sequestration of transcription factor activity by repetitive dna. J. Biol.
  Chem. 282~(29), 20868--20876.

\bibitem[{Ma et~al.(2009)Ma, Trusina, El-Samad, Lim, and Tang}]{ma}
Ma, W., Trusina, A., El-Samad, H., Lim, W.~A., Tang, C., 2009. Defining network
  topologies that can achieve biochemical adaptation. Cell 138~(4), 760--773.

\bibitem[{Mangan and Alon(2003)}]{mangan}
Mangan, S., Alon, U., 2003. Structure and function of the feed-forward loop
  network motif. Proc. Natl. Acad. Sci. U.S.A 100~(21), 11980--11985.

\bibitem[{Milo et~al.(2002)Milo, Shen-Orr, Itzkovitz, Kashtan, Chklovskii, and
  Alon}]{milo}
Milo, R., Shen-Orr, S., Itzkovitz, S., Kashtan, N., Chklovskii, D., Alon, U.,
  2002. Network motifs: Simple building blocks of complex networks. Science
  298~(5594), 824--827.

\bibitem[{Osella et~al.(2011)Osella, Bosia, Cor{\'a}, and Caselle}]{osella}
Osella, M., Bosia, C., Cor{\'a}, D., Caselle, M., 2011. The role of incoherent
  microrna-mediated feedforward loops in noise buffering. PLOS Comput. Biol.
  7~(3), 1--16.

\bibitem[{Pariat et~al.(1997)Pariat, Carillo, Molinari, Salvat, Deb{\"u}ssche,
  Bracco, Milner, and Piechaczyk}]{pariat}
Pariat, M., Carillo, S., Molinari, M., Salvat, C., Deb{\"u}ssche, L., Bracco,
  L., Milner, J., Piechaczyk, M., 1997. Proteolysis by calpains: a possible
  contribution to degradation of p53. Mol. Cell. Biol. 17~(5), 2806--2815.

\bibitem[{Rackauckas and Nie(2017)}]{rackauckas}
Rackauckas, C., Nie, Q., 2017. {DifferentialEquations.jl} -- a performant and
  feature-rich ecosystem for solving differential equations in {Julia}. J. Open
  Source Softw. 5~(1), 15.

\bibitem[{Reeves(2019)}]{reeves}
Reeves, G.~T., 2019. The engineering principles of combining a transcriptional
  incoherent feedforward loop with negative feedback. Journal of Biological
  Engineering 13~(1), 62.
\newline\urlprefix\url{https://doi.org/10.1186/s13036-019-0190-3}

\bibitem[{Rosenfeld et~al.(2002)Rosenfeld, Elowitz, and Alon}]{rosenfeld}
Rosenfeld, N., Elowitz, M.~B., Alon, U., 2002. Negative autoregulation speeds
  the response times of transcription networks. J. Mol. Biol. 323~(5),
  785--793.

\bibitem[{Santos-Zavaleta et~al.(2018)Santos-Zavaleta, S{\'a}nchez-P{\'e}rez,
  Salgado, Vel{\'a}zquez-Ram{\'\i}rez, Gama-Castro, Tierrafr{\'\i}a, Busby,
  Aquino, Fang, Palsson, Galagan, and Collado-Vides}]{regulondb10}
Santos-Zavaleta, A., S{\'a}nchez-P{\'e}rez, M., Salgado, H.,
  Vel{\'a}zquez-Ram{\'\i}rez, D.~A., Gama-Castro, S., Tierrafr{\'\i}a, V.~H.,
  Busby, S. J.~W., Aquino, P., Fang, X., Palsson, B.~O., Galagan, J.~E.,
  Collado-Vides, J., 2018. A unified resource for transcriptional regulation in
  escherichia coli k-12 incorporating high-throughput-generated binding data
  into regulondb version 10.0. BMC Biol. 16~(1), 91.

\bibitem[{Sepulchre and Ventura(2013)}]{sepulchre}
Sepulchre, J.-A., Ventura, A.~C., 2013. Intrinsic feedbacks in {MAPK} signaling
  cascades lead to bistability and oscillations. Acta. Biotheor. 61~(1),
  59--78.

\bibitem[{Shi et~al.(2017)Shi, Ma, Xiong, Zhang, and Tang}]{shi}
Shi, W., Ma, W., Xiong, L., Zhang, M., Tang, C., 2017. Adaptation with
  transcriptional regulation. Sci. Rep. 7, 42648.

\bibitem[{Siciliano et~al.(2013)Siciliano, Garzilli, Fracassi, Criscuolo,
  Ventre, and di~Bernardo}]{siciliano}
Siciliano, V., Garzilli, I., Fracassi, C., Criscuolo, S., Ventre, S.,
  di~Bernardo, D., 2013. mirnas confer phenotypic robustness to gene networks
  by suppressing biological noise. Nat. Commun. 4~(1), 2364.

\bibitem[{Stewart et~al.(2013)Stewart, Seymour, Pomiankowski, and
  Reuter}]{stewart}
Stewart, A.~J., Seymour, R.~M., Pomiankowski, A., Reuter, M., 2013.
  Under-dominance constrains the evolution of negative autoregulation in
  diploids. PLOS Comput. Biol. 9~(3), e1002992.

\bibitem[{{Wang} and {Belta}(2019)}]{wang}
{Wang}, J., {Belta}, C., 2019. Retroactivity affects the adaptive robustness of
  transcriptional regulatory networks. In: 2019 American Control Conference
  (ACC). Philadelphia, PA, USA, pp. 5396--5401.

\bibitem[{Wang et~al.(2018)Wang, Isaacson, and Belta}]{wang2018}
Wang, J., Isaacson, S.~A., Belta, C., 2018. Predictions of genetic circuit
  behaviors based on modular composition in transiently transfected mammalian
  cells. 2018 IEEE Life Sciences Conference (LSC), 85--88.

\bibitem[{Wang et~al.(2019)Wang, Isaacson, and Belta}]{wang_acs}
Wang, J., Isaacson, S.~A., Belta, C., 2019. Modeling genetic circuit behavior
  in transiently transfected mammalian cells. ACS Synth. Biol. 8~(4), 697--707.

\bibitem[{Wang et~al.(2016)Wang, Potoyan, and Wolynes}]{zhipeng}
Wang, Z., Potoyan, D.~A., Wolynes, P.~G., 2016. Molecular stripping, targets
  and decoys as modulators of oscillations in the nf-κb/iκbα/dna genetic
  network. J. R. Soc. Interface 13~(122), 20160606.

\end{thebibliography}



\section*{Supplementary material}

# Contents

# 1 Transparent Methods

## 1.1 Hill Function

Hill functions are commonly used to model transcriptional regulation in ODE models. Here, we consider the case where species $x_i$ is regulated by multiple TFs. Let $m_i$ be the number of parents of node $i$ and $M_i$ be the collection of all nonempty subsets of $\vec{p}_i$, i.e., $\{1,2,...,m_i\}$. Then under the assumption of independent binding, $H_i(\vec{p}_i)$ can be expressed as:

$$H_i(\vec{p}_i) = \frac{\sum_{X \in M_i} \pi_{i,X} \prod_{j \in X} \left(\frac{p_{ij}}{K_{ij}}\right)^{h_{ij}}}{1 + \sum_{X \in M_i} \prod_{j \in X} \left(\frac{p_{ij}}{K_{ij}}\right)^{h_{ij}}}, \tag{1}$$

where $p_{ij}$, $h_{ij}$, and $K_{ij}$ are counterparts of $p_i$, $h_i$, $K_i$ for the $j$-th parent of node $i$. Similar to Gyorgy and Del Vecchio (2014), here we assume that no parents of the same node are identical. $X$ corresponds to each complex formed by a different combination of TFs, and $\pi_{i,X}$ denotes the normalized production rate of species $x_i$ due to the corresponding complex.

Under the assumption of AND logic, the regulated gene is turned on only when all the corresponding activators aggregate and bind to the promoter. Let $M_{i,A} = \{j | j \in \{1,2,...,m_i\}$ and $p_{ij}$ is an activator$\}$ denote the complex formed by all the activators. Then

$$\pi_{i,X} = \begin{cases} 1, & \text{if } X = M_{i,A}, \\ 0, & \text{otherwise}. \end{cases} \tag{2}$$

## 1.2 Retroactivity Matrix

Under the assumption of AND logic, $R_i(\vec{p}_i)$ is a diagonal matrix, and the $k$-th entry on the diagonal $r_{ik}$ is (Gyorgy and Del Vecchio (2014)):

$$r_{ik} = \eta_i \frac{h_{ik}^2 p_{ik}^{h_{ik}-1}}{K_{ik}^{h_{ik}}} \left(1 + \left(\frac{p_{ik}}{K_{ik}}\right)^{h_{ik}}\right)^{-2}, \tag{3}$$

where $\eta_i$ stands for the number of downstream binding sites (DNA copy number) of node $i$. $p_{ik}$, $h_{ik}$, and $K_{ik}$ are the protein concentration, the Hill coefficient, and the dissociation coefficient of the $k$-th parent of node $i$.

## 1.3 Non-dimensionalization and Parameter Sampling

In order to reduce the dimensions of parameter space, we non-dimensionalized our models via methods shown in Cao et al. (2016). We rescaled the model parameters via the following equations:

$$\begin{aligned}
\tilde{x}_A &= \frac{x_A \delta_A}{\beta_A} & \tilde{x}_B &= \frac{x_B \delta_B}{\beta_B} & \tilde{x}_C &= \frac{x_C \delta_C}{\beta_C} \\
\tilde{K}_{AB} &= \frac{K_{AB} \delta_A}{\beta_A} & \tilde{K}_{AC} &= \frac{K_{AC} \delta_A}{\beta_A} & \tilde{K}_{BC} &= \frac{K_{BC} \delta_B}{\beta_B} \\
\tilde{K}_{AD_A} &= \frac{K_{AD_A} \delta_A}{\beta_A} & \tilde{K}_{BD_B} &= \frac{K_{BD_B} \delta_B}{\beta_B} & \tilde{K}_{CD_C} &= \frac{K_{CD_C} \delta_C}{\beta_C} \\
\tilde{\eta}_{AD_A} &= \frac{\eta_A}{K_{AD_A}} & \tilde{\eta}_{BD_B} &= \frac{\eta_B}{K_{BD_B}} & \tilde{\eta}_{CD_C} &= \frac{\eta_C}{K_{CD_C}}
\end{aligned} \tag{4}$$

For simplicity, we assumed proteins A, B, and C have equal decay rates, i.e., $\delta_A = \delta_B = \delta_C$. To non-dimensionalize time, we rescaled $t$ against the mean lifetime (equal to the reciprocal of the decay rate):

$$\tau = \frac{t}{\frac{1}{\delta}} = t \cdot \delta. \tag{5}$$

Non-dimensionlized models of IFFLs with and without retroactivity, are provided in Supplemental Information Section 1.6. To simplify our analysis, we assume the following for all simulations carried out in this work except Section 2.6 and Supplemental Information Section 1.19: each protein binds to its downstream binding sites, including both the functional target site and accessible ND sites, with equal affinity and equal cooperativity, i.e., $\tilde{K}_{AB} = \tilde{K}_{AC} = \tilde{K}_{AD_A}$, $\tilde{K}_{BC} = \tilde{K}_{BD_B}$, $h_{AB} = h_{AC} = h_{AD_A}$, and $h_{BC} = h_{BD_B}$. To ensure sufficient coverage of parameter space, we sampled the kinetic parameters $\tilde{K}_{XD_X}$ (X=A, B, C) spanning two orders of magnitude: $\tilde{K}_{XD_X} \in \{0.01, 0.03, 0.1, 0.3, 1.0\}$ and included both positive and negative cooperative binding, sampling Hill coefficients $h_{XD_X}$ ($X = A, B, C$) at 0.5, 1.0, and 2.0 (Mangan and Alon (2003)). The retroactivity coefficient $\tilde{\eta}$ is the total concentration of the accessible ND sites for a given TF divided by the corresponding dissociation constant (Wang and Belta (2019)). Without loss of generality, the basal fraction of the promoter that is active, $\gamma_X$ ($X = A, B, C$) is assumed to be $10^{-5}$.

## 1.4 Proof of Diagonality

Here, we prove that $V_i^T R_i(\vec{p}_i) V_i$ is a diagonal matrix. Let $V_i^T$ be an $n \times m$ matrix and $a_{kj}$ denote the $(k, j)$-th entry of $V_i^T$. Recall that each row of $V_i$ has only one non-zero entry by definition. Similarly, let $R_i(\vec{p}_i)$ be an $m \times m$ diagonal matrix and $x_{k'j'}$ denote the $(k', j')$-th entry of $R_i(\vec{p}_i)$. This implies that the product $V_i^T R_i(\vec{p}_i)$ is an $n \times m$ matrix. The $(k^*, j^*)$-th entry of the product $V_i^T R_i(\vec{p}_i)$ can be expressed as:

$$\sum_{u=1}^{m} a_{k^* u} x_{u j^*} = a_{k^* j^*} x_{j^* j^*},$$

as $R_i(\vec{p}_i)$ is diagonal. The $(\hat{k}, \hat{j})$-th entry of the product $V_i^T R_i(\vec{p}_i) V_i$ can be expressed as:

$$\sum_{v=1}^{n} \left( a_{\hat{k} v} x_{vv} \right) a_{\hat{j} v} = \sum_{v=1}^{n} \left( a_{\hat{k} v} a_{\hat{j} v} \right) x_{vv}.$$

Here $a_{\hat{k} v} a_{\hat{j} v} = 0$ for $\hat{k} \neq \hat{j}$ as otherwise there would be two non-zero entries in the same row of $V_i$, which contradicts the definition of $V_i$. Hence, the $(\hat{k}, \hat{j})$-th entry of the product $V_i^T R_i(\vec{p}_i) V_i$ is always zero if $\hat{k} \neq \hat{j}$. In other words, $V_i^T R_i(\vec{p}_i) V_i$ is a diagonal matrix.

## 1.5 ODE Models for IFFL

Without retroactivity, the ODE model for the I1-FFL is given as:

$$\frac{dx_A}{dt} = f_A = \beta_A \left[ (1 - \gamma_A) \frac{\left(\frac{x_I}{K_{IA}}\right)^{h_{IA}}}{1 + \left(\frac{x_I}{K_{IA}}\right)^{h_{IA}}} + \gamma_A \right] - \delta_A x_A$$

$$\frac{dx_B}{dt} = f_B = \beta_B \left[ (1 - \gamma_B) \frac{\left(\frac{x_A}{K_{AB}}\right)^{h_{AB}}}{1 + \left(\frac{x_A}{K_{AB}}\right)^{h_{AB}}} + \gamma_B \right] - \delta_B x_B$$

$$\frac{dx_C}{dt} = f_C = \beta_C \left[ (1 - \gamma_C) \frac{\left(\frac{x_A}{K_{AC}}\right)^{h_{AC}}}{\left(1 + \left(\frac{x_A}{K_{AC}}\right)^{h_{AC}}\right)\left(1 + \left(\frac{x_B}{K_{BC}}\right)^{h_{BC}}\right)} + \gamma_C \right] - \delta_C x_C.$$

I2-FFL:

$$\frac{dx_A}{dt} = f_A = \beta_A \left[ (1 - \gamma_A) \frac{\left(\frac{x_I}{K_{IA}}\right)^{h_{IA}}}{1 + \left(\frac{x_I}{K_{IA}}\right)^{h_{IA}}} + \gamma_A \right] - \delta_A x_A$$

$$\frac{dx_B}{dt} = f_B = \beta_B \left[ \frac{1 - \gamma_B}{1 + \left(\frac{x_A}{K_{AB}}\right)^{h_{AB}}} + \gamma_B \right] - \delta_B x_B$$

$$\frac{dx_C}{dt} = f_C = \beta_C \left[ \frac{1 - \gamma_C}{\left(1 + \left(\frac{x_A}{K_{AC}}\right)^{h_{AC}}\right)\left(1 + \left(\frac{x_B}{K_{BC}}\right)^{h_{BC}}\right)} + \gamma_C \right] - \delta_C x_C.$$

I3-FFL:

$$\frac{dx_A}{dt} = f_A = \beta_A \left[ (1-\gamma_A) \frac{\left(\frac{x_I}{K_{IA}}\right)^{h_{IA}}}{1+\left(\frac{x_I}{K_{IA}}\right)^{h_{IA}}} + \gamma_A \right] - \delta_A x_A$$

$$\frac{dx_B}{dt} = f_B = \beta_B \left[ (1-\gamma_B) \frac{\left(\frac{x_A}{K_{AB}}\right)^{h_{AB}}}{1+\left(\frac{x_A}{K_{AB}}\right)^{h_{AB}}} + \gamma_B \right] - \delta_B x_B$$

$$\frac{dx_C}{dt} = f_C = \beta_C \left[ (1-\gamma_C) \frac{\left(\frac{x_B}{K_{BC}}\right)^{h_{BC}}}{\left(1+\left(\frac{x_A}{K_{AC}}\right)^{h_{AC}}\right)\left(1+\left(\frac{x_B}{K_{BC}}\right)^{h_{BC}}\right)} + \gamma_C \right] - \delta_C x_C.$$

I4-FFL:

$$\frac{dx_A}{dt} = f_A = \beta_A \left[ (1-\gamma_A) \frac{\left(\frac{x_I}{K_{IA}}\right)^{h_{IA}}}{1+\left(\frac{x_I}{K_{IA}}\right)^{h_{IA}}} + \gamma_A \right] - \delta_A x_A$$

$$\frac{dx_B}{dt} = f_B = \beta_B \left[ \frac{1-\gamma_B}{1+\left(\frac{x_A}{K_{AB}}\right)^{h_{AB}}} + \gamma_B \right] - \delta_B x_B$$

$$\frac{dx_C}{dt} = f_C = \beta_C \left[ (1-\gamma_C) \frac{\left(\frac{x_A}{K_{AC}}\right)^{h_{AC}}\left(\frac{x_B}{K_{BC}}\right)^{h_{BC}}}{\left(1+\left(\frac{x_A}{K_{AC}}\right)^{h_{AC}}\right)\left(1+\left(\frac{x_B}{K_{BC}}\right)^{h_{BC}}\right)} + \gamma_C \right] - \delta_C x_C.$$

When only retroactivity on A is considered, the retroactivity matrix $R(\vec{x})$, which is the same for all four IFFLs, is calculated as (Gyorgy and Del Vecchio (2014)):

$$R(\vec{x}) = \begin{bmatrix} \eta_{D_A} \frac{h_{AD_A}^2 x_A^{h_{AD_A}-1}}{K_{AD_A}^{h_{AD_A}}} \left(1+\left(\frac{x_A}{K_{AD_A}}\right)^{h_{AD_A}}\right)^{-2} & 0 & 0 \\ 0 & 0 & 0 \\ 0 & 0 & 0 \end{bmatrix}.$$

When only retroactivity on B is considered, the retroactivity matrix $R(\vec{x})$, which is the same for all four IFFLs, is calculated as (Gyorgy and Del Vecchio (2014)):

$$R(\vec{x}) = \begin{bmatrix} 0 & 0 & 0 \\ 0 & \eta_{D_B} \frac{h_{BD_B}^2 x_B^{h_{BD_B}-1}}{K_{BD_B}^{h_{BD_B}}} \left(1+\left(\frac{x_B}{K_{BD_B}}\right)^{h_{BD_B}}\right)^{-2} & 0 \\ 0 & 0 & 0 \end{bmatrix}.$$

When only retroactivity on C is considered, the retroactivity matrix $R(\vec{x})$, which is the same for all four IFFLs, is calculated as (Gyorgy and Del Vecchio (2014)):

$$R(\vec{x}) = \begin{bmatrix} 0 & 0 & 0 \\ 0 & 0 & 0 \\ 0 & 0 & \eta_{D_C} \frac{h_{CD_C}^2 x_C^{h_{CD_C}-1}}{K_{CD_C}^{h_{CD_C}}} \left(1+\left(\frac{x_C}{K_{CD_C}}\right)^{h_{CD_C}}\right)^{-2} \end{bmatrix}.$$

## 1.6 Non-Dimensionalized ODE Models for IFFL

Without retroactivity, the non-dimensionalized ODE model for the I1-FFL is given as:

$$\frac{d\tilde{x}_A}{d\tau} = f_{\tilde{A}} = (1-\gamma_A) \frac{\left(\frac{x_I}{K_{IA}}\right)^{h_{IA}}}{1+\left(\frac{x_I}{K_{IA}}\right)^{h_{IA}}} + \gamma_A - \tilde{x}_A$$

$$\frac{d\tilde{x}_B}{d\tau} = f_{\tilde{B}} = (1-\gamma_B) \frac{\left(\frac{\tilde{x}_A}{\tilde{K}_{AB}}\right)^{h_{AB}}}{1+\left(\frac{\tilde{x}_A}{\tilde{K}_{AB}}\right)^{h_{AB}}} + \gamma_B - \tilde{x}_B$$

$$\frac{d\tilde{x}_C}{d\tau} = f_{\tilde{C}} = (1-\gamma_C) \frac{\left(\frac{\tilde{x}_A}{\tilde{K}_{AC}}\right)^{h_{AC}}}{\left(1+\left(\frac{\tilde{x}_A}{\tilde{K}_{AC}}\right)^{h_{AC}}\right)\left(1+\left(\frac{\tilde{x}_B}{\tilde{K}_{BC}}\right)^{h_{BC}}\right)} + \gamma_C - \tilde{x}_C.$$

I2-FFL:

$$\frac{d\tilde{x}_A}{d\tau} = f_{\tilde{A}} = (1-\gamma_A) \frac{\left(\frac{x_I}{K_{IA}}\right)^{h_{IA}}}{1+\left(\frac{x_I}{K_{IA}}\right)^{h_{IA}}} + \gamma_A - \tilde{x}_A$$

$$\frac{d\tilde{x}_B}{d\tau} = f_{\tilde{B}} = \frac{1-\gamma_B}{1+\left(\frac{\tilde{x}_A}{\tilde{K}_{AB}}\right)^{h_{AB}}} + \gamma_B - \tilde{x}_B$$

$$\frac{d\tilde{x}_C}{d\tau} = f_{\tilde{C}} = \frac{1-\gamma_C}{\left(1+\left(\frac{\tilde{x}_A}{\tilde{K}_{AC}}\right)^{h_{AC}}\right)\left(1+\left(\frac{\tilde{x}_B}{\tilde{K}_{BC}}\right)^{h_{BC}}\right)} + \gamma_C - \tilde{x}_C.$$

I3-FFL:

$$\frac{d\tilde{x}_A}{d\tau} = f_{\tilde{A}} = (1-\gamma_A) \frac{\left(\frac{x_I}{K_{IA}}\right)^{h_{IA}}}{1+\left(\frac{x_I}{K_{IA}}\right)^{h_{IA}}} + \gamma_A - \tilde{x}_A$$

$$\frac{d\tilde{x}_B}{d\tau} = f_{\tilde{B}} = (1-\gamma_B) \frac{\left(\frac{\tilde{x}_A}{\tilde{K}_{AB}}\right)^{h_{AB}}}{1+\left(\frac{\tilde{x}_A}{\tilde{K}_{AB}}\right)^{h_{AB}}} + \gamma_B - \tilde{x}_B$$

$$\frac{d\tilde{x}_C}{d\tau} = f_{\tilde{C}} = (1-\gamma_C) \frac{\left(\frac{\tilde{x}_B}{\tilde{K}_{BC}}\right)^{h_{BC}}}{\left(1+\left(\frac{\tilde{x}_A}{\tilde{K}_{AC}}\right)^{h_{AC}}\right)\left(1+\left(\frac{\tilde{x}_B}{\tilde{K}_{BC}}\right)^{h_{BC}}\right)} + \gamma_C - \tilde{x}_C.$$

I4-FFL:

$$\frac{d\tilde{x}_A}{d\tau} = f_{\tilde{A}} = (1-\gamma_A) \frac{\left(\frac{x_I}{K_{IA}}\right)^{h_{IA}}}{1+\left(\frac{x_I}{K_{IA}}\right)^{h_{IA}}} + \gamma_A - \tilde{x}_A$$

$$\frac{d\tilde{x}_B}{d\tau} = f_{\tilde{B}} = \frac{1-\gamma_B}{1+\left(\frac{\tilde{x}_A}{\tilde{K}_{AB}}\right)^{h_{AB}}} + \gamma_B - \tilde{x}_B$$

$$\frac{d\tilde{x}_C}{d\tau} = f_{\tilde{C}} = (1-\gamma_C) \frac{\left(\frac{\tilde{x}_A}{\tilde{K}_{AC}}\right)^{h_{AC}}\left(\frac{\tilde{x}_B}{\tilde{K}_{BC}}\right)^{h_{BC}}}{\left(1+\left(\frac{\tilde{x}_A}{\tilde{K}_{AC}}\right)^{h_{AC}}\right)\left(1+\left(\frac{\tilde{x}_B}{\tilde{K}_{BC}}\right)^{h_{BC}}\right)} + \gamma_C - \tilde{x}_C.$$

$R(\vec{\tilde{x}})$ with only retroactivity on A can be written as (Gyorgy and Del Vecchio (2014)):

$$R(\vec{\tilde{x}}) = \begin{bmatrix} \tilde{\eta}_{AD_A} h_{AD_A}^2 \left(\frac{\tilde{x}_A}{\tilde{K}_{AD_A}}\right)^{h_{AD_A}-1} \left(1 + \left(\frac{\tilde{x}_A}{\tilde{K}_{AD_A}}\right)^{h_{AD_A}}\right)^{-2} & 0 & 0 \\ 0 & 0 & 0 \\ 0 & 0 & 0 \end{bmatrix}.$$

$R(\vec{\tilde{x}})$ with only retroactivity on B can be written as (Gyorgy and Del Vecchio (2014)):

$$R(\vec{\tilde{x}}) = \begin{bmatrix} 0 & 0 & 0 \\ 0 & \tilde{\eta}_{BD_B} h_{BD_B}^2 \left(\frac{\tilde{x}_B}{\tilde{K}_{BD_B}}\right)^{h_{BD_B}-1} \left(1 + \left(\frac{\tilde{x}_B}{\tilde{K}_{BD_B}}\right)^{h_{BD_B}}\right)^{-2} & 0 \\ 0 & 0 & 0 \end{bmatrix}.$$

$R(\vec{\tilde{x}})$ with only retroactivity on C can be written as (Gyorgy and Del Vecchio (2014)):

$$R(\vec{\tilde{x}}) = \begin{bmatrix} 0 & 0 & 0 \\ 0 & 0 & 0 \\ 0 & 0 & \tilde{\eta}_{CD_C} h_{CD_C}^2 \left(\frac{\tilde{x}_C}{\tilde{K}_{CD_C}}\right)^{h_{CD_C}-1} \left(1 + \left(\frac{\tilde{x}_C}{\tilde{K}_{CD_C}}\right)^{h_{CD_C}}\right)^{-2} \end{bmatrix}.$$

## 1.7 Non-Dimensionalized ODE Models for Other Sign-Sensitive Response-Acceleration Motifs

Without retroactivity, the non-dimensionalized ODE model for a type-1 two-input circuit is given as:

$$\frac{d\tilde{x}_A}{d\tau} = f_{\tilde{A}} = (1-\gamma_A)\frac{\left(\frac{x_I}{K_{IA}}\right)^{h_{IA}}}{1+\left(\frac{x_I}{K_{IA}}\right)^{h_{IA}}} + \gamma_A - \tilde{x}_A$$

$$\frac{d\tilde{x}_{A_2}}{d\tau} = f_{\tilde{A}_2} = (1-\gamma_{A_2})\frac{\left(\frac{x_{I_2}}{K_{IA_2}}\right)^{h_{IA_2}}}{1+\left(\frac{x_{I_2}}{K_{IA_2}}\right)^{h_{IA_2}}} + \gamma_{A_2} - \tilde{x}_{A_2}$$

$$\frac{d\tilde{x}_B}{d\tau} = f_{\tilde{B}} = (1-\gamma_B)\frac{\left(\frac{\tilde{x}_{A_2}}{\tilde{K}_{A_2B}}\right)^{h_{A_2B}}}{1+\left(\frac{\tilde{x}_{A_2}}{\tilde{K}_{A_2B}}\right)^{h_{A_2B}}} + \gamma_B - \tilde{x}_B$$

$$\frac{d\tilde{x}_C}{d\tau} = f_{\tilde{C}} = (1-\gamma_C)\frac{\left(\frac{\tilde{x}_A}{\tilde{K}_{AC}}\right)^{h_{AC}}}{\left(1+\left(\frac{\tilde{x}_A}{\tilde{K}_{AC}}\right)^{h_{AC}}\right)\left(1+\left(\frac{\tilde{x}_B}{\tilde{K}_{BC}}\right)^{h_{BC}}\right)} + \gamma_C - \tilde{x}_C.$$

The non-dimensionalized ODE model for a type-4 two-input circuit is given as:

$$\frac{d\tilde{x}_A}{d\tau} = f_{\tilde{A}} = (1-\gamma_A)\frac{\left(\frac{x_I}{K_{IA}}\right)^{h_{IA}}}{1+\left(\frac{x_I}{K_{IA}}\right)^{h_{IA}}} + \gamma_A - \tilde{x}_A$$

$$\frac{d\tilde{x}_{A_2}}{d\tau} = f_{\tilde{A}_2} = (1-\gamma_{A_2})\frac{\left(\frac{x_{I_2}}{K_{IA_2}}\right)^{h_{IA_2}}}{1+\left(\frac{x_{I_2}}{K_{IA_2}}\right)^{h_{IA_2}}} + \gamma_{A_2} - \tilde{x}_{A_2}$$

$$\frac{d\tilde{x}_B}{d\tau} = f_{\tilde{B}} = (1-\gamma_B)\frac{1}{1+\left(\frac{\tilde{x}_{A_2}}{\tilde{K}_{A_2B}}\right)^{h_{A_2B}}} + \gamma_B - \tilde{x}_B$$

$$\frac{d\tilde{x}_C}{d\tau} = f_{\tilde{C}} = (1-\gamma_C)\frac{\left(\frac{\tilde{x}_A}{\tilde{K}_{AC}}\right)^{h_{AC}}\left(\frac{\tilde{x}_B}{\tilde{K}_{BC}}\right)^{h_{BC}}}{\left(1+\left(\frac{\tilde{x}_A}{\tilde{K}_{AC}}\right)^{h_{AC}}\right)\left(1+\left(\frac{\tilde{x}_B}{\tilde{K}_{BC}}\right)^{h_{BC}}\right)} + \gamma_C - \tilde{x}_C.$$

The non-dimensionalized ODE model for a negative autoregulated circuit is given as:

$$\frac{d\tilde{x}_A}{d\tau} = f_{\tilde{A}} = (1-\gamma_A)\frac{\left(\frac{x_I}{K_{IA}}\right)^{h_{IA}}}{1+\left(\frac{x_I}{K_{IA}}\right)^{h_{IA}}} + \gamma_A - \tilde{x}_A$$

$$\frac{d\tilde{x}_C}{d\tau} = f_{\tilde{C}} = (1-\gamma_C)\frac{\left(\frac{\tilde{x}_A}{\tilde{K}_{AC}}\right)^{h_{AC}}}{\left(1+\left(\frac{\tilde{x}_A}{\tilde{K}_{AC}}\right)^{h_{AC}}\right)\left(1+\left(\frac{\tilde{x}_C}{\tilde{K}_{CC}}\right)^{h_{CC}}\right)} + \gamma_C - \tilde{x}_C.$$

When only retroactivity on A is considered, the retroactivity matrix $R(\vec{\tilde{x}})$, which is the same for both type-1 and type-4 two-input circuits, is given as:

$$R(\vec{\tilde{x}}) = \begin{bmatrix} \tilde{\eta}_{AD_A} h_{AD_A}^2 \left(\frac{\tilde{x}_A}{\tilde{K}_{AD_A}}\right)^{h_{AD_A}-1} \left(1+\left(\frac{\tilde{x}_A}{\tilde{K}_{AD_A}}\right)^{h_{AD_A}}\right)^{-2} & 0 & 0 & 0 \\ 0 & 0 & 0 & 0 \\ 0 & 0 & 0 & 0 \\ 0 & 0 & 0 & 0 \end{bmatrix}.$$

When only retroactivity on A is considered, the retroactivity matrix $R(\vec{x})$ for a negative autoregulated circuit is given as:

$$R(\vec{x}) = \begin{bmatrix} \tilde{\eta}_{AD_A} h_{AD_A}^2 \left(\frac{\tilde{x}_A}{\tilde{K}_{AD_A}}\right)^{h_{AD_A}-1} \left(1 + \left(\frac{\tilde{x}_A}{\tilde{K}_{AD_A}}\right)^{h_{AD_A}}\right)^{-2} & 0 & 0 \\ 0 & 0 & 0 \\ 0 & 0 & 0 \end{bmatrix}.$$

When only retroactivity on C is considered, the retroactivity matrix $R(\vec{x})$ for a negative autoregulated circuit is given as:

$$R(\vec{x}) = \begin{bmatrix} 0 & 0 & 0 \\ 0 & 0 & 0 \\ 0 & 0 & \tilde{\eta}_{CD_C} h_{CD_C}^2 \left(\frac{\tilde{x}_C}{\tilde{K}_{CD_C}}\right)^{h_{CD_C}-1} \left(1 + \left(\frac{\tilde{x}_C}{\tilde{K}_{CD_C}}\right)^{h_{CD_C}}\right)^{-2} \end{bmatrix}.$$

## 1.8 Pulsing Behavior of I1-FFL at Different Levels of $\tilde{\eta}_{BD_B}$

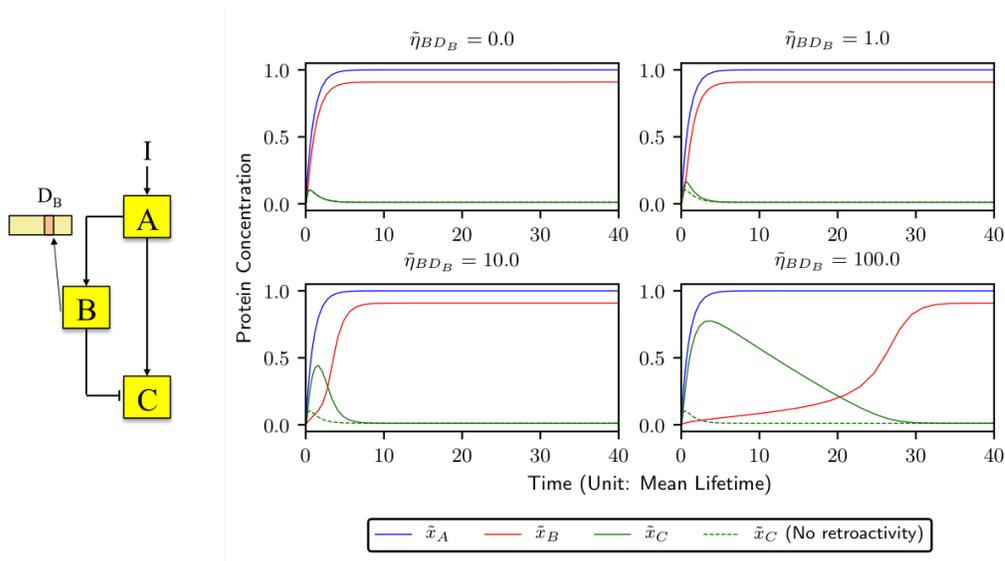

**Figure S1:** Increasing pulse amplitude due to increasing $\tilde{\eta}_{BD_B}$ in an I1-FFL. $\tilde{\eta}_{BD_B}$ increases in the order of top left, top right, bottom left, and bottom right. Values of the other parameters are: $\tilde{K}_{AB} = \tilde{K}_{AC} = \tilde{K}_{BC} = \tilde{K}_{BD_B} = 0.1$, $h_{AB} = h_{AC} = 1.0$, $h_{BC} = h_{BD_B} = 2.0$. For comparison, the green dashed curve represents the trajectory of $\tilde{x}_C$ when $\tilde{\eta}_{BD_B}$ equals 0.

## 1.9 Response Time of IFFLs at Different Levels of $\tilde{\eta}_{BD_B}$ and $\tilde{\eta}_{CD_C}$

|  |  | I1-FFL | | | | I4-FFL | | | |
|---|---|---|---|---|---|---|---|---|---|
|  |  | $h_{BD_B} = 0.5$ | | | | | | | |
| $\tilde{K}_{BD_B}$ | $\tilde{\eta}_{BD_B}$ | 0.0 | 1.0 | 10.0 | 100.0 | 0.0 | 1.0 | 10.0 | 100.0 |
| 0.01 |  | 0.31 | 0.30 | 0.23 | 0.13 | 0.75 | 0.75 | 0.75 | 0.75 |
| 0.03 |  | 0.37 | 0.36 | 0.25 | 0.19 | 0.68 | 0.68 | 0.68 | 0.68 |
| 0.1 |  | 0.46 | 0.43 | 0.30 | 0.26 | 0.59 | 0.59 | 0.59 | 0.58 |
| 0.3 |  | 0.54 | 0.50 | 0.37 | 0.35 | 0.51 | 0.51 | 0.51 | 0.50 |
| 1.0 |  | 0.64 | 0.58 | 0.48 | 0.47 | 0.44 | 0.44 | 0.43 | 0.43 |
|  |  | $h_{BD_B} = 1.0$ | | | | | | | |
| $\tilde{K}_{BD_B}$ | $\tilde{\eta}_{BD_B}$ | 0.0 | 1.0 | 10.0 | 100.0 | 0.0 | 1.0 | 10.0 | 100.0 |
| 0.01 |  | 0.04 | 0.04 | 0.04 | 0.04 | 0.83 | 0.83 | 0.83 | 0.83 |
| 0.03 |  | 0.08 | 0.07 | 0.07 | 0.07 | 0.70 | 0.70 | 0.70 | 0.70 |
| 0.1 |  | 0.16 | 0.15 | 0.13 | 0.13 | 0.48 | 0.48 | 0.48 | 0.48 |
| 0.3 |  | 0.30 | 0.28 | 0.26 | 0.26 | 0.30 | 0.30 | 0.30 | 0.30 |
| 1.0 |  | 0.54 | 0.51 | 0.48 | 0.47 | 0.19 | 0.19 | 0.19 | 0.19 |
|  |  | $h_{BD_B} = 2.0$ | | | | | | | |
| $\tilde{K}_{BD_B}$ | $\tilde{\eta}_{BD_B}$ | 0.0 | 1.0 | 10.0 | 100.0 | 0.0 | 1.0 | 10.0 | 100.0 |
| 0.01 |  | 0.001 | 0.001 | 0.001 | 0.001 | 0.91 | 0.91 | 0.91 | 0.91 |
| 0.03 |  | 0.01 | 0.01 | 0.01 | 0.01 | 0.82 | 0.82 | 0.82 | 0.82 |
| 0.1 |  | 0.04 | 0.04 | 0.04 | 0.04 | 0.42 | 0.42 | 0.42 | 0.42 |
| 0.3 |  | 0.13 | 0.13 | 0.13 | 0.13 | 0.13 | 0.13 | 0.13 | 0.13 |
| 1.0 |  | 0.50 | 0.50 | 0.50 | 0.49 | 0.04 | 0.04 | 0.04 | 0.04 |

**Table S1:** Response time of gene C in IFFL models with different values of $\tilde{K}_{BD_B}$, $h_{BD_B}$, and $\tilde{\eta}_{BD_B}$ (values rounded to two decimal places). Values of the other parameters are: $\tilde{K}_{AB} = \tilde{K}_{AC} = 0.1$, $h_{AB} = h_{AC} = 1.0$.

|  |  | I1-FFL | | | | I4-FFL | | | |
|---|---|---|---|---|---|---|---|---|---|
|  |  | $h_{CD_C} = 0.5$ | | | | | | | |
| $\tilde{K}_{CD_C}$ | $\tilde{\eta}_{CD_C}$ | 0.0 | 1.0 | 10.0 | 100.0 | 0.0 | 1.0 | 10.0 | 100.0 |
| 0.01 |  | 0.16 | 0.16 | 0.24 | 2.14 | 0.48 | 0.49 | 0.53 | 1.00 |
| 0.03 |  | 0.16 | 0.18 | 0.40 | 8.06 | 0.48 | 0.49 | 0.63 | 2.01 |
| 0.1 |  | 0.16 | 0.21 | 1.09 | 24.73 | 0.48 | 0.52 | 0.90 | 6.61 |
| 0.3 |  | 0.16 | 0.28 | 3.59 | 54.33 | 0.48 | 0.58 | 1.48 | 17.46 |
| 1.0 |  | 0.16 | 0.44 | 9.73 | 116.91 | 0.48 | 0.71 | 3.13 | 42.74 |
|  |  | $h_{CD_C} = 1.0$ | | | | | | | |
| $\tilde{K}_{CD_C}$ | $\tilde{\eta}_{CD_C}$ | 0.0 | 1.0 | 10.0 | 100.0 | 0.0 | 1.0 | 10.0 | 100.0 |
| 0.01 |  | 0.16 | 0.18 | 0.43 | 8.90 | 0.48 | 0.49 | 0.61 | 1.87 |
| 0.03 |  | 0.16 | 0.21 | 1.07 | 23.40 | 0.48 | 0.52 | 0.86 | 5.62 |
| 0.1 |  | 0.16 | 0.25 | 2.67 | 44.21 | 0.48 | 0.59 | 1.54 | 18.14 |
| 0.3 |  | 0.16 | 0.28 | 3.98 | 57.73 | 0.48 | 0.68 | 2.66 | 36.69 |
| 1.0 |  | 0.16 | 0.29 | 4.66 | 64.44 | 0.48 | 0.77 | 4.04 | 54.55 |
|  |  | $h_{CD_C} = 2.0$ | | | | | | | |
| $\tilde{K}_{CD_C}$ | $\tilde{\eta}_{CD_C}$ | 0.0 | 1.0 | 10.0 | 100.0 | 0.0 | 1.0 | 10.0 | 100.0 |
| 0.01 |  | 0.16 | 0.21 | 1.12 | 23.22 | 0.48 | 0.51 | 0.76 | 3.76 |
| 0.03 |  | 0.16 | 0.29 | 4.36 | 61.84 | 0.48 | 0.57 | 1.34 | 13.94 |
| 0.1 |  | 0.16 | 0.26 | 3.42 | 55.26 | 0.48 | 0.74 | 3.58 | 48.38 |
| 0.3 |  | 0.16 | 0.20 | 0.77 | 20.81 | 0.48 | 0.83 | 5.24 | 69.08 |
| 1.0 |  | 0.16 | 0.17 | 0.29 | 4.68 | 0.48 | 0.64 | 2.11 | 30.36 |

**Table S2:** Response time of gene C in IFFL models with different values of $\tilde{K}_{CD_C}$, $h_{CD_C}$, and $\tilde{\eta}_{CD_C}$ (values rounded to two decimal places). Values of the other parameters are: $\tilde{K}_{AB} = \tilde{K}_{AC} = \tilde{K}_{BC} = 0.1$, $h_{AB} = h_{AC} = h_{BC} = 1.0$.

## 1.10 Pulse Amplitude of IFFLs at Different Levels of $\tilde{\eta}_{BD_B}$ and $\tilde{\eta}_{CD_C}$

Here, a trajectory is considered to contain a pulse if the the trajectory maximum is larger than the pre-induction and post-induction steady states. A trajectory not satisfying this criteria is labeled as "NA".

|  |  | I1-FFL | | | | I4-FFL | | | |
|---|---|---|---|---|---|---|---|---|---|
|  |  | \multicolumn{8}{c}{$h = 0.5$} |
| $\tilde{K}_{BD_B}$ | $\tilde{\eta}_{BD_B}$ | 0.0 | 1.0 | 10.0 | 100.0 | 0.0 | 1.0 | 10.0 | 100.0 |
| 0.01 | | 0.09 | 0.09 | 0.10 | 0.19 | 0.71 | 0.71 | 0.71 | 0.71 |
| 0.03 | | 0.15 | 0.15 | 0.16 | 0.34 | 0.62 | 0.62 | 0.62 | 0.64 |
| 0.1 | | 0.23 | 0.23 | 0.26 | 0.59 | 0.51 | 0.51 | 0.52 | 0.57 |
| 0.3 | | 0.33 | 0.34 | 0.41 | 0.74 | 0.40 | 0.41 | 0.43 | 0.50 |
| 1.0 | | 0.47 | 0.47 | 0.60 | 0.84 | 0.29 | 0.30 | 0.34 | 0.41 |
|  |  | \multicolumn{8}{c}{$h = 1.0$} |
| $\tilde{K}_{BD_B}$ | $\tilde{\eta}_{BD_B}$ | 0.0 | 1.0 | 10.0 | 100.0 | 0.0 | 1.0 | 10.0 | 100.0 |
| 0.01 | | 0.03 | 0.03 | 0.06 | 0.28 | 0.83 | 0.83 | 0.83 | 9.84 |
| 0.03 | | 0.06 | 0.07 | 0.15 | 0.49 | 0.75 | 0.75 | 0.75 | 0.78 |
| 0.1 | | 0.14 | 0.16 | 0.32 | 0.70 | 0.60 | 0.60 | 0.63 | 0.72 |
| 0.3 | | 0.27 | 0.31 | 0.53 | 0.81 | 0.42 | 0.44 | 0.51 | 0.62 |
| 1.0 | | 0.49 | 0.53 | 0.72 | 0.87 | 0.22 | 0.25 | 0.33 | 0.42 |
|  |  | \multicolumn{8}{c}{$h = 2.0$} |
| $\tilde{K}_{BD_B}$ | $\tilde{\eta}_{BD_B}$ | 0.0 | 1.0 | 10.0 | 100.0 | 0.0 | 1.0 | 10.0 | 100.0 |
| 0.01 | | 0.01 | 0.02 | 0.09 | 0.41 | 0.90 | 0.90 | 0.90 | 0.90 |
| 0.03 | | 0.04 | 0.06 | 0.22 | 0.62 | 0.85 | 0.85 | 0.85 | 0.87 |
| 0.1 | | 0.10 | 0.16 | 0.44 | 0.78 | 0.71 | 0.72 | 0.77 | 0.84 |
| 0.3 | | 0.25 | 0.36 | 0.65 | 0.85 | 0.48 | 0.52 | 0.66 | 0.78 |
| 1.0 | | 0.56 | 0.65 | 0.81 | 0.89 | 0.15 | 0.22 | 0.36 | 0.43 |

**Table S3:** Pulse amplitude of gene C in models with different values of $\tilde{K}_{BD_B}$, $h_{BD_B}$, and $\tilde{\eta}_{BD_B}$ (values rounded to two decimal places). Values of the other parameters are: $\tilde{K}_{AB} = \tilde{K}_{AC} = 0.1$, $h_{AB} = h_{AC} = 1.0$.

|  |  | I1-FFL | | | | I4-FFL | | | |
|---|---|---|---|---|---|---|---|---|---|
|  |  | \multicolumn{8}{c}{$h = 0.5$} |
| $\tilde{K}_{CD_C}$ | $\tilde{\eta}_{CD_C}$ | 0.0 | 1.0 | 10.0 | 100.0 | 0.0 | 1.0 | 10.0 | 100.0 |
| 0.01 | | 0.14 | 0.14 | 0.13 | 0.09 | 0.60 | 0.60 | 0.59 | 0.55 |
| 0.03 | | 0.14 | 0.14 | 0.11 | 0.09 | 0.60 | 0.60 | 0.58 | 0.47 |
| 0.1 | | 0.14 | 0.13 | 0.09 | NA | 0.60 | 0.60 | 0.56 | 0.43 |
| 0.3 | | 0.14 | 0.12 | 0.09 | NA | 0.60 | 0.59 | 0.49 | 0.43 |
| 1.0 | | 0.14 | 0.10 | 0.09 | NA | 0.60 | 0.57 | 0.43 | NA |
|  |  | \multicolumn{8}{c}{$h = 1.0$} |
| $\tilde{K}_{CD_C}$ | $\tilde{\eta}_{CD_C}$ | 0.0 | 1.0 | 10.0 | 100.0 | 0.0 | 1.0 | 10.0 | 100.0 |
| 0.01 | | 0.14 | 0.14 | 0.11 | 0.09 | 0.60 | 0.60 | 0.59 | 0.49 |
| 0.03 | | 0.14 | 0.13 | 0.09 | NA | 0.60 | 0.60 | 0.57 | 0.43 |
| 0.1 | | 0.14 | 0.12 | 0.09 | NA | 0.60 | 0.59 | 0.49 | 0.43 |
| 0.3 | | 0.14 | 0.11 | 0.09 | NA | 0.60 | 0.57 | 0.43 | NA |
| 1.0 | | 0.14 | 0.10 | NA | NA | 0.60 | 0.54 | 0.43 | NA |
|  |  | \multicolumn{8}{c}{$h = 2.0$} |
| $\tilde{K}_{CD_C}$ | $\tilde{\eta}_{CD_C}$ | 0.0 | 1.0 | 10.0 | 100.0 | 0.0 | 1.0 | 10.0 | 100.0 |
| 0.01 | | 0.14 | 0.13 | 0.09 | 0.09 | 0.60 | 0.60 | 0.58 | 0.44 |
| 0.03 | | 0.14 | 0.12 | 0.09 | NA | 0.60 | 0.59 | 0.53 | 0.43 |
| 0.1 | | 0.14 | 0.10 | NA | NA | 0.60 | 0.58 | 0.43 | NA |
| 0.3 | | 0.14 | 0.11 | NA | NA | 0.60 | 0.52 | NA | NA |
| 1.0 | | 0.14 | 0.13 | NA | NA | 0.60 | 0.51 | NA | NA |

**Table S4:** Pulse amplitude of gene C in models with different values of $\tilde{K}_{CD_C}$, $h_{CD_C}$, and $\tilde{\eta}_{CD_C}$ (values rounded to two decimal places). Values of the other parameters are: $\tilde{K}_{AB} = \tilde{K}_{AC} = \tilde{K}_{BC} = 0.1$, $h_{AB} = h_{AC} = h_{BC} = 1.0$.

## 1.11 Response Time of IFFLs and Two-Input Circuits at Different Levels of $\tilde{\eta}_{AD_A}$

|  | I1-FFL | | | | Type-1 Two-Input Circuit | | | |
|---|---|---|---|---|---|---|---|---|
| | $h_{AD_A} = 0.5$ | | | | | | | |
| $\tilde{\eta}_{AD_A}$ \ $\tilde{K}_{AD_A}$ | 0.0 | 1.0 | 10.0 | 100.0 | 0.0 | 1.0 | 10.0 | 100.0 |
| 0.01 | 0.09 | 0.09 | 0.11 | 0.24 | 0.09 | 0.09 | 0.11 | 0.33 |
| 0.03 | 0.11 | 0.11 | 0.16 | 0.44 | 0.11 | 0.12 | 0.18 | 0.81 |
| 0.1 | 0.14 | 0.15 | 0.26 | 0.87 | 0.14 | 0.16 | 0.33 | 2.58 |
| 0.3 | 0.18 | 0.21 | 0.45 | 1.71 | 0.18 | 0.23 | 0.65 | 6.20 |
| 1.0 | 0.25 | 0.33 | 0.85 | 4.06 | 0.25 | 0.36 | 1.61 | 14.06 |
| | $h_{AD_A} = 1.0$ | | | | | | | |
| $\tilde{\eta}_{AD_A}$ \ $\tilde{K}_{AD_A}$ | 0.0 | 1.0 | 10.0 | 100.0 | 0.0 | 1.0 | 10.0 | 100.0 |
| 0.01 | 0.08 | 0.09 | 0.13 | 0.37 | 0.08 | 0.09 | 0.15 | 0.69 |
| 0.03 | 0.11 | 0.12 | 0.21 | 0.67 | 0.11 | 0.12 | 0.27 | 1.92 |
| 0.1 | 0.16 | 0.19 | 0.39 | 1.37 | 0.16 | 0.20 | 0.60 | 5.72 |
| 0.3 | 0.24 | 0.31 | 0.70 | 2.92 | 0.24 | 0.33 | 1.43 | 13.79 |
| 1.0 | 0.41 | 0.55 | 1.40 | 7.86 | 0.41 | 0.61 | 3.86 | 31.25 |
| | $h_{AD_A} = 2.0$ | | | | | | | |
| $\tilde{\eta}_{AD_A}$ \ $\tilde{K}_{AD_A}$ | 0.0 | 1.0 | 10.0 | 100.0 | 0.0 | 1.0 | 10.0 | 100.0 |
| 0.01 | 0.07 | 0.08 | 0.17 | 0.53 | 0.07 | 0.09 | 0.22 | 1.37 |
| 0.03 | 0.10 | 0.13 | 0.30 | 1.01 | 0.10 | 0.14 | 0.47 | 3.89 |
| 0.1 | 0.17 | 0.24 | 0.58 | 2.24 | 0.17 | 0.27 | 1.36 | 11.61 |
| 0.3 | 0.32 | 0.45 | 1.13 | 5.63 | 0.32 | 0.53 | 4.34 | 34.34 |
| 1.0 | 0.72 | 1.00 | 2.64 | 18.21 | 0.72 | 1.16 | 10.10 | 84.06 |

**Table S5:** Response time of gene C in I1-FFLs and type-1 two-input circuits with different values of $\tilde{K}_{AD_A}$, $h_{AD_A}$, and $\tilde{\eta}_{AD_A}$ (values rounded to two decimal places). $\tilde{K}_{BC} = 0.1$, $h_{BC} = 1.0$.

|  | I4-FFL | | | | Type-4 Two-Input Circuit | | | |
|---|---|---|---|---|---|---|---|---|
| | $h_{AD_A} = 0.5$ | | | | | | | |
| $\tilde{\eta}_{AD_A}$ \ $\tilde{K}_{AD_A}$ | 0.0 | 1.0 | 10.0 | 100.0 | 0.0 | 1.0 | 10.0 | 100.0 |
| 0.01 | 0.36 | 0.36 | 0.38 | 0.57 | 0.36 | 0.36 | 0.38 | 0.58 |
| 0.03 | 0.51 | 0.52 | 0.57 | 1.12 | 0.51 | 0.52 | 0.58 | 1.15 |
| 0.1 | 0.68 | 0.70 | 0.87 | 2.45 | 0.68 | 0.70 | 0.88 | 2.68 |
| 0.3 | 0.83 | 0.88 | 1.33 | 5.39 | 0.83 | 0.88 | 1.36 | 5.97 |
| 1.0 | 0.97 | 1.11 | 2.29 | 13.21 | 0.97 | 1.11 | 2.35 | 14.05 |
| | $h_{AD_A} = 1.0$ | | | | | | | |
| $\tilde{\eta}_{AD_A}$ \ $\tilde{K}_{AD_A}$ | 0.0 | 1.0 | 10.0 | 100.0 | 0.0 | 1.0 | 10.0 | 100.0 |
| 0.01 | 0.07 | 0.08 | 0.11 | 0.33 | 0.07 | 0.08 | 0.11 | 0.33 |
| 0.03 | 0.19 | 0.21 | 0.32 | 1.00 | 0.19 | 0.21 | 0.32 | 1.05 |
| 0.1 | 0.48 | 0.52 | 0.90 | 3.48 | 0.48 | 0.53 | 0.92 | 5.34 |
| 0.3 | 0.87 | 1.00 | 2.04 | 10.9 | 0.87 | 1.00 | 2.15 | 13.79 |
| 1.0 | 1.27 | 1.60 | 4.17 | 29.04 | 1.27 | 1.61 | 4.39 | 31.25 |
| | $h_{AD_A} = 2.0$ | | | | | | | |
| $\tilde{\eta}_{AD_A}$ \ $\tilde{K}_{AD_A}$ | 0.0 | 1.0 | 10.0 | 100.0 | 0.0 | 1.0 | 10.0 | 100.0 |
| 0.01 | 0.01 | 0.01 | 0.02 | 0.05 | 0.01 | 0.01 | 0.02 | 0.05 |
| 0.03 | 0.03 | 0.04 | 0.09 | 0.26 | 0.03 | 0.04 | 0.09 | 0.26 |
| 0.1 | 0.15 | 0.22 | 0.52 | 1.87 | 0.15 | 0.22 | 0.53 | 2.25 |
| 0.3 | 0.68 | 0.96 | 2.69 | 16.90 | 0.68 | 0.97 | 3.33 | 34.35 |
| 1.0 | 1.72 | 2.58 | 9.37 | 77.33 | 1.72 | 2.62 | 10.11 | 84.06 |

**Table S6:** Response time of gene C in I4-FFLs and type-4 two-input circuits with different values of $\tilde{K}_{AD_A}$, $h_{AD_A}$, and $\tilde{\eta}_{AD_A}$ (values rounded to two decimal places). $\tilde{K}_{BC} = 0.1$, $h_{BC} = 1.0$.

## 1.12 Pulse Amplitude of IFFLs and Two-Input Circuits at Different Levels of $\tilde{\eta}_{AD_A}$

Here, a trajectory is considered to contain a pulse if the the trajectory maximum is larger than the pre-induction and post-induction steady states. A trajectory not satisfying this criteria is labeled as "NA".

|  |  | I1-FFL |  |  |  | Type-1 Two-Input Circuit |  |  |  |
|---|---|---|---|---|---|---|---|---|---|
|  |  | \multicolumn{8}{c}{$h_{AD_A} = 0.5$} |
| $\tilde{K}_{AD_A}$ | $\tilde{\eta}_{AD_A}$ | 0.0 | 1.0 | 10.0 | 100.0 | 0.0 | 1.0 | 10.0 | 100.0 |
| 0.01 |  | 0.12 | 0.12 | 0.12 | 0.12 | 0.12 | 0.12 | 0.11 | 0.09 |
| 0.03 |  | 0.12 | 0.12 | 0.12 | 0.11 | 0.12 | 0.12 | 0.11 | 0.08 |
| 0.1 |  | 0.12 | 0.12 | 0.12 | 0.10 | 0.12 | 0.12 | 0.10 | 0.08 |
| 0.3 |  | 0.12 | 0.12 | 0.11 | 0.09 | 0.12 | 0.11 | 0.08 | 0.08 |
| 1.0 |  | 0.11 | 0.10 | 0.09 | 0.08 | 0.11 | 0.10 | 0.08 | 0.08 |
|  |  | \multicolumn{8}{c}{$h_{AD_A} = 1.0$} |
| $\tilde{K}_{AD_A}$ | $\tilde{\eta}_{AD_A}$ | 0.0 | 1.0 | 10.0 | 100.0 | 0.0 | 1.0 | 10.0 | 100.0 |
| 0.01 |  | 0.15 | 0.15 | 0.15 | 0.14 | 0.15 | 0.15 | 0.13 | 0.10 |
| 0.03 |  | 0.15 | 0.15 | 0.14 | 0.12 | 0.15 | 0.14 | 0.12 | 0.09 |
| 0.1 |  | 0.14 | 0.14 | 0.13 | 0.10 | 0.14 | 0.13 | 0.10 | 0.09 |
| 0.3 |  | 0.13 | 0.13 | 0.11 | 0.09 | 0.13 | 0.11 | 0.09 | 0.09 |
| 1.0 |  | 0.11 | 0.10 | 0.09 | NA | 0.11 | 0.09 | 0.08 | NA |
|  |  | \multicolumn{8}{c}{$h_{AD_A} = 2.0$} |
| $\tilde{K}_{AD_A}$ | $\tilde{\eta}_{AD_A}$ | 0.0 | 1.0 | 10.0 | 100.0 | 0.0 | 1.0 | 10.0 | 100.0 |
| 0.01 |  | 0.15 | 0.15 | 0.15 | 0.14 | 0.15 | 0.15 | 0.12 | 0.09 |
| 0.03 |  | 0.15 | 0.15 | 0.15 | 0.11 | 0.15 | 0.14 | 0.11 | 0.09 |
| 0.1 |  | 0.15 | 0.15 | 0.13 | 0.09 | 0.15 | 0.13 | 0.09 | 0.09 |
| 0.3 |  | 0.14 | 0.13 | 0.10 | 0.09 | 0.14 | 0.10 | 0.09 | NA |
| 1.0 |  | 0.10 | 0.10 | 0.08 | NA | 0.10 | 0.08 | NA | NA |

**Table S7:** Pulse amplitude of gene C in I1-FFLs and type-1 two-input circuits with different values of $\tilde{K}_{AD_A}$, $h_{AD_A}$, and $\tilde{\eta}_{AD_A}$ (values rounded to two decimal places). $\tilde{K}_{BC} = 0.1$, $h_{BC} = 1.0$.

|  |  | I4-FFL |  |  |  | Type-4 Two-Input Circuit |  |  |  |
|---|---|---|---|---|---|---|---|---|---|
|  |  | \multicolumn{8}{c}{$h_{AD_A} = 0.5$} |
| $\tilde{K}_{AD_A}$ | $\tilde{\eta}_{AD_A}$ | 0.0 | 1.0 | 10.0 | 100.0 | 0.0 | 1.0 | 10.0 | 100.0 |
| 0.01 |  | 0.57 | 0.57 | 0.57 | 0.57 | 0.57 | 0.57 | 0.57 | 0.55 |
| 0.03 |  | 0.57 | 0.57 | 0.57 | 0.57 | 0.57 | 0.57 | 0.57 | 0.52 |
| 0.1 |  | 0.55 | 0.55 | 0.55 | 0.54 | 0.55 | 0.55 | 0.54 | 0.53 |
| 0.3 |  | 0.50 | 0.50 | 0.50 | 0.50 | 0.50 | 0.50 | 0.50 | 0.50 |
| 1.0 |  | 0.41 | 0.41 | 0.41 | 0.42 | 0.41 | 0.41 | 0.41 | 0.42 |
|  |  | \multicolumn{8}{c}{$h_{AD_A} = 1.0$} |
| $\tilde{K}_{AD_A}$ | $\tilde{\eta}_{AD_A}$ | 0.0 | 1.0 | 10.0 | 100.0 | 0.0 | 1.0 | 10.0 | 100.0 |
| 0.01 |  | 0.62 | 0.62 | 0.62 | 0.62 | 0.62 | 0.62 | 0.61 | 0.54 |
| 0.03 |  | 0.62 | 0.62 | 0.62 | 0.60 | 0.62 | 0.61 | 0.60 | 0.37 |
| 0.1 |  | 0.60 | 0.60 | 0.60 | 0.57 | 0.60 | 0.59 | 0.54 | 0.43 |
| 0.3 |  | 0.56 | 0.56 | 0.55 | 0.54 | 0.56 | 0.55 | 0.54 | 0.54 |
| 1.0 |  | 0.42 | 0.42 | 0.42 | NA | 0.42 | 0.42 | 0.42 | NA |
|  |  | \multicolumn{8}{c}{$h_{AD_A} = 2.0$} |
| $\tilde{K}_{AD_A}$ | $\tilde{\eta}_{AD_A}$ | 0.0 | 1.0 | 10.0 | 100.0 | 0.0 | 1.0 | 10.0 | 100.0 |
| 0.01 |  | 0.62 | 0.62 | 0.62 | 0.62 | 0.62 | 0.62 | 0.61 | 0.43 |
| 0.03 |  | 0.62 | 0.62 | 0.62 | 0.60 | 0.62 | 0.62 | 0.57 | 0.16 |
| 0.1 |  | 0.62 | 0.62 | 0.62 | 0.56 | 0.62 | 0.60 | 0.42 | 0.09 |
| 0.3 |  | 0.60 | 0.60 | 0.57 | 0.54 | 0.60 | 0.56 | 0.42 | NA |
| 1.0 |  | 0.42 | 0.42 | NA | NA | 0.42 | 0.42 | NA | NA |

**Table S8:** Pulse amplitude of gene C in I4-FFLs and type-4 two-input circuits with different values of $\tilde{K}_{AD_A}$, $h_{AD_A}$, and $\tilde{\eta}_{AD_A}$ (values rounded to two decimal places). $\tilde{K}_{BC} = 0.1$, $h_{BC} = 1.0$.

## 1.13 Comparing the Pulsing Behavior of I1-FFLs and I4-FFLs at Different Levels of $\tilde{\eta}_{BD_B}$

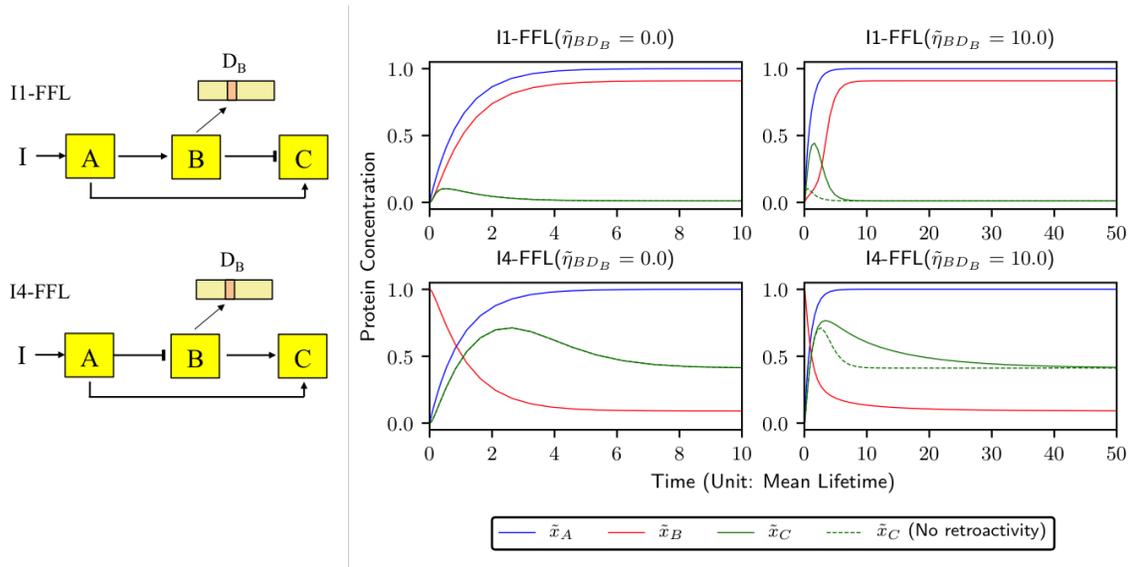

**Figure S2:** The effect of retroactivity $\tilde{\eta}_{BD_B}$ on pulse amplitude is more pronounced in an I1-FFL (top row) than in an I4-FFL (bottom row). Values of the parameters are: $\tilde{K}_{AB} = \tilde{K}_{AC} = \tilde{K}_{BC} = \tilde{K}_{BD_B} = 0.1$, $h_{AB} = h_{AC} = 1.0$, $h_{BC} = h_{BD_B} = 2.0$. For comparison, the green dashed curves represent the trajectories of $\tilde{x}_C$ when $\tilde{\eta}_{BD_B}$ equals 0.

15

## 1.14 Response Time of IFFLs with OR Logic at Different Levels of $\tilde{\eta}_{BD_B}$

Under the assumption of OR logic, the non-dimensionalized ODE model for an I1-FFL without retroactivity is given as:

$$\frac{d\tilde{x}_A}{d\tau} = f_{\tilde{A}} = (1-\gamma_A)\frac{\left(\frac{x_I}{K_{IA}}\right)^{h_{IA}}}{1+\left(\frac{x_I}{K_{IA}}\right)^{h_{IA}}} + \gamma_A - \tilde{x}_A$$

$$\frac{d\tilde{x}_B}{d\tau} = f_{\tilde{B}} = (1-\gamma_B)\frac{\left(\frac{\tilde{x}_A}{\tilde{K}_{AB}}\right)^{h_{AB}}}{1+\left(\frac{\tilde{x}_A}{\tilde{K}_{AB}}\right)^{h_{AB}}} + \gamma_B - \tilde{x}_B$$

$$\frac{d\tilde{x}_C}{d\tau} = f_{\tilde{C}} = (1-\gamma_C)\left[\frac{\left(\frac{\tilde{x}_A}{\tilde{K}_{AC}}\right)^{h_{AC}}}{1+\left(\frac{\tilde{x}_A}{\tilde{K}_{AC}}\right)^{h_{AC}}} + \frac{1}{1+\left(\frac{\tilde{x}_B}{\tilde{K}_{BC}}\right)^{h_{BC}}}\right] + \gamma_C - \tilde{x}_C.$$

Under the assumption of OR logic, the non-dimensionalized ODE model for an I4-FFL without retroactivity is given as:

$$\frac{d\tilde{x}_A}{d\tau} = f_{\tilde{A}} = (1-\gamma_A)\frac{\left(\frac{x_I}{K_{IA}}\right)^{h_{IA}}}{1+\left(\frac{x_I}{K_{IA}}\right)^{h_{IA}}} + \gamma_A - \tilde{x}_A$$

$$\frac{d\tilde{x}_B}{d\tau} = f_{\tilde{B}} = (1-\gamma_B)\frac{1}{1+\left(\frac{\tilde{x}_A}{\tilde{K}_{AB}}\right)^{h_{AB}}} + \gamma_B - \tilde{x}_B$$

$$\frac{d\tilde{x}_C}{d\tau} = f_{\tilde{C}} = (1-\gamma_C)\left[\frac{\left(\frac{\tilde{x}_A}{\tilde{K}_{AC}}\right)^{h_{AC}}}{1+\left(\frac{\tilde{x}_A}{\tilde{K}_{AC}}\right)^{h_{AC}}} + \frac{\left(\frac{\tilde{x}_B}{\tilde{K}_{BC}}\right)^{h_{BC}}}{1+\left(\frac{\tilde{x}_B}{\tilde{K}_{BC}}\right)^{h_{BC}}}\right] + \gamma_C - \tilde{x}_C.$$

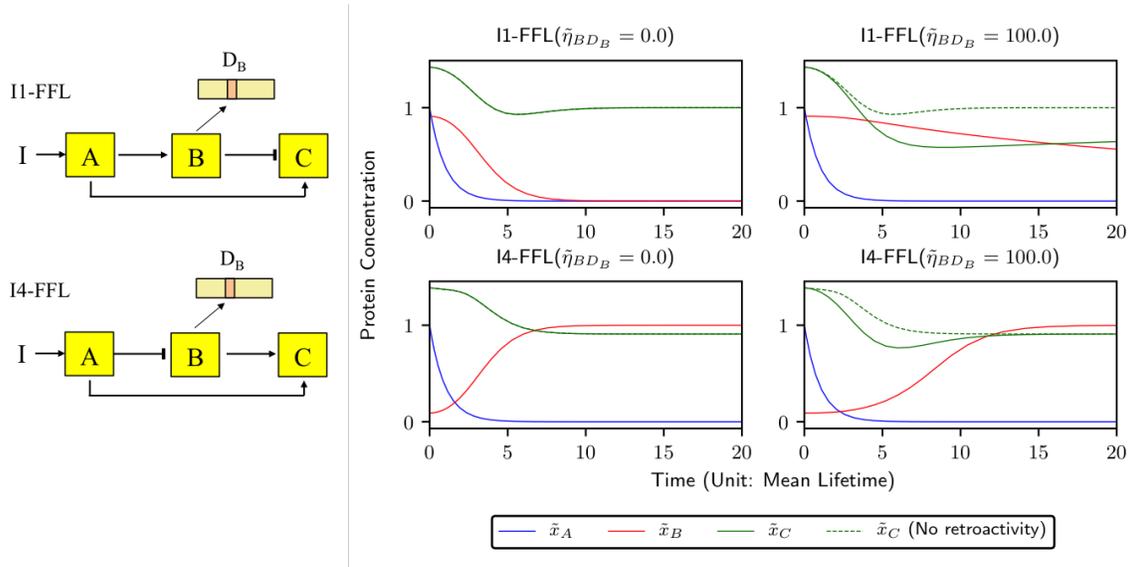

**Figure S3:** In response to an OFF step, the effect of retroactivity $\tilde{\eta}_{BD_B}$ on response times is more pronounced in an I4-FFL (bottom row) than in an I1-FFL (top row) under the assumption of OR logic. Values of the parameters are: $\tilde{K}_{AB} = \tilde{K}_{AC} = 0.1$, $\tilde{K}_{BC} = \tilde{K}_{BD_B} = 1.0$, $h_{AB} = h_{AC} = h_{BC} = h_{BD_B} = 1.0$. For comparison, the green dashed curves represent the trajectories of $\tilde{x}_C$ when $\tilde{\eta}_{BD_B}$ equals 0.

|  | I1-FFL | | | | I4-FFL | | | |
|---|---|---|---|---|---|---|---|---|
| | $h_{BD_B} = 0.5$ | | | | | | | |
| $\tilde{K}_{BD_B}$ \ $\tilde{\eta}_{BD_B}$ | 0.0 | 1.0 | 10.0 | 100.0 | 0.0 | 1.0 | 10.0 | 100.0 |
| 0.03 | 1.22 | 1.22 | 1.22 | 1.23 | 3.56 | 3.56 | 3.55 | 3.45 |
| 0.1 | 1.55 | 1.55 | 1.55 | 1.54 | 3.56 | 3.55 | 3.5 | 3.07 |
| 0.3 | 1.93 | 1.93 | 1.93 | 1.9 | 3.51 | 3.49 | 3.3 | 2.84 |
| 1.0 | 2.37 | 2.36 | 2.34 | 2.29 | 3.44 | 3.38 | 3.05 | 2.83 |
| 3.0 | 2.71 | 2.7 | 2.65 | 2.61 | 3.4 | 3.3 | 3.02 | 2.93 |
| | $h_{BD_B} = 1.0$ | | | | | | | |
| $\tilde{K}_{BD_B}$ \ $\tilde{\eta}_{BD_B}$ | 0.0 | 1.0 | 10.0 | 100.0 | 0.0 | 1.0 | 10.0 | 100.0 |
| 0.03 | NA | NA | NA | NA | 3.71 | 3.7 | 3.68 | 3.43 |
| 0.1 | 0.34 | 0.34 | 0.34 | 0.34 | 3.96 | 3.93 | 3.51 | 2.5 |
| 0.3 | 1.44 | 1.44 | 1.42 | 1.39 | 3.7 | 3.45 | 2.37 | 2.14 |
| 1.0 | 2.48 | 2.44 | 2.34 | 2.29 | 3.16 | 2.91 | 2.53 | 2.45 |
| 3.0 | 3.1 | 3.04 | 2.93 | 2.89 | 3.23 | 3.12 | 2.96 | 2.92 |
| | $h_{BD_B} = 2.0$ | | | | | | | |
| $\tilde{K}_{BD_B}$ \ $\tilde{\eta}_{BD_B}$ | 0.0 | 1.0 | 10.0 | 100.0 | 0.0 | 1.0 | 10.0 | 100.0 |
| 0.03 | NA | NA | NA | NA | 3.61 | 3.6 | 3.6 | 3.5 |
| 0.1 | NA | NA | NA | NA | 4.71 | 4.65 | 2.71 | 2.15 |
| 0.3 | 0.3 | 0.3 | 0.3 | 0.3 | 5.68 | 1.14 | 0.99 | 0.97 |
| 1.0 | 2.88 | 2.59 | 2.39 | 2.35 | 2.45 | 2.36 | 2.27 | 2.24 |
| 3.0 | 3.42 | 3.36 | 3.26 | 3.24 | 3.26 | 3.24 | 3.22 | 3.2 |

**Table S9:** Response time of gene C in IFFL models under the assumption of OR logic with different values of $\tilde{K}_{BD_B}$, $h_{BD_B}$, and $\tilde{\eta}_{BD_B}$ (values rounded to two decimal places). Values of the other parameters are: $\tilde{K}_{AB} = \tilde{K}_{AC} = 0.1$, $h_{AB} = h_{AC} = 1.0$. In response to an OFF step, $\tilde{x}_C$ transitions from a high pre-stimulus state to a low post-stimulus state in I1-FFLs, I4-FLLs, and their simple regulation counterparts. NA represents cases where the post-induction steady state and the mid-point are larger than the pre-induction steady state.

## 1.15 Supplemental Information for Figures 4(a) and (b)

| | | Response Time | | | $\arg\max_t b(\tilde{x}_B)$ | | |
|---|---|---|---|---|---|---|---|
| $\tilde{K}_{BD_B}$ \ $\tilde{\eta}_{BD_B}$ | | 1.0 | 10.0 | 100.0 | 1.0 | 10.0 | 100.0 |
| 0.01 | | 0.001 | 0.001 | 0.001 | 0.0543 | 0.1483 | 0.7416 |
| 0.03 | | 0.0098 | 0.0098 | 0.0098 | 0.1058 | 0.3187 | 1.9157 |
| 0.1 | | 0.037 | 0.037 | 0.037 | 0.2394 | 0.8322 | 6.0376 |
| 0.3 | | 0.1326 | 0.1322 | 0.1321 | 0.579 | 2.3339 | 19.2472 |
| 1.0 | | 0.5004 | 0.4952 | 0.4934 | 2.2718 | 11.0335 | 98.8473 |

**Table S10:** Response time and time at which $b(\tilde{x}_B)$ attains its maximum for the I1-FFL model with $h_{BC} = 2$.

| | I1-FFL | | | I4-FFL | | |
|---|---|---|---|---|---|---|
| $\tilde{K}_{BD_B}$ \ $\tilde{\eta}_{BD_B}$ | 1.0 | 10.0 | 100.0 | 1.0 | 10.0 | 100.0 |
| 0.01 | Y | Y | Y | N | N | N |
| 0.03 | Y | Y | Y | N | N | N |
| 0.1 | Y | Y | Y | Y | Y | Y |
| 0.3 | Y | Y | Y | Y | Y | Y |
| 1.0 | N | N | N | N | N | N |

**Table S11:** Whether $\tilde{x}_B$ transitions from a value lower (higher) than $\tilde{K}_{BC}$ to a value higher (lower) than $\tilde{K}_{BC}$ for an I1-FFL (I4-FFL) model with $h_{BC} = 2$. Y stands for yes; N stands for no.

## 1.16 Proof of the Effects of $\tilde{\eta}_{BD_B}$ on Response Time and Pulse Amplitude in IFFLs

Here we show that for any parameters, increases in retroactivity on node B, $\tilde{\eta}_{BD_B}$, in an I1-FFL lead to a decrease in response time and an increase in pulse amplitude.

According to Supplemental Information Section 1.6, $\frac{d\tilde{x}_A}{d\tau}$, $\frac{d\tilde{x}_B}{d\tau}$, and $\frac{d\tilde{x}_C}{d\tau}$ in an I1-FFL where only $\tilde{\eta}_{BD_B}$ is allowed to vary can be expressed as:

$$\begin{bmatrix} \frac{d\tilde{x}_A}{d\tau} \\ \frac{d\tilde{x}_B}{d\tau} \\ \frac{d\tilde{x}_C}{d\tau} \end{bmatrix} = \begin{bmatrix} \frac{1}{1+r_{AD_A}(\tilde{x}_A)} & 0 & 0 \\ 0 & \frac{1}{1+r_{BD_B}(\tilde{x}_B)} & 0 \\ 0 & 0 & \frac{1}{1+r_{CD_C}(\tilde{x}_C)} \end{bmatrix} \begin{bmatrix} f_{\tilde{A}} \\ f_{\tilde{B}} \\ f_{\tilde{C}} \end{bmatrix}, \tag{6}$$

$$f_{\tilde{A}} = \frac{\left(\frac{x_I}{K_{IA}}\right)^{h_{IA}}}{1+\left(\frac{x_I}{K_{IA}}\right)^{h_{IA}}}(1-\gamma_A) + \gamma_A - \tilde{x}_A$$

$$f_{\tilde{B}} = \frac{\left(\frac{\tilde{x}_A}{\tilde{K}_{AB}}\right)^{h_{AB}}}{1+\left(\frac{\tilde{x}_A}{\tilde{K}_{AB}}\right)^{h_{AB}}}(1-\gamma_B) + \gamma_B - \tilde{x}_B \tag{7}$$

$$f_{\tilde{C}} = \frac{\left(\frac{\tilde{x}_A}{\tilde{K}_{AC}}\right)^{h_{AC}}}{\left(1+\left(\frac{\tilde{x}_A}{\tilde{K}_{AC}}\right)^{h_{AC}}\right)\left(1+\left(\frac{\tilde{x}_B}{\tilde{K}_{BC}}\right)^{h_{BC}}\right)}(1-\gamma_C) + \gamma_C - \tilde{x}_C,$$

$$r_{AD_A}(\tilde{x}_A) = \tilde{\eta}_{AD_A} h_{AD_A}^2 \left(\frac{\tilde{x}_A}{\tilde{K}_{AD_A}}\right)^{h_{AD_A}-1} \left(1+\left(\frac{\tilde{x}_A}{\tilde{K}_{AD_A}}\right)^{h_{AD_A}}\right)^{-2}$$

$$r_{BD_B}(\tilde{x}_B) = \tilde{\eta}_{BD_B} h_{BD_B}^2 \left(\frac{\tilde{x}_B}{\tilde{K}_{BD_B}}\right)^{h_{BD_B}-1} \left(1+\left(\frac{\tilde{x}_B}{\tilde{K}_{BD_B}}\right)^{h_{BD_B}}\right)^{-2} \tag{8}$$

$$r_{CD_C}(\tilde{x}_C) = \tilde{\eta}_{CD_C} h_{CD_C}^2 \left(\frac{\tilde{x}_C}{\tilde{K}_{CD_C}}\right)^{h_{CD_C}-1} \left(1+\left(\frac{\tilde{x}_C}{\tilde{K}_{CD_C}}\right)^{h_{CD_C}}\right)^{-2},$$

where $\gamma_A, \gamma_B, \gamma_C \in (0,1)$ such that $\tilde{x}_A, \tilde{x}_B, \tilde{x}_C \in (0,1)$.

Let $\vec{\tilde{x}}_1$ and $\vec{\tilde{x}}_2$ denote the concentrations of A, B, and C in two I1-FFL models (i.e., $\vec{\tilde{x}}_1 = [\tilde{x}_{A_1}, \tilde{x}_{B_1}, \tilde{x}_{C_1}]$, $\vec{\tilde{x}}_2 = [\tilde{x}_{A_2}, \tilde{x}_{B_2}, \tilde{x}_{C_2}]$), in which all parameters are held identical except that node B is connected to different numbers of downstream targets such that retroactivity coefficient $\tilde{\eta}_{BD_B}$ equals $\tilde{\eta}_{BD_{B_1}}$ and $\tilde{\eta}_{BD_{B_2}}$, respectively. At $\tau = 0$, $x_I$ undergoes a stepwise increase and is kept constant afterwards. The initial values of $\tilde{x}_A$, $\tilde{x}_{B_i}$ ($i = 1, 2$), and $\tilde{x}_C$ are the corresponding steady values before the increase in $x_I$. Without loss of generality, we assume $\tilde{\eta}_{BD_{B_1}} < \tilde{\eta}_{BD_{B_2}}$. We now show that $\forall \tau > 0$, $\tilde{x}_{C_1}(\tau) < \tilde{x}_{C_2}(\tau)$, i.e., the concentration of C in I1-FFL model #1 is less than the concentration of C in I1-FFL model #2 at all positive times.

First, we prove the following two lemmas:

**Lemma 1.** *Let $\frac{d\tilde{x}}{d\tau} = g(\tilde{x})(c - \tilde{x})$. If $0 < \tilde{x}(0) < c$, and $g(\tilde{x})$ is positive and smooth for all $\tilde{x} \in (0, \infty)$, then $\frac{d\tilde{x}}{d\tau} > 0$ for all $\tau \geq 0^+$.*

*Proof.* It is clear that $\tilde{x}$ has a unique steady state equal to $c$ for $\tilde{x} \in (0, \infty)$. Because $\frac{d\tilde{x}}{d\tau} > 0$ for $\tilde{x} < c$ and $0 < \tilde{x}(0) < c$, we have $\frac{d\tilde{x}}{d\tau} > 0$ for all $\tau \geq 0^+$. □

**Lemma 2.** *Suppose that a smooth function $h(\tau)$ defined on $[0, \infty)$ satisfies the following properties: (i) there exists a positive integer $k$ such that $\frac{d^k h}{d\tau^k}(0^+) > 0$ and $\frac{d^i h}{d\tau^i}(0^+) = 0$ for all $i = 0, 1, 2, ..., k-1$; (ii) for any $\tau^*$ in $(0, \infty)$ where $h(\tau^*) = 0$ we always have $\frac{dh}{d\tau}(\tau^*) > 0$. Then $h(\tau) > 0$ for all $\tau > 0$.*

*Proof.* Property (i) of $h(\tau)$ implies that $h(\tau) > 0$ on some interval $(0, \delta)$. Let $(0, T)$ be the largest interval where $h(\tau) > 0$. We claim that $T = \infty$. If $T < \infty$, then by continuity, $h(T) = 0$. We immediately arrive at a contradiction as property (ii) implies that $h(\tau)$ for $\tau$ near but less than T cannot be decreasing. $\square$

**Theorem 1.** $\forall \tau > 0$, $\tilde{x}_{C_1}(\tau) < \tilde{x}_{C_2}(\tau)$.

*Proof.* We begin by showing that $\tilde{x}_A$ and $\tilde{x}_B$ are monotonically increasing in time. Based on (7), we know $0 < \tilde{x}_A(0) < \tilde{x}_{A_{ss}}$ for nonzero $x_I$. From Lemma 1 it follows that $\frac{d\tilde{x}_A}{d\tau} > 0$ for all $\tau \geq 0^+$.

Let $H_{\tilde{A}}(\tilde{x}_A) = \frac{\left(\frac{\tilde{x}_A}{\tilde{K}_{AB}}\right)^{h_{AB}}}{1 + \left(\frac{\tilde{x}_A}{\tilde{K}_{AB}}\right)^{h_{AB}}}$. Assume that there exists $\tau^* \geq 0^+$ at which $f_{\tilde{B}}(\tau^*) = 0$. Using that $f_{\tilde{B}}(\tau^*) = 0$, $\frac{dH_A(\tilde{x}_A)}{d\tilde{x}_A} > 0$ for $\tilde{x}_A > 0$, and $\frac{d\tilde{x}_A}{d\tau} > 0$ for $\tau \geq 0^+$, we get

$$\begin{aligned}
\frac{d^2 \tilde{x}_B}{d\tau^2}\bigg|_{\tau=\tau^*} &= \frac{d}{d\tau}\left[\frac{1}{1 + r_{BD_B}(\tilde{x}_B)} f_{\tilde{B}}\right]\bigg|_{\tau=\tau^*} \\
&= \frac{d}{d\tau}\left[\frac{1}{1 + r_{BD_B}(\tilde{x}_B)}\right] \cdot f_{\tilde{B}}\bigg|_{\tau=\tau^*} + \frac{1}{1 + r_{BD_B}(\tilde{x}_B)} \cdot \frac{df_{\tilde{B}}}{d\tau}\bigg|_{\tau=\tau^*} \\
&= \frac{1}{1 + r_{BD_B}(\tilde{x}_B)} \cdot \frac{df_{\tilde{B}}}{d\tau}\bigg|_{\tau=\tau^*} \\
&= \frac{1}{1 + r_{BD_B}(\tilde{x}_B)} \cdot \left[\frac{dH_A(\tilde{x}_A)}{d\tau}(1 - \gamma_B) - \frac{d\tilde{x}_B}{d\tau}\right]\bigg|_{\tau=\tau^*} \\
&= \frac{1}{1 + r_{BD_B}(\tilde{x}_B)} \frac{dH_A(\tilde{x}_A)}{d\tilde{x}_A} \frac{d\tilde{x}_A}{d\tau}(1 - \gamma_B)\bigg|_{\tau=\tau^*} > 0.
\end{aligned}$$

Because (i) $\frac{d\tilde{x}_B}{d\tau}\big|_{\tau=0^+} = 0$, $\frac{d^2 \tilde{x}_B}{d\tau^2}\big|_{\tau=0^+} > 0$ (ii) $\frac{d^2 \tilde{x}_B}{d\tau^2}\big|_{\tau=\tau^*} > 0$ wherever $\frac{d\tilde{x}_B}{d\tau}\big|_{\tau=\tau^*} = 0$ and $\tau^* > 0$, based on Lemma 2 we know that $\frac{d\tilde{x}_B}{d\tau} > 0$ for all $\tau > 0$.

Next, we will show that $\forall \tau > 0$, $\tilde{x}_{B_1}(\tau) > \tilde{x}_{B_2}(\tau)$. Let $w_B(\tau) = \tilde{x}_{B_1}(\tau) - \tilde{x}_{B_2}(\tau)$. Based on (7), we know that $\tilde{x}_{B_1}(0^+) = \tilde{x}_{B_2}(0^+)$, i.e., $w_B(0^+) = 0$. Consider any $\tau^* \geq 0^+$ at which $w_B(\tau^*) = 0$, i.e., $\tilde{x}_{B_1}(\tau^*) = \tilde{x}_{B_2}(\tau^*)$. Because $\tilde{\eta}_{BD_{B_1}} < \tilde{\eta}_{BD_{B_2}}$, based on (8) we know $\frac{1}{1 + r_{BD_{B_1}}(\tilde{x}_{B_1})}\big|_{\tau=\tau^*} > \frac{1}{1 + r_{BD_{B_2}}(\tilde{x}_{B_2})}\big|_{\tau=\tau^*}$. Hence,

$$\begin{aligned}
\frac{dw_B}{d\tau}\bigg|_{\tau=\tau^*} &= \frac{d}{d\tau}\left[\tilde{x}_{B_1} - \tilde{x}_{B_2}\right]\bigg|_{\tau=\tau^*} \\
&= \left[\frac{1}{1 + r_{BD_{B_1}}(\tilde{x}_{B_1})} f_{\tilde{B}_1} - \frac{1}{1 + r_{BD_{B_2}}(\tilde{x}_{B_2})} f_{\tilde{B}_2}\right]\bigg|_{\tau=\tau^*} \\
&= \left(\frac{1}{1 + r_{BD_{B_1}}(\tilde{x}_B)} - \frac{1}{1 + r_{BD_{B_2}}(\tilde{x}_B)}\right) f_{\tilde{B}}\bigg|_{\tau=\tau^*} \\
&\begin{cases} = 0, & \text{if } \tau^* = 0^+ \quad \text{as } f_{\tilde{B}}\big|_{\tau=0^+} = 0 \\ > 0, & \text{if } \tau^* > 0 \quad \text{as } \frac{d\tilde{x}_B}{d\tau} > 0 \text{ for } \tau > 0 \implies f_{\tilde{B}}\big|_{\tau=\tau^*} > 0. \end{cases}
\end{aligned}$$

If $\tau^* = 0^+$, we can further show that

$$\frac{d^2 w_B}{d\tau^2}\bigg|_{\tau=0^+} = \frac{d^2}{d\tau^2}\left[\tilde{x}_{B_1} - \tilde{x}_{B_2}\right]\bigg|_{\tau=0^+} = \left(\frac{1}{1 + r_{BD_{B_1}}(\tilde{x}_B)} - \frac{1}{1 + r_{BD_{B_2}}(\tilde{x}_B)}\right) \frac{dH_A(\tilde{x}_A)}{d\tilde{x}_A} \frac{d\tilde{x}_A}{d\tau}(1 - \gamma_B)\bigg|_{\tau=0^+} > 0.$$

Now because (i) $w_B(0^+) = 0$, $\frac{dw_B}{d\tau}\big|_{\tau=0^+} = 0$, $\frac{d^2 w_B}{d\tau^2}\big|_{\tau=0^+} > 0$ (ii) $\frac{dw_B}{d\tau}\big|_{\tau=\tau^*} > 0$ wherever $w_B(\tau^*) = 0$ and $\tau^* > 0$, based on Lemma 2 we know that $w_B(\tau) > 0$, i.e., $\tilde{x}_{B_1}(\tau) > \tilde{x}_{B_2}(\tau)$ for all $\tau > 0$.

Finally, we will show that $\forall \tau > 0$, $\tilde{x}_{C_1}(\tau) < \tilde{x}_{C_2}(\tau)$. Let $w_C(\tau) = \tilde{x}_{C_2}(\tau) - \tilde{x}_{C_1}(\tau)$. Based on (7), we know $\tilde{x}_{C_1}(0^+) = \tilde{x}_{C_2}(0^+)$, i.e., $w_C(0^+) = 0$. Consider any $\tau^* \geq 0^+$ at which $w_C(\tau^*) = 0$, i.e., $\tilde{x}_{C_1}(\tau^*) = \tilde{x}_{C_2}(\tau^*)$. Let $\tilde{x}_C = \tilde{x}_{C_1}(\tau^*) = \tilde{x}_{C_2}(\tau^*)$.

$$\left.\frac{dw_C}{d\tau}\right|_{\tau=\tau^*} = \frac{1}{1 + r_{CD_C}(\tilde{x}_C)} \left.(f_{\tilde{C}_2} - f_{\tilde{C}_1})\right|_{\tau=\tau^*}.$$

If $\tau^* > 0^+$, then $\tilde{x}_{B_1}(\tau^*) > \tilde{x}_{B_2}(\tau^*)$, which based on (7) indicates that $f_{\tilde{C}_2}(\tau^*) > f_{\tilde{C}_1}(\tau^*)$. In this case, $\left.\frac{dw_C}{d\tau}\right|_{\tau=\tau^*} > 0$.

Now consider the case where $\tau^* = 0^+$. Because $\tilde{x}_{B_1}(0^+) = \tilde{x}_{B_2}(0^+)$ and $\tilde{x}_{C_1}(0^+) = \tilde{x}_{C_2}(0^+)$, we know that $\forall n \in \mathbb{N}$, $\left.\frac{\partial^n}{\partial \tilde{x}_i^n}\left(\frac{d\tilde{x}_{C_1}}{d\tau}\right)\right|_{\tau=0^+} = \left.\frac{\partial^n}{\partial \tilde{x}_i^n}\left(\frac{d\tilde{x}_{C_2}}{d\tau}\right)\right|_{\tau=0^+}$ ($i = A, B_1, B_2, C_1, C_2$). Using the chain rule we can further show that

$$\begin{aligned}
\left.\frac{d^2 w_C}{d\tau^2}\right|_{\tau=0^+} &= \left.\frac{d}{d\tau}\left(\frac{d\tilde{x}_{C_2}}{d\tau}\right) - \frac{d}{d\tau}\left(\frac{d\tilde{x}_{C_1}}{d\tau}\right)\right|_{\tau=0^+} \\
&= \frac{\partial}{\partial \tilde{x}_A}\left(\frac{d\tilde{x}_{C_2}}{d\tau}\right)\frac{d\tilde{x}_A}{d\tau} + \frac{\partial}{\partial \tilde{x}_{B_2}}\left(\frac{d\tilde{x}_{C_2}}{d\tau}\right)\frac{d\tilde{x}_{B_2}}{d\tau} + \frac{\partial}{\partial \tilde{x}_{C_2}}\left(\frac{d\tilde{x}_{C_2}}{d\tau}\right)\frac{d\tilde{x}_{C_2}}{d\tau} \\
&\quad - \frac{\partial}{\partial \tilde{x}_A}\left(\frac{d\tilde{x}_{C_1}}{d\tau}\right)\frac{d\tilde{x}_A}{d\tau} - \frac{\partial}{\partial \tilde{x}_{B_1}}\left(\frac{d\tilde{x}_{C_1}}{d\tau}\right)\frac{d\tilde{x}_{B_1}}{d\tau} - \left.\frac{\partial}{\partial \tilde{x}_{C_1}}\left(\frac{d\tilde{x}_{C_1}}{d\tau}\right)\frac{d\tilde{x}_{C_1}}{d\tau}\right|_{\tau=0^+} \\
&= \left.\frac{\partial}{\partial \tilde{x}_A}\left(\frac{d\tilde{x}_{C_2}}{d\tau}\right)\frac{d\tilde{x}_A}{d\tau} - \frac{\partial}{\partial \tilde{x}_A}\left(\frac{d\tilde{x}_{C_1}}{d\tau}\right)\frac{d\tilde{x}_A}{d\tau}\right|_{\tau=0^+} = 0,
\end{aligned}$$

$$\begin{aligned}
\left.\frac{d^3 w_C}{d\tau^3}\right|_{\tau=0^+} &= \left.\frac{d^2}{d\tau^2}\left(\frac{d\tilde{x}_{C_2}}{d\tau}\right) - \frac{d^2}{d\tau^2}\left(\frac{d\tilde{x}_{C_1}}{d\tau}\right)\right|_{\tau=0^+} \\
&= \frac{d}{d\tau}\left[\frac{\partial}{\partial \tilde{x}_A}\left(\frac{d\tilde{x}_{C_2}}{d\tau}\right)\frac{d\tilde{x}_A}{d\tau}\right] + \frac{d}{d\tau}\left[\frac{\partial}{\partial \tilde{x}_{B_2}}\left(\frac{d\tilde{x}_{C_2}}{d\tau}\right)\frac{d\tilde{x}_{B_2}}{d\tau}\right] + \frac{d}{d\tau}\left[\frac{\partial}{\partial \tilde{x}_{C_2}}\left(\frac{d\tilde{x}_{C_2}}{d\tau}\right)\frac{d\tilde{x}_{C_2}}{d\tau}\right] \\
&\quad - \frac{d}{d\tau}\left[\frac{\partial}{\partial \tilde{x}_A}\left(\frac{d\tilde{x}_{C_1}}{d\tau}\right)\frac{d\tilde{x}_A}{d\tau}\right] - \frac{d}{d\tau}\left[\frac{\partial}{\partial \tilde{x}_{B_1}}\left(\frac{d\tilde{x}_{C_1}}{d\tau}\right)\frac{d\tilde{x}_{B_1}}{d\tau}\right] - \left.\frac{d}{d\tau}\left[\frac{\partial}{\partial \tilde{x}_{C_1}}\left(\frac{d\tilde{x}_{C_1}}{d\tau}\right)\frac{d\tilde{x}_{C_1}}{d\tau}\right]\right|_{\tau=0^+} \\
&= \frac{\partial}{\partial \tilde{x}_A}\left(\frac{d\tilde{x}_{C_2}}{d\tau}\right)\frac{d}{d\tau}\left(\frac{d\tilde{x}_A}{d\tau}\right) + \frac{d}{d\tau}\left[\frac{\partial}{\partial \tilde{x}_A}\left(\frac{d\tilde{x}_{C_2}}{d\tau}\right)\right]\frac{d\tilde{x}_A}{d\tau} + \frac{\partial}{\partial \tilde{x}_{B_2}}\left(\frac{d\tilde{x}_{C_2}}{d\tau}\right)\frac{d}{d\tau}\left(\frac{d\tilde{x}_{B_2}}{d\tau}\right) \\
&\quad + \frac{d}{d\tau}\left[\frac{\partial}{\partial \tilde{x}_{B_2}}\left(\frac{d\tilde{x}_{C_2}}{d\tau}\right)\right]\frac{d\tilde{x}_{B_2}}{d\tau} + \frac{\partial}{\partial \tilde{x}_{C_2}}\left(\frac{d\tilde{x}_{C_2}}{d\tau}\right)\frac{d}{d\tau}\left(\frac{d\tilde{x}_{C_2}}{d\tau}\right) + \frac{d}{d\tau}\left[\frac{\partial}{\partial \tilde{x}_{C_2}}\left(\frac{d\tilde{x}_{C_2}}{d\tau}\right)\right]\frac{d\tilde{x}_{C_2}}{d\tau} \\
&\quad - \frac{\partial}{\partial \tilde{x}_A}\left(\frac{d\tilde{x}_{C_1}}{d\tau}\right)\frac{d}{d\tau}\left(\frac{d\tilde{x}_A}{d\tau}\right) - \frac{d}{d\tau}\left[\frac{\partial}{\partial \tilde{x}_A}\left(\frac{d\tilde{x}_{C_1}}{d\tau}\right)\right]\frac{d\tilde{x}_A}{d\tau} - \frac{\partial}{\partial \tilde{x}_{B_1}}\left(\frac{d\tilde{x}_{C_1}}{d\tau}\right)\frac{d}{d\tau}\left(\frac{d\tilde{x}_{B_1}}{d\tau}\right) \\
&\quad - \frac{d}{d\tau}\left[\frac{\partial}{\partial \tilde{x}_{B_1}}\left(\frac{d\tilde{x}_{C_1}}{d\tau}\right)\right]\frac{d\tilde{x}_{B_1}}{d\tau} - \frac{\partial}{\partial \tilde{x}_{C_1}}\left(\frac{d\tilde{x}_{C_1}}{d\tau}\right)\frac{d}{d\tau}\left(\frac{d\tilde{x}_{C_1}}{d\tau}\right) - \left.\frac{d}{d\tau}\left[\frac{\partial}{\partial \tilde{x}_{C_1}}\left(\frac{d\tilde{x}_{C_1}}{d\tau}\right)\right]\frac{d\tilde{x}_{C_1}}{d\tau}\right|_{\tau=0^+} \\
&= \frac{\partial}{\partial \tilde{x}_A}\left(\frac{d\tilde{x}_{C_2}}{d\tau}\right)\frac{d^2\tilde{x}_A}{d\tau^2} + \frac{\partial^2}{\partial \tilde{x}_A^2}\left(\frac{d\tilde{x}_{C_2}}{d\tau}\right)\left(\frac{d\tilde{x}_A}{d\tau}\right)^2 + \frac{\partial}{\partial \tilde{x}_{B_2}}\left(\frac{d\tilde{x}_{C_2}}{d\tau}\right)\frac{d^2\tilde{x}_{B_2}}{d\tau^2} + \frac{\partial}{\partial \tilde{x}_{C_2}}\left(\frac{d\tilde{x}_{C_2}}{d\tau}\right)\frac{d^2\tilde{x}_{C_2}}{d\tau^2} \\
&\quad - \frac{\partial}{\partial \tilde{x}_A}\left(\frac{d\tilde{x}_{C_1}}{d\tau}\right)\frac{d^2\tilde{x}_A}{d\tau^2} - \frac{\partial^2}{\partial \tilde{x}_A^2}\left(\frac{d\tilde{x}_{C_1}}{d\tau}\right)\left(\frac{d\tilde{x}_A}{d\tau}\right)^2 - \frac{\partial}{\partial \tilde{x}_{B_1}}\left(\frac{d\tilde{x}_{C_1}}{d\tau}\right)\frac{d^2\tilde{x}_{B_1}}{d\tau^2} - \left.\frac{\partial}{\partial \tilde{x}_{C_1}}\left(\frac{d\tilde{x}_{C_1}}{d\tau}\right)\frac{d^2\tilde{x}_{C_1}}{d\tau^2}\right|_{\tau=0^+} \\
&= \left.\frac{\partial}{\partial \tilde{x}_{B_2}}\left(\frac{d\tilde{x}_{C_2}}{d\tau}\right)\frac{d^2\tilde{x}_{B_2}}{d\tau^2} - \frac{\partial}{\partial \tilde{x}_{B_1}}\left(\frac{d\tilde{x}_{C_1}}{d\tau}\right)\frac{d^2\tilde{x}_{B_1}}{d\tau^2}\right|_{\tau=0^+}.
\end{aligned}$$

Then because $\left.\frac{\partial}{\partial \tilde{x}_{B_1}}\left(\frac{d\tilde{x}_{C_1}}{d\tau}\right)\right|_{\tau=0^+} = \left.\frac{\partial}{\partial \tilde{x}_{B_2}}\left(\frac{d\tilde{x}_{C_2}}{d\tau}\right)\right|_{\tau=0^+} < 0$ and $\left.\frac{d^2 w_B}{d\tau^2}\right|_{\tau=0^+} > 0$, we know $\left.\frac{d^3 w_C}{d\tau^3}\right|_{\tau=0^+} > 0$.

Now as (i) $w_C(0^+) = 0$, $\left.\frac{dw_C}{d\tau}\right|_{\tau=0^+} = 0$, $\left.\frac{d^2 w_C}{d\tau^2}\right|_{\tau=0^+} = 0$, $\left.\frac{d^3 w_C}{d\tau^3}\right|_{\tau=0^+} > 0$ (ii) $\left.\frac{dw_C}{d\tau}\right|_{\tau=\tau^*} > 0$ wherever $w_C(\tau^*) = 0$ and $\tau^* > 0$, based on Lemma 2 we know that $w_C(\tau) > 0$, i.e., $\tilde{x}_{C_1}(\tau) < \tilde{x}_{C_2}(\tau)$ for all $\tau > 0$. $\square$

**Theorem 2.** $RT_{\tilde{x}_{C_1}} > RT_{\tilde{x}_{C_2}}$ (RT: response time).

*Proof.* Based on the previous lemmas and previous theorem, we know that $\tilde{x}_C$ has a unique steady state $\tilde{x}_{C_{ss}}$. Note also that neither $\tilde{x}_{C_{ss}}$ nor $\tilde{x}_C(0)$ depends on the choice of $\tilde{\eta}_{XD_X}$ ($X = A, B, C$).

Based on Theorem 1, we know that $\tilde{x}_{C_2}$ is larger than $\tilde{x}_{C_1}$. This means when $\tilde{x}_{C_1}$ reaches the midpoint between $\tilde{x}_C(0)$ and $\tilde{x}_{C_{ss}}$ (for biological implications, we only consider $\tilde{x}_C(0) < \tilde{x}_{C_{ss}}$), $\tilde{x}_{C_2}$ has reached a value larger than the midpoint. By continuity of $\tilde{x}_C$, we know that $\tilde{x}_{C_2}$ must have reached the midpoint earlier than $\tilde{x}_{C_1}$. This in turn implies that the response time of $\tilde{x}_{C_1}$ is larger than the response time of $\tilde{x}_{C_2}$. □

## 1.17 Proof of the Effects of Intermodular Retroactivity on Response Time and Pulse Amplitude in IFFLs

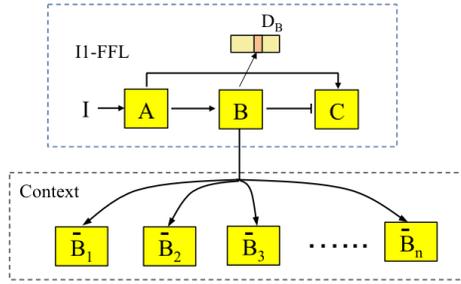

**Figure S4:** An I1-FFL connected to $n$ additional circuits, i.e., its context.

We now consider the model of an IFFL where node B is connected to, and hence, serves as the input to other circuits (Figure S4). Using the method shown in Gyorgy and Del Vecchio (2014), we can derive a model for an IFFL in which intermodular retroactivity is accounted for. $\frac{d\tilde{x}_A}{d\tau}$, $\frac{d\tilde{x}_B}{d\tau}$, and $\frac{d\tilde{x}_C}{d\tau}$ in an I1-FFL where only node B is connected to $n$ additional modules (nodes) can be expressed as:

$$\begin{bmatrix} \frac{d\tilde{x}_A}{d\tau} \\ \frac{d\tilde{x}_B}{d\tau} \\ \frac{d\tilde{x}_C}{d\tau} \end{bmatrix} = \begin{bmatrix} 1 & 0 & 0 \\ 0 & \frac{1+r_{BD_B}(\tilde{x}_B)}{1+r_{BD_B}(\tilde{x}_B)+\Sigma_{i=1}^{n}S_{B^i}(\tilde{x}_B)} & 0 \\ 0 & 0 & 1 \end{bmatrix} \begin{bmatrix} g_{\tilde{A}} \\ g_{\tilde{B}} \\ g_{\tilde{C}} \end{bmatrix} \qquad (9)$$

where

$$S_{B^i}(\tilde{x}_B) = \tilde{\eta}_{B^i} h_{B^i}^2 \left(\frac{\tilde{x}_B}{\tilde{K}_{B^i}}\right)^{h_{B^i}-1} \left(1 + \left(\frac{\tilde{x}_B}{\tilde{K}_{B^i}}\right)^{h_{B^i}}\right)^{-2} \qquad (10)$$

and

$$\begin{aligned} g_{\tilde{A}} &= \frac{1}{1 + r_{AD_A}(\tilde{x}_A)} f_{\tilde{A}} \\ g_{\tilde{B}} &= \frac{1}{1 + r_{BD_B}(\tilde{x}_B)} f_{\tilde{B}} \\ g_{\tilde{C}} &= \frac{1}{1 + r_{BD_B}(\tilde{x}_B)} f_{\tilde{C}}, \end{aligned} \qquad (11)$$

where $r_{AD_A}$, $r_{BD_B}$, $r_{CD_C}$, $f_{\tilde{A}}$, $f_{\tilde{B}}$, and $f_{\tilde{C}}$ are defined the same as in (6).

Let $\vec{\tilde{x}}_1$ and $\vec{\tilde{x}}_2$ denote the concentrations of $A$, $B$, and $C$ in two I1-FFL models (i.e., $\vec{\tilde{x}}_1 = [\tilde{x}_{A_1}, \tilde{x}_{B_1}, \tilde{x}_{C_1}]$, $\vec{\tilde{x}}_2 = [\tilde{x}_{A_2}, \tilde{x}_{B_2}, \tilde{x}_{C_2}]$), in which all parameters are held identical except that node B is connected to different numbers of binding sites in the $k$-th module of its context ($1 \le k \le n$). $\tilde{\eta}_{B^k}$ equals $\tilde{\eta}_{B_1^k}$ and $\tilde{\eta}_{B_2^k}$, respectively. At $\tau = 0$, $x_I$ undergoes a stepwise increase and is kept constant afterwards. The initial values of $\tilde{x}_A$, $\tilde{x}_{B_i}$ ($i = 1, 2$), and $\tilde{x}_C$ are the corresponding steady values before the increase in $x_I$. Without loss of generality, we assume that for one given $k$, $\tilde{\eta}_{B_1^k} < \tilde{\eta}_{B_2^k}$. We now show that $\forall \tau > 0$, $\tilde{x}_{C_1}(\tau) < \tilde{x}_{C_2}(\tau)$, i.e., the concentration of $C$ in I1-FFL model #1 is less than the concentration of $C$ in I1-FFL model #2 at all positive times.

We begin by showing that $\tilde{x}_A$ and $\tilde{x}_B$ are monotonically increasing in time. From the proof of Theorem 1, we already know that $\frac{d\tilde{x}_A}{d\tau} > 0$ for all $\tau \ge 0^+$. Assume that there exists $\tau^* \ge 0^+$ at which $\frac{d\tilde{x}_B}{d\tau} = 0$, i.e., $g_{\tilde{B}} = 0$. Using that $\frac{dg_{\tilde{B}}}{d\tau}\bigg|_{\tau=\tau^*} > 0$ wherever $g_{\tilde{B}}\big|_{\tau=\tau^*} = 0$ and $\tau^* \ge 0^+$ from the proof of Theorem 1, we get

23

$$\left.\frac{d^2\tilde{x}_B}{d\tau^2}\right|_{\tau=\tau^*} = \frac{1+r_{BD_B}(\tilde{x}_B)}{1+r_{BD_B}(\tilde{x}_B)+\Sigma_{i=1}^n S_{B^i}(\tilde{x}_B)} \left.\frac{dg_{\tilde{B}}}{d\tau}\right|_{\tau=\tau^*} > 0.$$

Then because(i) $\left.\frac{d\tilde{x}_B}{d\tau}\right|_{\tau=0^+} = 0$, $\left.\frac{d^2\tilde{x}_B}{d\tau^2}\right|_{\tau=0^+} > 0$ (ii) $\left.\frac{d^2\tilde{x}_B}{d\tau^2}\right|_{\tau=\tau^*} > 0$ wherever $\left.\frac{d\tilde{x}_B}{d\tau}\right|_{\tau=\tau^*} = 0$ and $\tau^* > 0$, based on Lemma 2 we know that $\frac{d\tilde{x}_B}{d\tau} > 0$ for all $\tau > 0$.

Next, we will show that $\forall \tau > 0$, $\tilde{x}_{B_1}(\tau) > \tilde{x}_{B_2}(\tau)$. Let $w_B(\tau) = \tilde{x}_{B_1}(\tau) - \tilde{x}_{B_2}(\tau)$. Based on (7) and (9), we know that $\tilde{x}_{B_1}(0^+) = \tilde{x}_{B_2}(0^+)$, i.e., $w_B(0^+) = 0$. Consider any $\tau^* \geq 0^+$ at which $w_B(\tau^*) = 0$, i.e., $\tilde{x}_{B_1}(\tau^*) = \tilde{x}_{B_2}(\tau^*) = \tilde{x}_B(\tau^*)$. Because $\tilde{\eta}_{B_1^k} < \tilde{\eta}_{B_2^k}$, based on (10) we know $\left.S_{B_1^k}(\tilde{x}_B)\right|_{\tau=\tau^*} < \left.S_{B_2^k}(\tilde{x}_B)\right|_{\tau=\tau^*}$. Hence,

$$\left.\frac{dw_B}{d\tau}\right|_{\tau=\tau^*} = \left( \frac{1+r_{BD_B}(\tilde{x}_B)}{1+r_{BD_B}(\tilde{x}_B)+\Sigma_{i=1,i\neq k}^n S_{B^i}(\tilde{x}_B)+S_{B_1^k}(\tilde{x}_B)} - \frac{1+r_{BD_B}(\tilde{x}_B)}{1+r_{BD_B}(\tilde{x}_B)+\Sigma_{i=1,i\neq k}^n S_{B^i}(\tilde{x}_B)+S_{B_2^k}(\tilde{x}_B)} \right) g_{\tilde{B}}\bigg|_{\tau=\tau^*}$$

$$\begin{cases} = 0, & \text{if } \tau^* = 0^+ \quad \text{as } g_{\tilde{B}}|_{\tau=0^+} = 0 \\ > 0, & \text{if } \tau^* > 0 \quad \text{as } \frac{d\tilde{x}_B}{d\tau} > 0 \text{ for } \tau > 0 \implies g_{\tilde{B}}|_{\tau=\tau^*} > 0. \end{cases}$$

If $\tau^* = 0^+$, we can further show that

$$\left.\frac{d^2 w_B}{d\tau^2}\right|_{\tau=0^+} = \left( \frac{1+r_{BD_B}(\tilde{x}_B)}{1+r_{BD_B}(\tilde{x}_B)+\Sigma_{i=1,i\neq k}^n S_{B^i}(\tilde{x}_B)+S_{B_1^k}(\tilde{x}_B)} - \frac{1+r_{BD_B}(\tilde{x}_B)}{1+r_{BD_B}(\tilde{x}_B)+\Sigma_{i=1,i\neq k}^n S_{B^i}(\tilde{x}_B)+S_{B_2^k}(\tilde{x}_B)} \right)$$
$$\cdot \left.\frac{dg_{\tilde{B}}}{d\tau}\right|_{\tau=0^+} > 0.$$

Now because (i) $w_B(0^+) = 0$, $\left.\frac{dw_B}{d\tau}\right|_{\tau=0^+} = 0$, $\left.\frac{d^2 w_B}{d\tau^2}\right|_{\tau=0^+} > 0$ (ii) $\left.\frac{dw_B}{d\tau}\right|_{\tau=\tau^*} > 0$ wherever $w_B(\tau^*) = 0$ and $\tau^* > 0$, based on Lemma 2 we know that $w_B(\tau) > 0$, i.e., $\tilde{x}_{B_1}(\tau) > \tilde{x}_{B_2}(\tau)$ for all $\tau > 0$.

Finally, following the rest of the proof for Theorem 1, we know that $\tilde{x}_{C_1}(\tau) < \tilde{x}_{C_2}(\tau)$ for all $\tau > 0$. Then, similar to the proof of Theorem 2, we conclude that the response time of $\tilde{x}_{C_1}$ is larger than the response time of $\tilde{x}_{C_2}$.

## 1.18 Proof of the Effects of $\tilde{\eta}_{AD_A}$ and $\tilde{\eta}_{CD_C}$ on Response Time in a Negative Autoregulated Circuit

In this section, we show that for any parameters, response time always increases in a negative autoregulated circuit if retroactivity on node A or C, $\tilde{\eta}_{AD_A}$ or $\tilde{\eta}_{CD_C}$, increases.

$\frac{d\tilde{x}_A}{d\tau}$ and $\frac{d\tilde{x}_C}{d\tau}$ in a negative autoregulated circuit where only $\tilde{\eta}_{AD_A}$ may be allowed to vary can be expressed as:

$$\begin{bmatrix} \frac{d\tilde{x}_A}{d\tau} \\ \frac{d\tilde{x}_C}{d\tau} \end{bmatrix} = \begin{bmatrix} \frac{1}{1+r_{AD_A}(\tilde{x}_A)} & 0 \\ 0 & \frac{1}{1+r_{CD_C}(\tilde{x}_C)} \end{bmatrix} \begin{bmatrix} f_{\tilde{A}} \\ f_{\tilde{C}} \end{bmatrix}, \tag{12}$$

$$f_{\tilde{A}} = \frac{\left(\frac{x_I}{K_{IA}}\right)^{h_{IA}}}{1+\left(\frac{x_I}{K_{IA}}\right)^{h_{IA}}}(1-\gamma_A) + \gamma_A - \tilde{x}_A$$

$$f_{\tilde{C}} = \frac{\left(\frac{\tilde{x}_A}{\tilde{K}_{AC}}\right)^{h_{AC}}}{\left(1+\left(\frac{\tilde{x}_A}{\tilde{K}_{AC}}\right)^{h_{AC}}\right)\left(1+\left(\frac{\tilde{x}_C}{\tilde{K}_{CC}}\right)^{h_{CC}}\right)}(1-\gamma_C) + \gamma_C - \tilde{x}_C, \tag{13}$$

$$r_{AD_A}(\tilde{x}_A) = \tilde{\eta}_{AD_A} h_{AD_A}^2 \left(\frac{\tilde{x}_A}{\tilde{K}_{AD_A}}\right)^{h_{AD_A}-1} \left(1+\left(\frac{\tilde{x}_A}{\tilde{K}_{AD_A}}\right)^{h_{AD_A}}\right)^{-2}$$

$$r_{CD_C}(\tilde{x}_C) = \tilde{\eta}_{CD_C} h_{CD_C}^2 \left(\frac{\tilde{x}_C}{\tilde{K}_{CD_C}}\right)^{h_{CD_C}-1} \left(1+\left(\frac{\tilde{x}_C}{\tilde{K}_{CD_C}}\right)^{h_{CD_C}}\right)^{-2}, \tag{14}$$

where $\gamma_A, \gamma_C \in (0,1)$ such that $\tilde{x}_A, \tilde{x}_C \in (0,1)$.

Let $\vec{\tilde{x}}_1$ and $\vec{\tilde{x}}_2$ denote the concentrations of $A$ and $C$ in two negative autoregulated circuit models (i.e., $\vec{\tilde{x}}_1 = [\tilde{x}_{A_1}, \tilde{x}_{C_1}]$, $\vec{\tilde{x}}_2 = [\tilde{x}_{A_2}, \tilde{x}_{C_2}]$), in which all parameters are held identical except that node A is connected to different numbers of downstream targets such that retroactivity coefficient $\tilde{\eta}_{AD_A}$ equals $\tilde{\eta}_{AD_{A_1}}$ and $\tilde{\eta}_{AD_{A_2}}$, respectively. At $\tau = 0$, $x_I$ undergoes a stepwise increase and is kept constant afterwards. The initial values of $\tilde{x}_{A_i}$ and $\tilde{x}_{C_i}$ ($i = 1, 2$) are the corresponding steady values before the increase in $x_I$. Without loss of generality, we assume $\tilde{\eta}_{AD_{A_1}} < \tilde{\eta}_{AD_{A_2}}$. Now we will show that $\forall \tau > 0$, $\tilde{x}_{C_1}(\tau) > \tilde{x}_{C_2}(\tau)$, i.e., the concentration of $C$ in I1-FFL model #1 is larger than the concentration of $C$ in I1-FFL model #2 at all positive times.

**Theorem 3.** $\forall \tau > 0$, $\tilde{x}_{C_1} > \tilde{x}_{C_2}$ and $RT_{\tilde{x}_{C_1}} < RT_{\tilde{x}_{C_2}}$ (RT: response time).

*Proof.* We begin by showing that $\tilde{x}_A$ and $\tilde{x}_C$ are monotonically increasing in time. Based on (12), we know $\tilde{x}_A(0) < \tilde{x}_{A_{ss}}$ for nonzero $x_I$. From Lemma 1 it follows that $\frac{d\tilde{x}_A}{d\tau} > 0$ for all $\tau \geq 0^+$.

Let $H_{\tilde{A}}(\tilde{x}_A) = \frac{\left(\frac{\tilde{x}_A}{\tilde{K}_{AC}}\right)^{h_{AC}}}{1+\left(\frac{\tilde{x}_A}{\tilde{K}_{AC}}\right)^{h_{AC}}}$ and $H_{\tilde{C}}(\tilde{x}_C) = \frac{1}{1+\left(\frac{\tilde{x}_C}{\tilde{K}_{CC}}\right)^{h_{CC}}}$. Assume that there exists $\tau^* \geq 0^+$ at which $f_{\tilde{C}}(\tau^*) = 0$. Because $f_{\tilde{C}}(\tau^*) = 0$, $\frac{dH_A(\tilde{x}_A)}{d\tilde{x}_A} > 0$ for $\tilde{x}_A > 0$, and $\frac{d\tilde{x}_A}{d\tau} > 0$ for $\tau \geq 0^+$, we get

$$\begin{aligned} \left.\frac{d^2\tilde{x}_C}{d\tau^2}\right|_{\tau=\tau^*} &= \frac{1}{1+r_{CD_C}(\tilde{x}_C)} \left.\frac{df_{\tilde{C}}}{d\tau}\right|_{\tau=\tau^*} + \frac{d}{d\tau}\left[\frac{1}{1+r_{CD_C}(\tilde{x}_C)}\right] f_{\tilde{C}}\bigg|_{\tau=\tau^*} \\ &= \frac{1}{1+r_{CD_C}(\tilde{x}_C)} \left[\frac{d[H_A(\tilde{x}_A)H_C(\tilde{x}_C)]}{d\tau}(1-\gamma_C) - \frac{d\tilde{x}_C}{d\tau}\right]\bigg|_{\tau=\tau^*} \\ &= \frac{1}{1+r_{CD_C}(\tilde{x}_C)} \left[\frac{dH_A(\tilde{x}_A)}{d\tilde{x}_A}\frac{d\tilde{x}_A}{d\tau}H_C(\tilde{x}_C) + \frac{dH_C(\tilde{x}_C)}{d\tilde{x}_C}\frac{d\tilde{x}_C}{d\tau}H_A(\tilde{x}_A)\right](1-\gamma_C)\bigg|_{\tau=\tau^*} \\ &= \frac{1}{1+r_{CD_C}(\tilde{x}_C)} \frac{dH_A(\tilde{x}_A)}{d\tilde{x}_A}\frac{d\tilde{x}_A}{d\tau}H_C(\tilde{x}_C)(1-\gamma_C)\bigg|_{\tau=\tau^*} > 0. \end{aligned}$$

Because (i) $\left.\frac{d\tilde{x}_C}{d\tau}\right|_{\tau=0^+} = 0$, $\left.\frac{d^2\tilde{x}_C}{d\tau^2}\right|_{\tau=0^+} > 0$ (ii) $\left.\frac{d^2\tilde{x}_C}{d\tau^2}\right|_{\tau=\tau^*} > 0$ wherever $\left.\frac{d\tilde{x}_C}{d\tau}\right|_{\tau=\tau^*} = 0$ and $\tau^* > 0$, based on Lemma 2 we know that $\frac{d\tilde{x}_C}{d\tau} > 0$ for all $\tau > 0$.

Next, we will show that $\forall \tau > 0$, $\tilde{x}_{A_1} > \tilde{x}_{A_2}$. Let $w_A(\tau) = \tilde{x}_{A_1}(\tau) - \tilde{x}_{A_2}(\tau)$. Based on (13), we know $\tilde{x}_{A_1}(0^+) = \tilde{x}_{A_2}(0^+)$, i.e., $w_A(0^+) = 0$. Consider any $\tau^* \geq 0^+$ at which $w_A(\tau^*) = 0$, i.e., $\tilde{x}_{A_1}(\tau^*) = \tilde{x}_{A_2}(\tau^*)$. Since $\tilde{\eta}_{AD_{A_1}} < \tilde{\eta}_{AD_{A_2}}$, based on (14) we know $\left.\frac{1}{1+r_{AD_{A_1}}(\tilde{x}_{A_1})}\right|_{\tau=\tau^*} > \left.\frac{1}{1+r_{AD_{A_2}}(\tilde{x}_{A_2})}\right|_{\tau=\tau^*}$. Hence,

$$\left.\frac{dw_A}{d\tau}\right|_{\tau=\tau^*} = \left.\frac{1}{1+r_{AD_{A_1}}(\tilde{x}_{A_1})}f_{\tilde{A}_1} - \frac{1}{1+r_{AD_{A_2}}(\tilde{x}_{A_2})}f_{\tilde{A}_2}\right|_{\tau=\tau^*}$$

$$= \left.\left[\frac{1}{1+r_{AD_{A_1}}(\tilde{x}_A)} - \frac{1}{1+r_{AD_{A_2}}(\tilde{x}_A)}\right]f_{\tilde{A}}\right|_{\tau=\tau^*} > 0.$$

Now because (i) $w_A(0^+) = 0$, $\left.\frac{dw_A}{d\tau}\right|_{\tau=0^+} > 0$ (ii) $\left.\frac{dw_A}{d\tau}\right|_{\tau=\tau^*} > 0$ wherever $w_A(\tau^*) = 0$ and $\tau^* > 0$, based on Lemma 2 we know that $w_A(\tau) > 0$, i.e., $\tilde{x}_{A_1}(\tau) > \tilde{x}_{A_2}(\tau)$ for all $\tau > 0$.

Finally, we will show that $\forall \tau > 0$, $\tilde{x}_{C_1} > \tilde{x}_{C_2}$. Let $w_C(\tau) = \tilde{x}_{C_1}(\tau) - \tilde{x}_{C_2}(\tau)$. Based on (13), we know that $\tilde{x}_{C_1}(0^+) = \tilde{x}_{C_2}(0^+)$, i.e., $w_C(0^+) = 0$. Consider any $\tau^* \geq 0^+$ at which $w_C(\tau^*) = 0$, i.e., $\tilde{x}_{C_1}(\tau^*) = \tilde{x}_{C_2}(\tau^*)$. Because $\tilde{x}_{A_1}(\tau) > \tilde{x}_{A_2}(\tau)$ for all $\tau > 0$, we know $H_A(\tilde{x}_{A_1}) > H_A(\tilde{x}_{A_2})$ for all $\tau > 0$. Thus,

$$\left.\frac{dw_C}{d\tau}\right|_{\tau=\tau^*} = \left.\frac{d}{d\tau}\left[\tilde{x}_{C_1} - \tilde{x}_{C_2}\right]\right|_{\tau=\tau^*}$$

$$= \left.\frac{1}{1+r_{CD_{C_1}}(\tilde{x}_{C_1})}f_{\tilde{C}_1} - \frac{1}{1+r_{CD_{C_2}}(\tilde{x}_{C_2})}f_{\tilde{C}_2}\right|_{\tau=\tau^*}$$

$$= \left.\frac{1}{1+r_{CD_C}(\tilde{x}_C)}\left(f_{\tilde{C}_1} - f_{\tilde{C}_2}\right)\right|_{\tau=\tau^*}$$

$$\begin{cases} = 0, & \text{if } \tau^* = 0^+ \\ > 0, & \text{if } \tau^* > 0. \end{cases}$$

If $\tau^* = 0^+$, then using $\frac{dH_A(\tilde{x}_A)}{d\tilde{x}_A} > 0$ for $\tilde{x}_A > 0$, $\left.\frac{dw_A}{d\tau}\right|_{\tau=0^+} > 0$, and $\left.\frac{dw_C}{d\tau}\right|_{\tau=0^+} = 0$, we can further show that

$$\left.\frac{d^2 w_C}{d\tau^2}\right|_{\tau=0^+} = \left.\frac{d^2}{d\tau^2}\left[\tilde{x}_{C_1} - \tilde{x}_{C_2}\right]\right|_{\tau=0^+}$$

$$= \frac{1}{1+r_{CD_C}(\tilde{x}_{C_1})}\frac{dH_A(\tilde{x}_{A_1})}{d\tilde{x}_{A_1}}\frac{d\tilde{x}_{A_1}}{d\tau}H_C(\tilde{x}_{C_1})(1-\gamma_C)$$

$$- \left.\frac{1}{1+r_{CD_C}(\tilde{x}_{C_2})}\frac{dH_A(\tilde{x}_{A_2})}{d\tilde{x}_{A_2}}\frac{d\tilde{x}_{A_2}}{d\tau}H_C(\tilde{x}_{C_2})(1-\gamma_C)\right|_{\tau=0^+}$$

$$= \left.\frac{1}{1+r_{CD_C}(\tilde{x}_C)}\frac{dH_A(\tilde{x}_A)}{d\tilde{x}_A}\left(\frac{d\tilde{x}_{A_1}}{d\tau} - \frac{d\tilde{x}_{A_2}}{d\tau}\right)H_C(\tilde{x}_C)(1-\gamma_C)\right|_{\tau=0^+} > 0.$$

As (i) $w_C(0^+) = 0$, $\left.\frac{dw_C}{d\tau}\right|_{\tau=0^+} = 0$, $\left.\frac{d^2 w_C}{d\tau^2}\right|_{\tau=0^+} > 0$ (ii) $\left.\frac{dw_C}{d\tau}\right|_{\tau=\tau^*} > 0$ wherever $w_C(\tau^*) = 0$ and $\tau^* > 0$, based on Lemma 2 we know that $w_C(\tau) > 0$, i.e., $\tilde{x}_{C_1}(\tau) > \tilde{x}_{C_2}(\tau)$ for all $\tau > 0$.

Then, similar to the proof of Theorem 2, we conclude that the response time of $\tilde{x}_{C_1}$ is shorter than the response time of $\tilde{x}_{C_2}$. □

Now we consider $\frac{d\tilde{x}_A}{d\tau}$ and $\frac{d\tilde{x}_C}{d\tau}$ in a negative autoregulated circuit where only $\tilde{\eta}_{CD_C}$ may be allowed to vary. Let $\vec{\tilde{x}}_1$ and $\vec{\tilde{x}}_2$ denote the concentrations of $A$ and $C$ in two negative autoregulated circuit models (i.e., $\vec{\tilde{x}}_1 = [\tilde{x}_{A_1}, \tilde{x}_{C_1}]$, $\vec{\tilde{x}}_2 = [\tilde{x}_{A_2}, \tilde{x}_{C_2}]$), in which all parameters are held identical except that node C is connected to different numbers of downstream targets such that retroactivity coefficient $\tilde{\eta}_{CD_C}$ equals $\tilde{\eta}_{CD_{C_1}}$ and $\tilde{\eta}_{CD_{C_2}}$, respectively. At $\tau = 0$, $x_I$ undergoes a stepwise increase and is kept constant afterwards. The initial values of $\tilde{x}_{A_i}$ and $\tilde{x}_{C_i}$ ($i = 1, 2$) are the corresponding steady values before the increase in $x_I$. Without loss of generality, we assume $\tilde{\eta}_{CD_{C_1}} < \tilde{\eta}_{CD_{C_2}}$. Now we will show that $\forall \tau > 0$, $\tilde{x}_{C_1}(\tau) > \tilde{x}_{C_2}(\tau)$, i.e., the concentration of $C$ in I1-FFL model #1 is larger than the concentration of $C$ in I1-FFL model #2 at all positive times.

**Theorem 4.** $\forall \tau > 0$, $\tilde{x}_{C_1} > \tilde{x}_{C_2}$ and $RT_{\tilde{x}_{C_1}} < RT_{\tilde{x}_{C_2}}$ (RT: response time).

*Proof.* Similar to before, we have that $\tilde{x}_A$ and $\tilde{x}_C$ are monotonically increasing in time, i.e., (i) $\forall \tau \geq 0^+$, $\frac{d\tilde{x}_A}{d\tau} > 0$ (ii) $\forall \tau > 0$, $\frac{d\tilde{x}_C}{d\tau} > 0$ (see proof of Theorem 3).

Next, we will show that $\forall \tau > 0$, $\tilde{x}_{C_1}(\tau) > \tilde{x}_{C_2}(\tau)$. We define $w_C(\tau)$ and $\tau^*$ similarly as in the proof of Theorem 3 such that $w_C(\tau^*) = 0$. Because $\tilde{\eta}_{CD_{C_1}} < \tilde{\eta}_{CD_{C_2}}$, we have $\frac{1}{1+r_{CD_{C_1}}(\tilde{x}_{C_1})}\Big|_{\tau=\tau^*} > \frac{1}{1+r_{CD_{C_2}}(\tilde{x}_{C_2})}\Big|_{\tau=\tau^*}$. Thus,

$$\frac{dw_C}{d\tau}\bigg|_{\tau=\tau^*} = \left(\frac{1}{1+r_{CD_{C_1}}(\tilde{x}_C)} - \frac{1}{1+r_{CD_{C_2}}(\tilde{x}_C)}\right) f_C \bigg|_{\tau=\tau^*}$$

$$\begin{cases} = 0, & \text{if } \tau^* = 0^+ \\ > 0, & \text{if } \tau^* > 0. \end{cases}$$

If $\tau^* = 0^+$, then using $\frac{dH_A(\tilde{x}_A)}{d\tilde{x}_A} > 0$ for $\tilde{x}_A > 0$ and $\frac{d\tilde{x}_A}{d\tau} > 0$ for $\tau \geq 0^+$, we can further show that

$$\frac{d^2 w_C}{d\tau^2}\bigg|_{\tau=0^+} = \left[\frac{1}{1+r_{CD_{C_1}}(\tilde{x}_{C_1})} - \frac{1}{1+r_{CD_{C_2}}(\tilde{x}_{C_2})}\right] \frac{dH_A(\tilde{x}_A)}{d\tilde{x}_A} \frac{d\tilde{x}_A}{d\tau} H_C(\tilde{x}_C)(1-\gamma_C)\bigg|_{\tau=0^+} > 0.$$

Now because (i) $w_C(0^+) = 0$, $\frac{dw_C}{d\tau}\big|_{\tau=0^+} = 0$, $\frac{d^2 w_C}{d\tau^2}\big|_{\tau=0^+} > 0$ (ii) $\frac{dw_C}{d\tau}\big|_{\tau=\tau^*} > 0$ wherever $w_C(\tau^*) = 0$ and $\tau^* > 0$, based on Lemma 2 we know that $w_C(\tau) > 0$, i.e., $\tilde{x}_{C_1}(\tau) > \tilde{x}_{C_2}(\tau)$ for all $\tau > 0$. Then similar to the proof of Theorem 2, we conclude that the response time of $\tilde{x}_{C_1}$ is shorter than the response time of $\tilde{x}_{C_2}$. □

## 1.19 IFFL Acceleration Persists in the Absence of Parameter Isometry

We generated random kinetic parameters via Latin hypercube sampling. To ensure an even distribution over the large space, $\tilde{K}_X$ ($X = AB, AC, BC$) and $\delta_X$ ($X = A, B, C$) were sampled uniformly on a log scale from the same ranges of values used in Cao et al. (2016) and Shi et al. (2017): $\tilde{K}_X \sim 0.001 - 1$ and $\delta_X \sim 0.01 - 1$. Hill coefficients $h_X$ ($X = AB, AC, BC$) were sampled uniformly from a linear interval starting at 0.5 and ending at 2, including both positive and negative cooperativity.

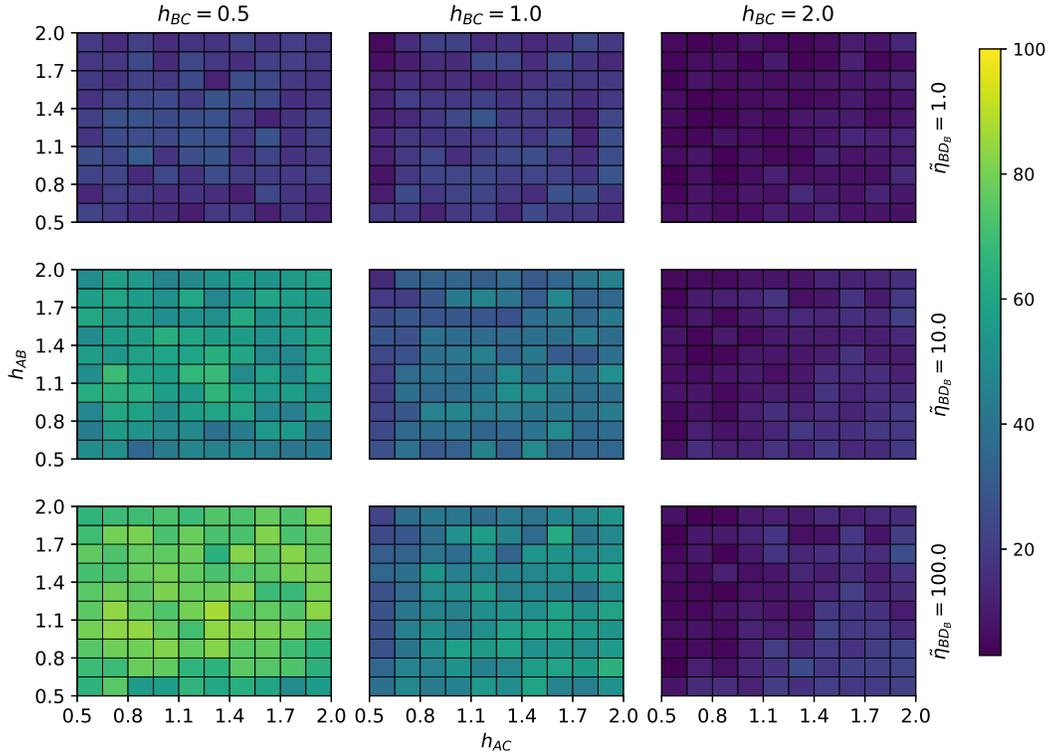

**Figure S5:** Percent of I1-FFL trajectories whose relative response time is less than 90% of the model in the absence of retroactivity calculated based on systematically exploring parameter space as described above, assuming $0.8 h_{BC} \leq h_{BD_B} \leq 1.2 h_{BC}$ and $0.8 \tilde{K}_{BC} \leq \tilde{K}_{BD_B} \leq 1.2 \tilde{K}_{BC}$. The trajectories are separated evenly by $h_{AB}$ and $h_{AC}$ into 100 voxels, the color of which represents the percent of trajectories whose relative response time is less than 90% of the model in the absence of retroactivity out of the 100 simulated trajectories falling into that bin. Here, relative response time is defined as the ratio of the response time of the model to the response time of the model without retroactivity.

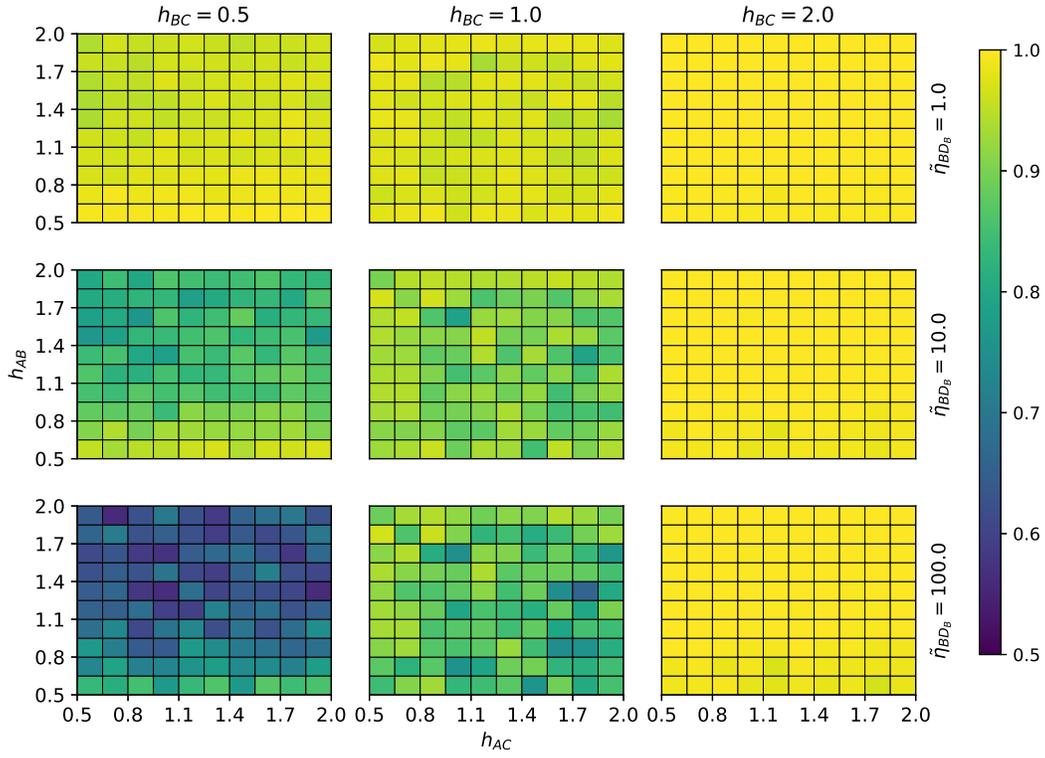

**Figure S6:** Median relative response time of I1-FFL trajectories calculated based on systematically exploring parameter space as described above, assuming $h_{BC} = h_{BD_B}$ and $\tilde{K}_{BC} = \tilde{K}_{BD_B}$. The trajectories are separated evenly by $h_{AB}$ and $h_{AC}$ into 100 voxels, the color of which represents the median relative response time of the 100 simulated trajectories falling into that bin. Here, relative response time is defined as the ratio of the response time of the model to the response time of the model without retroactivity.

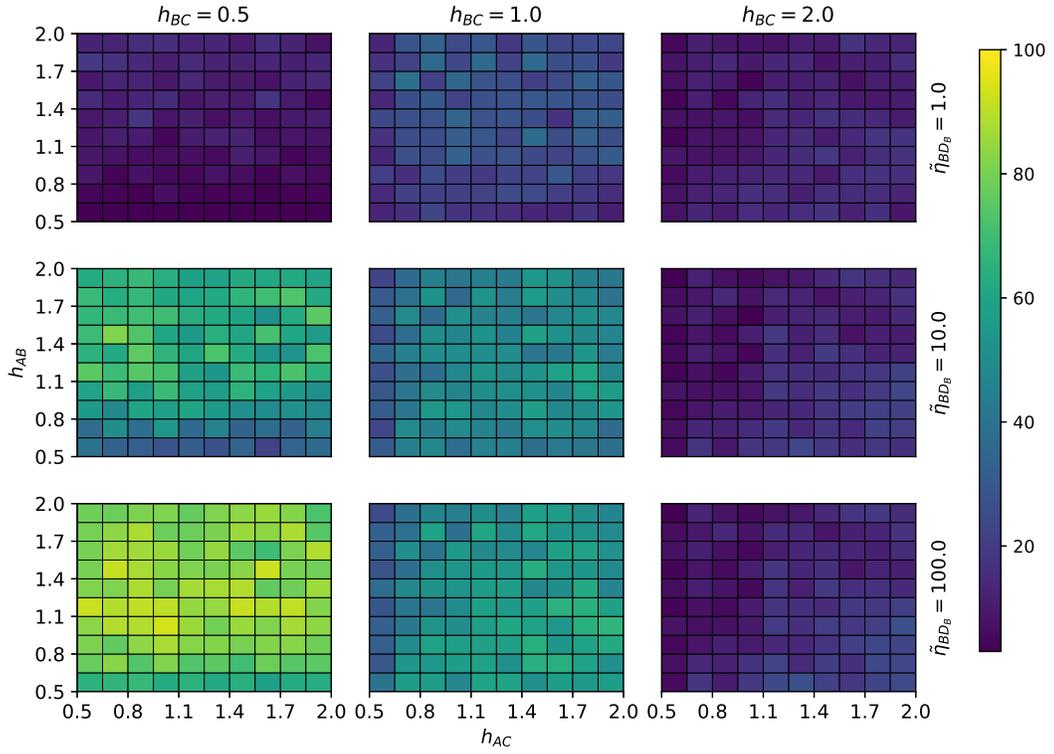

**Figure S7:** Percent of I1-FFL trajectories whose relative response time is less than 90% of the model in the absence of retroactivity calculated based on systematically exploring parameter space as described above, assuming $h_{BC} = h_{BD_B}$ and $\tilde{K}_{BC} = \tilde{K}_{BD_B}$. The trajectories are separated evenly by $h_{AB}$ and $h_{AC}$ into 100 voxels, the color of which represents the percent of trajectories whose relative response time is less than 90% of the model in the absence of retroactivity out of the 100 simulated trajectories falling into that bin. Here, relative response time is defined as the ratio of the response time of the model to the response time of the model without retroactivity.

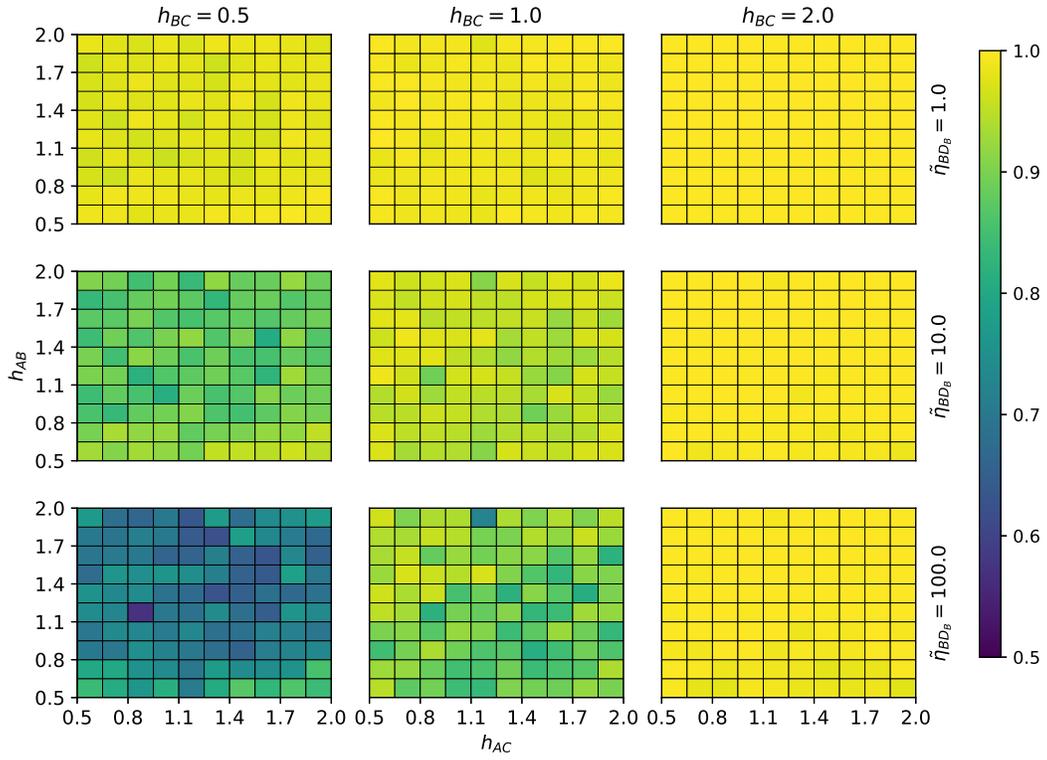

**Figure S8:** Median relative response time of I1-FFL trajectories calculated based on systematically exploring parameter space as described above, assuming $0.5 h_{BC} \leq h_{BD_B} \leq 1.5 h_{BC}$ and $0.5 \tilde{K}_{BC} \leq \tilde{K}_{BD_B} \leq 1.5 \tilde{K}_{BC}$. The trajectories are separated evenly by $h_{AB}$ and $h_{AC}$ into 100 voxels, the color of which represents the median relative response time of the 100 simulated trajectories falling into that bin. Here, relative response time is defined as the ratio of the response time of the model to the response time of the model without retroactivity.

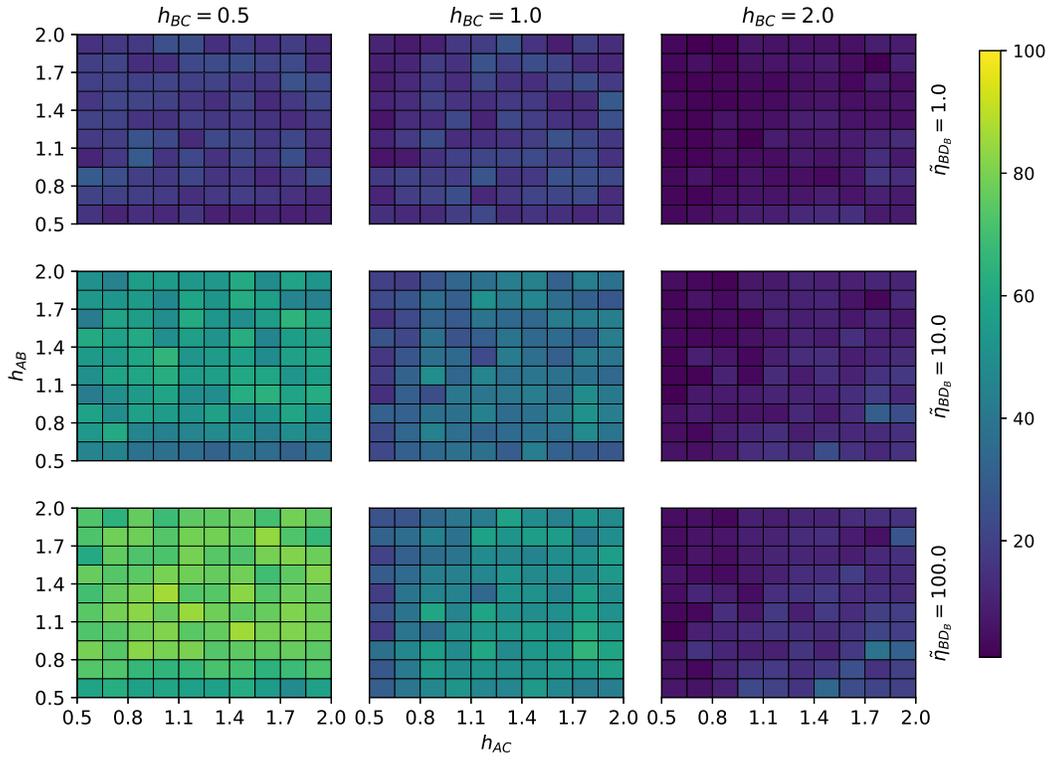

**Figure S9:** Percent of I1-FFL trajectories whose relative response time is less than 90% of the model in the absence of retroactivity calculated based on systematically exploring parameter space as described above, assuming $0.5h_{BC} \leq h_{BD_B} \leq 1.5h_{BC}$ and $0.5\tilde{K}_{BC} \leq \tilde{K}_{BD_B} \leq 1.5\tilde{K}_{BC}$. The trajectories are separated evenly by $h_{AB}$ and $h_{AC}$ into 100 voxels, the color of which represents the percent of trajectories whose relative response time is less than 90% of the model in the absence of retroactivity out of the 100 simulated trajectories falling into that bin. Here, relative response time is defined as the ratio of the response time of the model to the response time of the model without retroactivity.

## 1.20 Simulated Synthetic IFFL

The I2-FFL model shown in Figure 6 is given as:

$$\frac{dx_L}{dt} = f_L = \beta_L \left[ (1-\gamma_L) \frac{\left(\frac{x_I}{K_{IL}}\right)^{h_{IL}}}{1+\left(\frac{x_I}{K_{IL}}\right)^{h_{IL}}} + \gamma_L \right] - \delta_L x_L$$

$$\frac{dx_T}{dt} = f_T = \frac{1}{1 + \eta_{D_T} \frac{h_{TE}^2 x_T^{h_{TE}-1}}{K_{TE}^{h_{TE}}} \left(1+\left(\frac{x_T}{K_{TE}}\right)^{h_{TE}}\right)^{-2}} \left\{ \beta_T \left[ \frac{1-\gamma_T}{1+\left(\frac{x_L}{K_{LT}}\right)^{h_{LT}}} + \gamma_T \right] - \delta_T x_T \right\}$$

$$\frac{dx_E}{dt} = f_E = \beta_E \left[ \frac{1-\gamma_E}{\left(1+\left(\frac{x_L}{K_{LE}}\right)^{h_{LE}}\right)\left(1+\left(\frac{x_T}{K_{TE}}\right)^{h_{TE}}\right)} + \gamma_E \right] - \delta_E x_E,$$

where $L$, $T$, and $E$ represent LmrA, TAL21, and EYFP, respectively. TAL21 is assumed to bind to pUAS-Rep2 (promoter) and $D_T$ (decoy sites) with the same affinity and cooperativity, as they share the same operator binding sites. $\eta_{D_T}$ is the concentration of the decoy sites of TAL21. The kinetic parameters used in the model are taken from Supplementary Figure 3 in Wang et al. (2019): $\beta_L = 5 \times 10^4$ MEFL/hr, $\gamma_L = 1 \times 10^{-4}$, $\delta_L = 9.5$/hr, $\beta_T = 1.66 \times 10^5$ MEFL/hr, $\gamma_T = 2.36 \times 10^{-5}$, $K_{LT} = 3.73 \times 10^5$ MEFL, $h_{LT} = 0.59$, $\delta_T = 1.5$/hr, $\beta_E = 4.68 \times 10^5$ MEFL/hr, $\gamma_E = 1.1 \times 10^{-3}$, $K_{LE} = 3.73 \times 10^5$ MEFL, $h_{LE} = 0.59$, $K_{TE} = 2.9 \times 10^5$ MEFL, $h_{TE} = 0.72$, $\delta_E = 0.3$/hr. Production rates $\beta_X$ ($X = L, T, E$) are ten times the values given in Wang et al. (2019) as the rates given in Wang et al. (2019) represent the cell subpopulation with the lowest production rates. $\delta_X$ ($X = L, T, E$) are increased for the sake of a faster response time. Biologically, this can be achieved by adding degradation tags to the proteins.

## 1.21 Two-node Negative Feedback Loops

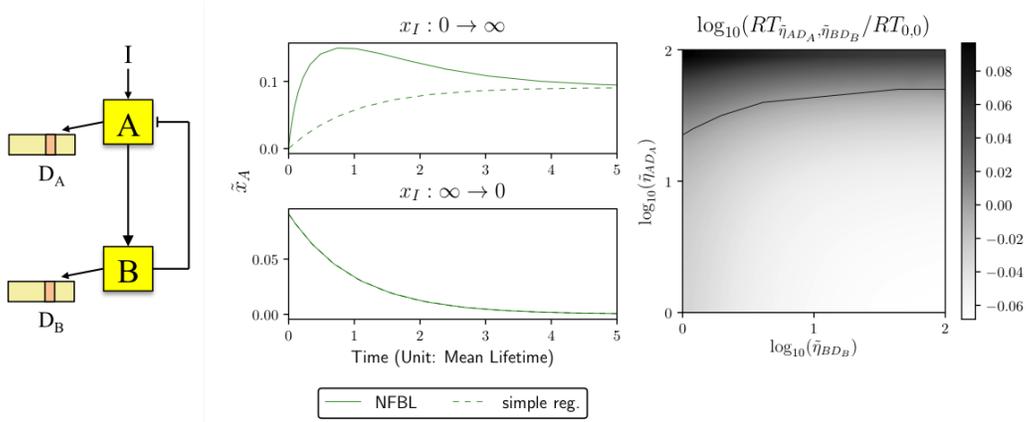

**Figure S10:** Two-node negative feedback loops (NFBLs). Left: diagram. Middle: sign-sensitive response acceleration of a two-node NFBL in the absence of retroactivity, i.e., $\tilde{\eta}_{AD_A} = \tilde{\eta}_{BD_B} = 0$. The response of $\tilde{x}_A$ is accelerated in response to an ON step, not in response to an OFF step. "Simple reg." represents a simple circuit where A is activated by an external inducer I without additional regulation. The "simple reg." model achieves the same steady state as the NFBL model. Right: response times of the two-node NFBL model at different levels of $\tilde{\eta}_{AD_A}$ and $\tilde{\eta}_{BD_B}$ compared to that of the model with no retroactivity in response to an ON step. The black curve, which we refer to as the "iso-response-time" curve, represents values of $\tilde{\eta}_{XD_X}$ ($X = A, B$) at which the response time is the same as the response time of the model with no retroactivity. Values of parameters used for making the middle and the right panels are: $\tilde{K}_{AB} = \tilde{K}_{AD_A} = 0.001$, $\tilde{K}_{BA} = \tilde{K}_{BD_B} = 1.0$, $h_{AB} = h_{AD_A} = h_{BA} = h_{BD_B} = 1.0$.

Model for the two-node NFBL is:

$$\begin{bmatrix} \frac{d\tilde{x}_A}{d\tau} \\ \frac{d\tilde{x}_B}{d\tau} \end{bmatrix} = \begin{bmatrix} \frac{1}{1+r_{AD_A}(\tilde{x}_A)} & 0 \\ 0 & \frac{1}{1+r_{BD_B}(\tilde{x}_B)} \end{bmatrix} \begin{bmatrix} f_{\tilde{A}} \\ f_{\tilde{B}} \end{bmatrix}, \tag{15}$$

$$f_{\tilde{A}} = (1-\gamma_A) \frac{\left(\frac{x_I}{K_{IA}}\right)^{h_{IA}}}{1+\left(\frac{x_I}{K_{IA}}\right)^{h_{IA}}} \frac{1}{1+\left(\frac{\tilde{x}_B}{\tilde{K}_{BA}}\right)^{h_{BA}}} + \gamma_A - \tilde{x}_A$$

$$f_{\tilde{B}} = (1-\gamma_B) \frac{\left(\frac{\tilde{x}_A}{\tilde{K}_{AB}}\right)^{h_{AB}}}{1+\left(\frac{\tilde{x}_A}{\tilde{K}_{AB}}\right)^{h_{AB}}} + \gamma_B - \tilde{x}_B, \tag{16}$$

$$r_{AD_A}(\tilde{x}_A) = \tilde{\eta}_{AD_A} h_{AD_A}^2 \left(\frac{\tilde{x}_A}{\tilde{K}_{AD_A}}\right)^{h_{AD_A}-1} \left(1+\left(\frac{\tilde{x}_A}{\tilde{K}_{AD_A}}\right)^{h_{AD_A}}\right)^{-2}$$

$$r_{BD_B}(\tilde{x}_B) = \tilde{\eta}_{BD_B} h_{BD_B}^2 \left(\frac{\tilde{x}_B}{\tilde{K}_{BD_B}}\right)^{h_{BD_B}-1} \left(1+\left(\frac{\tilde{x}_B}{\tilde{K}_{BD_B}}\right)^{h_{BD_B}}\right)^{-2}. \tag{17}$$

## 1.22 Significance of Motifs

Following the method outlined in Alon (2007), we compared the number of the times an IFFL is observed in real networks to the number of times an IFFL is expected in a randomized network . We began by computing the number of times an IFFL is expected to appear in a randomized ER network. Let $G$ denote a network (graph) consisting of $E$ edges and $N$ nodes. The probability of an edge in a given direction with the correct interaction type between a pair of nodes is (Alon (2007)):

$$p = E/N^2 * k, \tag{18}$$

where $k$ is the probability that a given edge is positive (activation) or negative (inhibition).

According to Alon (2007), the average number of occurrences of an IFFL in the randomized ER network is approximately equal to the number of ways of choosing $n$ nodes out of $N$ times the probability to get $g$ edges with correct interaction types in the correct places:

$$<N_G> = N^n p^g, \tag{19}$$

where both $n$ and $g$ equal 3, since an IFFL contains three nodes and three edges. For convenience of notations, we denote the number of occurrences of an IFFL in real networks by $\hat{N}_G$.

We searched the Regulon database v10.0 (Santos-Zavaleta et al. (2018)) and the TRRUST database v2 (Han et al. (2018)) for TF-gene interactions in the *E. coli* (Regulon), mouse (TRRUST), and human (TTRUST) TRNs. The number of genes (nodes), number of edges (interactions), percentage of activation, percentage of inhibition, and the number of IFFLs are listed in Table S12. Plugging these values into (19), we obtained $<N_G>$.

The comparison between real and randomized networks is shown in Table S12. The number of occurrences of an IFFL in a real *E. coli*, mouse, and human TRN is approximately 118.68, 85.61, and 161.96 times the number of occurrences of an IFFL in a randomized *E. coli*, mouse, and human TRN.

Following the same method as above, we compared the number of real and randomized two-node negative feedback loops (NFBLs) in different organisms. The number of occurrences of a two-node NFBL in a real *E. coli*, mouse, and human TRN is approximately 4.82, 18.02, and 18.44 times the number of occurrences of a two-node NBFL in a randomized *E. coli*, mouse, and human TRN (Table S13).

In addition, we found 86 out of 154 inhibitors in *E. coli* (Regulon) are negatively auto-regulated, whereas only 5 out of 448 inhibitors in mouse (TTRUST) and 4 out of 470 inhibitors in human (TTRUST) are auto-repressors. The number of occurrences of a negative autoregulatory loop in a real *E. coli*, mouse, and human TRN is approximately 104.88, 6.94, and 4.30 times the number of occurrences of a negative autoregulatory loop in a randomized *E. coli*, mouse, and human TRN (Table S14).

|  | *E. coli* | mouse | human |
|---|---|---|---|
| $N$ | 2870 | 1858 | 2072 |
| $E$ | 8149 | 4197 | 5071 |
| $k_+$ | 0.71 | 0.68 | 0.62 |
| $k_-$ | 0.29 | 0.32 | 0.38 |
| $\hat{N}_G$ | 1258 | 470 | 1171 |
| $<N_G>$ | 10.60 | 5.49 | 7.23 |
| $\frac{\hat{N}_G}{<N_G>}$ | 118.68 | 85.61 | 161.96 |

**Table S12:** Number of IFFLs in real and randomized *E. coli*, mouse, and human TRNs.

|   | *E. coli* | mouse | human |
|---|---|---|---|
| $N$ | 2870 | 1858 | 2072 |
| $E$ | 8149 | 4197 | 5071 |
| $k_+$ | 0.71 | 0.68 | 0.62 |
| $k_-$ | 0.29 | 0.32 | 0.38 |
| $\hat{N}_G$ | 8 | 20 | 26 |
| $<N_G>$ | 1.66 | 1.11 | 1.41 |
| $\frac{\hat{N}_G}{<N_G>}$ | 4.82 | 18.02 | 18.44 |

**Table S13:** Number of two-node NFBLs in real and randomized *E. coli*, mouse, and human TRNs.

|   | *E. coli* | mouse | human |
|---|---|---|---|
| $N$ | 2870 | 1858 | 2072 |
| $E$ | 8149 | 4197 | 5071 |
| $k_-$ | 0.29 | 0.32 | 0.38 |
| $\hat{N}_G$ | 86 | 5 | 4 |
| $<N_G>$ | 0.82 | 0.72 | 0.93 |
| $\frac{\hat{N}_G}{<N_G>}$ | 104.88 | 6.94 | 4.30 |

**Table S14:** Number of negative autoregulatory loops in real and randomized *E. coli*, mouse, and human TRNs.



\end{document}